\documentclass[aps,prd,nofootinbib,showpacs,preprintnumbers,longbibliography,amssymb,11pt]{revtex4-1} 
\pdfoutput=1

\usepackage[sort&compress]{natbib}
\usepackage{slashed}

\usepackage{graphicx}
\usepackage{epsfig}
\usepackage{dcolumn}
\usepackage{bm}
\usepackage{amsmath}
\usepackage{color}

\graphicspath{{./Figures/}}

\newcommand{\eqnref}[1]{Eq.~(\ref{eqn:#1})}

\def\slashchar#1{\setbox0=\hbox{$#1$}           
   \dimen0=\wd0                                 
   \setbox1=\hbox{/} \dimen1=\wd1               
   \ifdim\dimen0>\dimen1                        
      \rlap{\hbox to \dimen0{\hfil/\hfil}}      
      #1                                        
   \else                                        
      \rlap{\hbox to \dimen1{\hfil$#1$\hfil}}   
      /                                         
   \fi}

\begin{document}

\preprint{MITP/16-023}

\title{Angular observables for spin discrimination in boosted diboson
  final states}

\author{Malte Buschmann}
\email{buschmann@uni-mainz.de}
\affiliation{PRISMA Cluster of Excellence \& Mainz Institute for
  Theoretical Physics, Johannes Gutenberg University, 55099 Mainz,
  Germany}

\author{Felix Yu}
\email{yu001@uni-mainz.de}
\affiliation{PRISMA Cluster of Excellence \& Mainz Institute for
  Theoretical Physics, Johannes Gutenberg University, 55099 Mainz,
  Germany}


\begin{abstract}
We investigate the prospects for spin determination of a heavy diboson
resonance using angular observables.  Focusing in particular on
boosted fully hadronic final states, we detail both the differences in
signal efficiencies and distortions of differential distributions
resulting from various jet substructure techniques.  We treat the
2~TeV diboson excess as a case study, but our results are generally
applicable to any future discovery in the diboson channel.
Scrutinizing ATLAS and CMS analyses at 8~TeV and 13~TeV, we find that
the specific cuts employed in these analyses have a tremendous impact
on the discrimination power between different signal hypotheses.  We
discuss modified cuts that can offer a significant boost to spin
sensitivity in a post-discovery era.  Even without altered cuts, we
show that CMS, and partly also ATLAS, will be able to distinguish
between spin 0, 1, or 2 new physics diboson resonances at the
$2\sigma$ level with 30~fb$^{-1}$ of 13~TeV data, for our 2~TeV case
study.
\end{abstract}


\maketitle
\tableofcontents

\section{Introduction}
\label{sec:Introduction}
The resumption of the Large Hadron Collider (LHC) with proton-proton
collisions at 13 TeV has reignited the excitement for a possible
discovery of new physics.  The higher energies afforded by the
increase in energy during Run 2 also place additional importance on
the need for robust analysis tools to enable such discoveries in the
hadronic enviroment of the LHC.  One such suite of analysis techniques
is the maturing field of jet substructure~\cite{Butterworth:2008iy,
  Abdesselam:2010pt, Altheimer:2012mn, Altheimer:2013yza,
  Adams:2015hiv}, which take advantage of large Lorentz boosts of
decaying Standard Model (SM) or new physics (NP) particles to reveal
their underlying partonic constituents.  Jet substructure tools are
also invaluable for mitigating pile-up backgrounds at the LHC,
allowing the ATLAS and CMS experiments to use primary vertex
information and jet substructure methods to discard pile-up
contamination of jets resulting from the hard scattering process of
interest~\cite{ATLAS:2012am, Aad:2013gja}.

The special utility of jet substructure techniques as new physics
discovery tools was recently highlighted in the ATLAS 8 TeV search for
electroweak diboson resonances in fully hadronic final
states~\cite{Aad:2015owa}.  In this analysis, ATLAS observed a
2.5$\sigma$ global significance deviation at about 2 TeV in the
reconstructed $WZ$ invariant mass distribution.  The corresponding CMS
8 TeV analysis~\cite{Khachatryan:2014hpa} does not preclude a possible
signal at ATLAS, partly because the two experiments use different
reconstruction methods for tagging boosted, hadronically decaying $W$
and $Z$ candidates.  The most recent 13 TeV results from
ATLAS~\cite{ATLAS-CONF-2015-073} and CMS~\cite{CMS-PAS-EXO-15-002} in
the same fully hadronic diboson decay, however, show no evidence for a
continued excess.

If the excess is a new physics signal, numerous studies are needed to
characterize the resonance and measure the underlying new physics
Lagrangian.  First, for self-consistency, the signal must also begin
to show up in the semi-leptonic and fully leptonic diboson decays.
Observing the excess in these decays is also critical, though, because
the exclusive rates for the semi-leptonic and fully leptonic modes
will help diagnose the underlying $W^+ W^-$ vs.~$W^\pm Z$ vs.~$ZZ$
nature of the purported resonance, which is difficult to disentangle
using only hadronic diboson decays.  Currently, ATLAS has searches for
electroweak diboson resonances with 8 TeV data in the $\ell \nu \ell
\ell$ channel~\cite{Aad:2014pha}, $\ell \ell jj$
channel~\cite{Aad:2014xka}, and the $\ell \nu j j$
channel~\cite{Aad:2015ufa}, which have been combined with the fully
hadronic search in Ref.~\cite{Aad:2015ipg}.  In addition, CMS has
searches with 8 TeV data in the $\ell \nu \ell \ell$
channel~\cite{Khachatryan:2014xja} and $\ell \nu j j$ and $\ell \ell j
j$ channels~\cite{Khachatryan:2014gha}.  We remark, however, that the
2 TeV excess seen by ATLAS in the fully hadronic channel is only
marginally probed by the analyses targetting semi-leptonic diboson
decays, after rescaling the signals that fit the excess by the
appropriate leptonic branching fractions~\cite{Brehmer:2015dan}.

The current situation with 13 TeV data seems to favor the
interpretation that the 2 TeV excess was instead a statistical
fluctuation, although the data is not conclusive.  Both ATLAS and CMS
have retooled their fully hadronic diboson resonance
analyses~\cite{ATLAS-CONF-2015-073, CMS-PAS-EXO-15-002} to focus on
the multi-TeV regime, adopting different jet substructure methods than
those used previously during the 8 TeV run.  CMS and ATLAS also search
in the $\ell \nu jj$ channel~\cite{CMS-PAS-EXO-15-002,
  ATLAS-CONF-2015-075}, respectively, and ATLAS also has performed
analyses in the $\ell \ell jj$ channel~\cite{ATLAS-CONF-2015-071} as
well as the $\nu \nu jj$ channel~\cite{ATLAS-CONF-2015-068}.  Although
the integrated luminosity at 13 TeV is only 3.2 fb$^{-1}$ for ATLAS
and 2.6 fb$^{-1}$ for CMS, in comparison to the 20 fb$^{-1}$ datasets
for each experiment at 8 TeV, naive parton luminosity rescaling from 8
TeV to 13 TeV for the simplest new physics explanations of the 2 TeV
excess point to ATLAS and CMS being at the edge of NP exclusion
sensitivity (see Figure 8 of~\cite{ATLAS-CONF-2015-073}, Figure 4
of~\cite{ATLAS-CONF-2015-075}, Figure 4 of~\cite{ATLAS-CONF-2015-068},
and Figures 9 and 10 of~\cite{CMS-PAS-EXO-15-002}).  

Beyond the self-consistency requirement to observe the diboson excess
in leptonic channels, various new physics models also predict a new
dijet resonance as well as $VH$ resonances, where $V$ is a massive
electroweak boson and $H$ is the Higgs boson~\cite{Dobrescu:2015qna,
  Dobrescu:2015yba, Brehmer:2015cia, Dobrescu:2015jvn,
  Anchordoqui:2015uea}.  The corresponding dijet resonance searches
from ATLAS 8 TeV data~\cite{Aad:2014aqa}, CMS 8 TeV
data~\cite{Khachatryan:2015sja}, ATLAS 13 TeV
data~\cite{ATLAS:2015nsi} and CMS 13 TeV
data~\cite{Khachatryan:2015dcf}, as well as $WH$ and $ZH$ resonance
searches with 8 TeV ATLAS data~\cite{Aad:2015yza}, 8 TeV CMS
data~\cite{Khachatryan:2015ywa, Khachatryan:2015bma,
  Khachatryan:2016yji}, and 13 TeV ATLAS
data~\cite{ATLAS-CONF-2015-074}, have all variously been statistically
consistent with the SM background expectation, which then provide
important model-dependent constraints on new physics interpretations
of the 2 TeV excess.

Given the experimental situation, many papers have delved into the
model-building details and phenomenological questions that reconcile
the original excess with the currently available experimental data.
Spin-0 explanations are discussed in context of a Higgs
singlet~\cite{Chen:2015cfa}, a two Higgs doublet
model~\cite{Chen:2015xql, Omura:2015nwa, Chao:2015eea,
  Sierra:2015zma}, sparticles~\cite{Petersson:2015rza,
  Allanach:2015blv} or composite scalars
\cite{Chiang:2015lqa,Cacciapaglia:2015nga}.  Spin-1 proposals include
composite vector resonances~\cite{Fukano:2015hga, Franzosi:2015zra,
  Thamm:2015csa, Bian:2015ota, Fritzsch:2015aca, Lane:2015fza,
  Low:2015uha, Fukano:2015zua}, generic and effective field theory
(EFT) models~\cite{Cacciapaglia:2015eea, Allanach:2015hba,
  Bian:2015hda, Bhattacherjee:2015svr} as well as heavy $W'$
resonances~\cite{Xue:2015wha, Dobrescu:2015qna, Dobrescu:2015yba,
  Gao:2015irw, Brehmer:2015cia, Heeck:2015qra, Dev:2015pga,
  Deppisch:2015cua, Aydemir:2015nfa, Awasthi:2015ota, Ko:2015uma,
  Collins:2015wua, Dobrescu:2015jvn, Aguilar-Saavedra:2015iew,
  Aydemir:2015oob, Evans:2015cqq, Das:2016akd}, $Z'$
resonances~\cite{Hisano:2015gna, Alves:2015mua, Anchordoqui:2015uea,
  Faraggi:2015iaa, Li:2015yya, Wang:2015sxe, Allanach:2015gkd,
  Feng:2015rzn} or both~\cite{Cheung:2015nha, Cao:2015lia,
  Abe:2015jra, Abe:2015uaa, Fukano:2015uga, Appelquist:2015vdl,
  Das:2015ysz}.  Other NP scenarios include
glueballs~\cite{Sanz:2015zha}, excited composite
objects~\cite{Terazawa:2015bsa}, and in generic and EFT
models~\cite{Aguilar-Saavedra:2015rna, Kim:2015vba, Liew:2015osa,
  Arnan:2015csa, Fichet:2015yia, Sajjad:2015urz}.

Although the new physics situation with 13 TeV data is less attractive
because the initial dataset does not confirm the excess, the
experimental sensitivity with the current luminosity is nonetheless
insufficient to make a final conclusion for the original excess.  Thus
the question about whether the excess is a real signal will simply
have to wait for more integrated luminosity.

Apart from the excitement over the original ATLAS diboson excess,
however, we are motivated to consider how jet substructure techniques
can be used as post-discovery tools for resonance signal
discrimination.  After the Higgs discovery in 2012, the ATLAS and CMS
collaborations began comprehensive Higgs characterization programs,
which aim to measure the couplings, mass, width, spin, parity,
production modes, and decay modes of the Higgs boson.  In particular,
much of the spin and parity information about the 125 GeV Higgs boson
comes from angular correlations in the $h \to 4 \ell$
decay~\cite{Aad:2015mxa, Aad:2015gba, Khachatryan:2014kca,
  Khachatryan:2014jba}, where the Higgs candidate can be fully
reconstructed and all angular observables can be studied.

For the case of a possible 2~TeV resonance $X$, the exact same
analytic formalism for spin characterization used for $h \to 4
\ell$~\cite{Cabibbo:1965zzb, Dell'Aquila:1985ve, Dell'Aquila:1985vc,
  Nelson:1986ki, Gao:2010qx, Bolognesi:2012mm} applies to $X \to VV
\to 4q$~\cite{Kim:2015vba}, which naturally opens up the possibility
of designing a jet substructure analysis that targets spin and
possibly parity characterization of the $X$ resonance.  The $X \to VV
\to 4q$ situation is more difficult, however, because it is a priori
unknown how well the angular correlations in the final state quarks
are preserved after the important effects from showering and
hadronization, detector resolution, jet clustering, and hadronic $W$
and $Z$ boson tagging are included.  In contrast, the $h \to 4 \ell$
decay can be analyzed without the complications from quantum
chromodynamics (QCD) and only need to account for virtual $\gamma^*/Z$
interference and mild detector effects~\cite{Chen:2012jy,
  Chen:2014pia, Chen:2014hqs}.  Our study provides a thorough
investigation of these important and difficult complications, and we
connect distortions in angular observables with specific jet
substructure cuts.  Our results show significant differences between
the ATLAS and CMS 8~TeV and 13~TeV analyses regarding post-discovery
signal discrimination.  They also provide useful templates for
understanding the differences in sensitivity of the current jet
substructure methods to tranversely or longitudinally polarized
electroweak gauge bosons.  We also make projections for how well the
current slate of diboson reconstruction methods will perform with 30
fb$^{-1}$ of LHC 13~TeV integrated luminosity.  The next obvious
course of action would be to design a jet substructure method
optimized for both signal significance and post-discovery spin
discrimination using the extracted subjets.  We leave such work for
the future and instead focus on determining the viability of existing
jet substructure techniques with regards to spin determination.

In Section~\ref{sec:framework}, we review the angular analysis
framework for characterizing a resonance decay.  We also review the
broad classes of jet substructure methods and general challenge of
reconstructing angular correlations in the fully hadronic final state
and the hadronic environment.  In Section~\ref{sec:4qcuts}, we detail
the 2~TeV case study signal benchmarks, review the 8 TeV and 13 TeV
ATLAS and CMS fully hadronic boosted diboson decay selection criteria,
and show the differential distributions after implementing these
analyses.  We also identify specific jet substructure cuts to their
effects on the differential distributions.  We evaluate the
semileptonic analyses in Section~\ref{sec:semileptonic} in a similar
manner, highlighting the new distortions that arise when considering
semileptonic final states.  We present our expectations for model
discrimination with 30 fb$^{-1}$ of LHC 13~TeV data in
Section~\ref{sec:projections} and briefly discuss improvements in jet
substructure analyses targetting signal discrimination.  We conclude
in Section~\ref{sec:conclusion}.  In
Appendix~\ref{app:ATLAS13TeV_bkgd}, we discuss the inclusive
background determination for the ATLAS 13~TeV analysis neeeded in our
13~TeV, 30 fb$^{-1}$ projections.

\section{Reconstructing Angular Correlations in 
$pp \to X$, $X \to V_1 V_2 \to 4q$}
\label{sec:framework}

\subsection{General framework}
\label{subsec:angles_definition}
In this section, we review the general framework for studying angular
correlations of a resonance $X$ decaying to two intermediate vector
bosons that subsequently decay to four light quarks.  We will work in
the $X$ rest frame and orient the incoming partons along the
$+\hat{z}$ and $-\hat{z}$ axes as usual. We also neglect the masses of
our final state particles, which reduces the nominal sixteen final
state four-momentum components to twelve.  Four-momentum conservation
in the rest frame of the resonance further reduces the number of
independent components to eight.  Finally, the overall system can be
freely rotated about the $+\hat{z}$ axis, so we can completely
characterize the kinematics of the system with seven independent
variables, which are five angles and the two intermediate vector
masses.  If the resonance mass is not known, it also counts as an
independent quantity.  Finally, if the final state particles are not
massless, then their four masses also have to be used as independent
variables.

\begin{figure}[tb!]
\begin{center}
\includegraphics[scale=0.1]{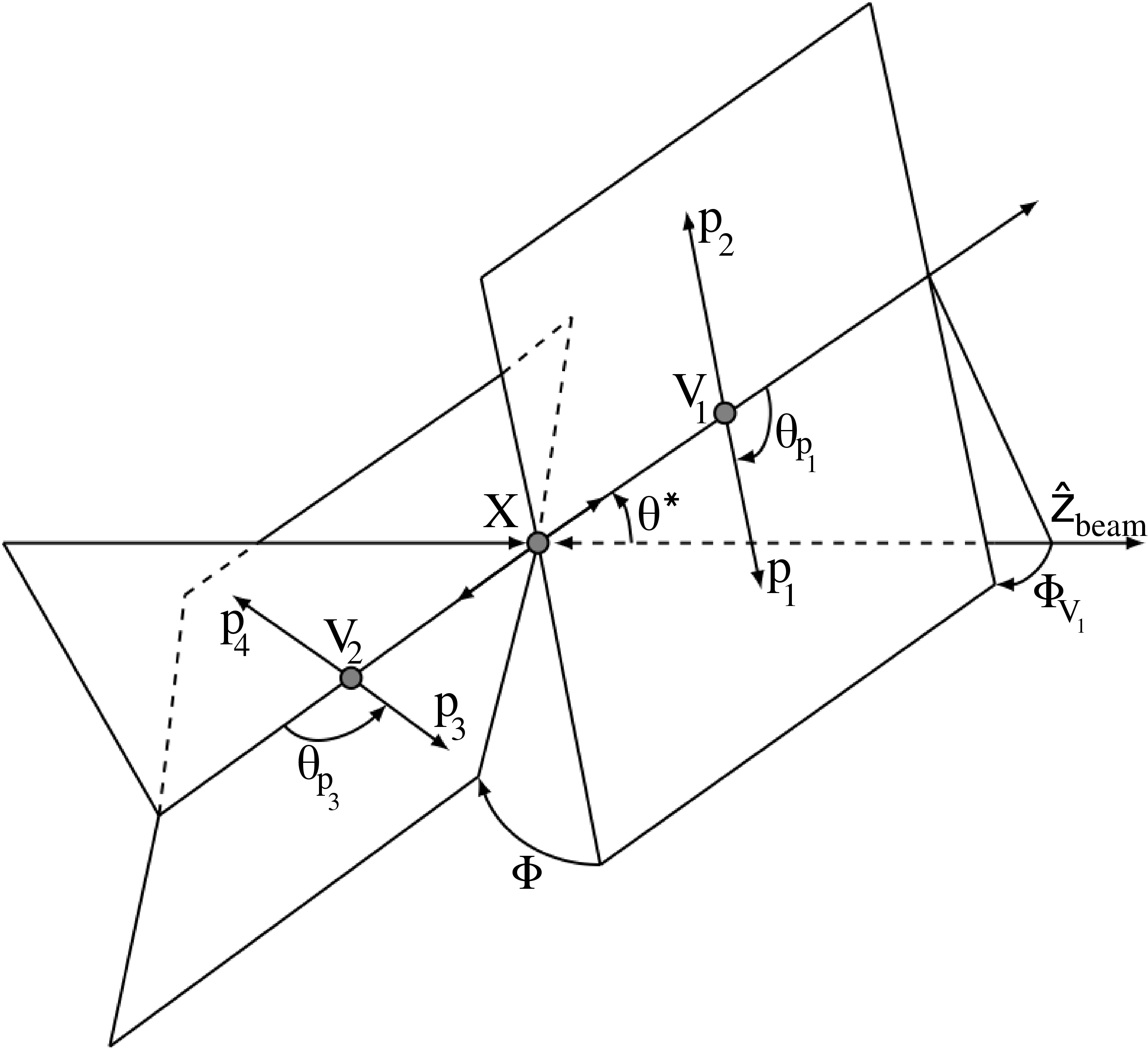}\\
\caption{Representation of the Cabibbo--Maksymowicz--Dell'Aquila--Nelson angles defined in
  Eq.~\ref{eq:Nangles}.}
\label{fig:Nelson}
\end{center}
\end{figure}

The five angles, known as the
Cabibbo--Maksymowicz--Dell'Aquila--Nelson
angles~\cite{Cabibbo:1965zzb, Dell'Aquila:1985ve, Dell'Aquila:1985vc,
  Nelson:1986ki}, the two intermediate vector masses, and the
resonance mass are hence completely sufficient to describe the
kinematics of the $pp \to X \to V_1V_2 \to (p_1 p_2) (p_3 p_4)$.
These angles are shown in Fig.~\ref{fig:Nelson} and are given
by
\begin{align}					
\cos{\theta_{p_1}} &= -\hat{p}_{p_1} \cdot \hat{p}_{V_2} \ , &
\Phi_{V_1} = \frac{ \vec{p}_{V_1} \cdot( \hat{n}_1 \times \hat{n}_\text{sc})}{
\left|\vec{p}_{V_1} \cdot( \hat{n}_1 \times \hat{n}_\text{sc}) \right|} 
\arccos( \hat{n}_1 \cdot \hat{n}_\text{sc}) \ , \notag \\
\cos{ \theta_{p_3}} &= -\hat{p}_{p_3} \cdot \hat{p}_{V_1} \ , &
\Phi = \frac{\vec{p}_{V_1} \cdot( \hat{n}_1 \times \hat{n}_2)}{
\left| \vec{p}_{V_1} \cdot( \hat{n}_1 \times \hat{n}_2) \right|}
\arccos( -\hat{n}_1 \cdot \hat{n}_2) \ , \notag \\
\cos{\theta^*} &= \hat{p}_{V_1} \cdot \hat{z}_{\text{beam}} \ ,
\label{eq:Nangles}
\end{align}
where $V_1$ and $V_2$ are the two bosons, $X$ is the resonance, $\hat
z_\text{beam}$ is the direction of the beam axis and
\begin{align}
\hat{n}_1 = \frac{ \vec{p}_{p_1} \times \vec{p}_{p_2}}{ 
\left| \vec{p}_{p_1} \times \vec{p}_{p_2} \right|} \ , \quad
\hat{n}_2 = \frac{ \vec{p}_{p_3} \times \vec{p}_{p_4}}{
\left| \vec{p}_{p_3} \times \vec{p}_{p_4} \right|} \ , \text{ and } 
\hat{n}_\text{sc} = \frac{ \hat{z}_{\text{beam}} \times \vec{p}_{p_1}}{
\left| \hat{z}_{\text{beam}} \times \vec{p}_{p_1} \right|} \ .
\end{align}
The intermediate vectors $V_1$ and $V_2$ are reconstructed by $p_{V_1}
= p_{p_1} + p_{p_2}$, $p_{V_2} = p_{p_3} + p_{p_4}$, and the resonance
$X$ is formed by $p_X = p_{V_1} + p_{V_2}$.  The angle
$\cos{\theta_{p_1}}$ ($\cos{\theta_{p_3}}$) is calculated with the
respective four-momenta boosted into the rest frame of particle $V_1$
($V_2$), whereas all other angles are computed in the rest frame of
particle $X$.  Additionally, we define the angle $\Psi = \Phi_{V_1} +
\Phi/2$ to supersede $\Phi_{V_1}$, where $\Psi$ is the average
azimuthal angle of the two decay planes.

Resonances with different spins will produce different angular
correlations among the decay products.  A full set of analytic
expressions for different resonance hypotheses and the subsequent
angular correlations in the $X \to V_1 V_2 \to$ 4 fermion final state
can be found in Ref.~\cite{Bolognesi:2012mm}, which we do not
reproduce here.  We have verified the analytic expressions in
Ref.~\cite{Bolognesi:2012mm} by comparing to parton level Monte Carlo
results for different resonant spin hypotheses.  Our full discussion
of Monte Carlo signal samples and analysis of angular correlations
analysis is given in Section~\ref{sec:4qcuts}.

\subsection{Phenomenology of jet substructure}
While the angles defined in Fig.~\ref{fig:Nelson} underpin any
analysis aimed at spin characterization of a given resonance, the
corresponding differential distributions are expected to be smeared
and skewed after accounting for showering and hadronization, detector
resolution effects, jet clustering methods, and jet substructure cuts.
Of these effects, the distortions introduced by jet clustering methods
and jet substructure cuts are the most pernicious.

The usual goal for jet substructure techniques is to isolate the
partonic constituents of a given wide angle jet that captures the
decay products of a boosted parent, like a $W$, $Z$, $h$, or $t$
resonance.  As a result, different methods have been developed to
maximize the tagging efficiency of these parent particles while
simultaneously minimizing the mistag rate from QCD or other
backgrounds~\cite{Altheimer:2012mn, Altheimer:2013yza}.  In this
endeavor, angular observables have played an implicit role to help
improve the overall tagging efficiency of a given parent particle over
the QCD background, but on the other hand, recovering the full phase
space of resonance decay products will be key for post-discovery
signal discrimination.  Moreover, understanding how angular
observables are distorted by jet substructure cuts is also necessary
to optimize signal hypothesis testing in a post-discovery scenario.

To this end, we review the main jet substructure methods to extract 
subjets from fat jets, as well as jet substructure techniques used for 
background discrimination.  Variants of these methods are all used, 
as we will see, in the most recent ATLAS and CMS 8~TeV and 13~TeV
analyses~\cite{Aad:2015owa, Khachatryan:2014hpa, ATLAS-CONF-2015-073,
  CMS-PAS-EXO-15-002}.
  \newline

\paragraph*{\bf Mass-drop filter technique\\}
The jet grooming procedure used in the 8~TeV ATLAS
analysis~\cite{Aad:2015owa} is known as mass-drop
filtering~\cite{Butterworth:2008iy}.  An original fat jet,
reconstructed with the Cambridge-Aachen (C/A) cluster
algorithm~\cite{Dokshitzer:1997in}, is ``unclustered'' in reverse
order.  Each step of the unclustering gives a pair of subjets that is
tested for both mass-drop and momentum balance conditions.  The
procedure is stopped if the two conditions are satisfied.

The mass-drop criterion requires each subjet to satisfy $\mu_i \equiv
m_i / m_0 \leq \mu_f$ for a given parameter $\mu_f$, where $m_i$ is
the subjet mass and $m_0$ is the original jet mass.  The 8~TeV ATLAS
hadronic and semi-leptonic diboson searches use $\mu_f = 1$, which
effectively means no mass-drop cut is applied.

The subjet momentum balance condition imposes a minimum threshold on
the relative $p_T$ and $\Delta R$ of each subjet, according to
\begin{align}
\sqrt{y} = \text{ min}(p_{T_1},\ p_{T_2}) 
\frac{ \Delta R}{m_0} \geq \sqrt{y_\text{min}} \ ,
\label{eqn:y_f}
\end{align}
where $p_{T_i}$ is the transverse momentum of each subjet $j_i$,
$\Delta R = \sqrt{ \left(\Delta \phi \right)^2 + \left( \Delta \eta
  \right)^2}$ is their angular distance, and $\sqrt{y_\text{min}}$ is
a parameter controlling the threshold.  To see how~\eqnref{y_f} acts
as a cut on the subjet momentum balance, we rewrite~\eqnref{y_f} using
\begin{align}
m_0^2 = 2 p_{T_1} p_{T_2} 
\left( \cosh( \Delta \eta) - \cos( \Delta \phi) \right)
\approx p_{T_1} p_{T_2} (\Delta R)^2 \ ,
\label{eqn:m0pTpTdR}
\end{align}
which holds as long as the rapidity difference $\Delta \eta$ and
azimuthal separation $\Delta \phi$ are small.  Using this
approximation, we see that the $\sqrt{y_\text{min}}$ cut is indeed a
subjet momentum balance cut as advertised,
\begin{align}
y \approx \frac{ 
\left\{ \text{ min}(p_{T_1},\ p_{T_2}) \right\}^2 }{p_{T_1} p_{T_2}}
=  \frac{p_{T, \text{ min}}}{p_{T, \text{ max}}} \geq y_\text{min} \ .
\label{eqn:ypTfrac}
\end{align}
At each stage of the unclustering, if the pair of subjets under
consideration satisfies $\sqrt{y} \geq \sqrt{y_{\text{min}}}$, the
procedure terminates and the total four-momentum of the subjets are
used as the $W$ or $Z$ boson candidate.  If the subjets fail the cut,
the softer subjet is discarded and the unclustering procedure
continues.
\newline

\paragraph*{\bf Pruning\\}
In contrast to mass-drop filtering, which recursively compares subjets
to the original fat jet kinematics, the jet pruning
method~\cite{Ellis:2009su, Ellis:2009me}, which is used in the 8~TeV CMS
analysis~\cite{Khachatryan:2014hpa}, tests each stage of the
reclustering for sufficient {\it hardness} and discards soft
recombinations.  In this way, each stage of the reclustering offers an
opportunity to remove constituents from the final jet, instead of
simply incorporating the soft contamination into the widest subjets.

Concretely, in the jet pruning method, the constituents of a fat jet
are reclustered using the C/A algorithm if they are sufficiently
balanced in transverse momentum and sufficiently close in $\Delta R$.
The transverse momentum balance condition is dictated by a minimum
requirement on the {\it hardness} $z$, defined by
\begin{align}
  z = \text{ min} \left( \frac{p_{T_i}}{p_{T_p}}, \frac{p_{T_j}}{p_{T_p}} 
\right) \ ,
\end{align}
where $p_{T_p}$ is the sum of the tranverse momentum of the psuedojets
$i$ and $j$.  Note that $z$ is related the momentum fraction $y$
from~\eqnref{ypTfrac} via
\begin{align}
  y \approx \frac{p_{T, \text{ min}}}{p_{T, \text{ max}}} = \frac{z}{1 - z} \ .
\end{align}
In addition to having sufficient hardness, the two pseudojets must also
be closer in $\Delta R$ than a parameter $D_\text{cut}$, given by
\begin{align}
  \Delta R_{ij} > D_\text{cut} \equiv \frac{m_\text{orig}}{p_{T, \text{ orig}}} \ ,
\end{align}
where $m_{\text{orig}}$ and $p_{T, \text{ orig}}$ are the invariant
mass and transverse momentum of the original fat jet.  If either the
hardness or the $\Delta R$ cut fails, then the softer $p_T$ pseudojet
is discarded.  The C/A reclustering procedure continues until all the
constituents of the original fat jet are included or discarded.
\newline

\paragraph*{\bf $N$-subjettiness\\}
The $N$-subjettiness variable~\cite{Thaler:2010tr, Thaler:2011gf} is
used by CMS in their 8 TeV and 13 TeV
analyses~\cite{Khachatryan:2014hpa, CMS-PAS-EXO-15-002} to help
suppress QCD multi-jet backgrounds and improve selection of hadronic
$W$ and $Z$ candidates.  The $N$-subjettiness is defined as
\begin{align}
 \tau_N = \frac{1}{d_0} \sum \limits_k p_{T_k} \text{ min} 
(\Delta R_{1,k}, \ldots, \Delta R_{N,k}) \ ,
\end{align}
where $p_{T_k}$ is the transverse momentum of the $k$th constituent of
the original jet and $\Delta R_{n, k}$ is the angular distance to the
$n$th subjet axis.  The set of $N$ subjets is determined by
reclustering all jet constituents of the unpruned jet with the $k_T$
algorithm and halting the reclustering when $N$ distinguishable
pseudojets are formed.  Here, $d_0 \equiv \sum_k p_{T_k} R_0$ is a
normalization factor for $\tau_N$, where $R_0$ is the cone size of the
original fat jet.  For the boosted hadronic $W$ and $Z$ analyses, the
ratio $\tau_{21} = \tau_2 / \tau_1$ is computed, where the signal $W$
and $Z$ candidates tend toward lower $\tau_{21}$ values, whereas the
QCD background peaks at higher values.
\newline

\paragraph*{\bf Trimming\\}
The 13 TeV ATLAS analysis~\cite{ATLAS-CONF-2015-073} was reoptimized
for multi-TeV scale diboson sensitivity and adopts the trimming
procedure~\cite{Krohn:2009th} instead of the earlier mass-drop
filtering technique.  Trimming takes a large radius fat jet and
reclusters the constituents with the $k_T$ cluster
algorithm~\cite{Catani:1993hr} using distance parameter $R = 0.2$.  Of
the resulting set of subjets, those kept must satisfy
\begin{align}
\frac{p_{T_j}}{p_{T_J}} > z_\text{min} \ ,
\end{align}
where $j$ denotes the subjet and $J$ the original fat jet.  The
four-momentum sum of all remaining subjets is used as a $W$ or $Z$
candidate.  For an ideal $W$ or $Z$ decay, with exactly two final
subjets, the above condition translate directly to the same balance
criteria as the filtering technique, 
\begin{align}
  y \approx \frac{p_{T, \text{ min}}}{p_{T, \text{ max}}} \geq y_{\text{min}} = 
\frac{z_{\text{min}}}{1 - z_{\text{min}}} \ .
\end{align}
Note, however, that this algorithm does not consider pairs of subjets
as the pruning or filtering techniques do.  Thus, it is possible to
obtain more than two subjets and hence additional cuts, such as energy
correlation function cuts, are needed to determine whether the trimmed
jet has a two-prong substructure.
\newline

\paragraph*{\bf Energy Correlation Functions\\}
The ATLAS 13 TeV analysis~\cite{ATLAS-CONF-2015-073} uses energy
correlation functions~\cite{Larkoski:2013eya, Larkoski:2014gra,
  Larkoski:2015kga} to characterize the number of hard subjets in
their set of trimmed jets.  The relevant 1-point, 2-point and 3-point
energy correlation functions are
\begin{align}
  e_1^{(\beta)}&=\sum\limits_{1\leq i\leq n_J}
		  p_{T_i} \ ,\notag\\
  e_2^{(\beta)}&=\sum\limits_{1\leq i<j\leq n_J}
		  p_{T_i}p_{T_j}\Delta R_{ij}^\beta \ ,\notag\\
  e_3^{(\beta)}&=\sum\limits_{1\leq i<j<k\leq n_J}
		  p_{T_i}p_{T_j}p_{T_k}
                  \Delta R_{ij}^\beta\Delta R_{ik}^\beta\Delta R_{jk}^\beta \ ,
\end{align}
where the sums are performed over jet constituents and $\beta$ is a
parameter weighting the angular separations of constituents against
their $p_T$ fractions.  Since the sums are performed over jet
constituents, the energy correlation functions are independent of any
jet algorithm.  An upper limit is set on the ratio of the function
\begin{align}
 D_2^{(\beta)} = \frac{e_3^{(\beta)} \left( e_1^{(\beta)} \right)^3}{ 
\left( e_2^{(\beta)} \right)^3} \ ,
\end{align}
where the ATLAS collaboration uses $\beta = 1$ in their 13 TeV
analysis.

\section{Angular observables in the $4q$ final state: the 2~TeV case study}
\label{sec:4qcuts}

\subsection{Signal benchmarks}
\label{subsec:signals}
We consider spin-0, spin-1 $W'$, spin-1 $Z'$, spin-1 $W_R$, and spin-2
new physics resonances as possible candidates for the 2 TeV excess
from the ATLAS 8 TeV analysis~\cite{Aad:2015owa}.  The spin-0
possibility is an ad-hoc real scalar model built from the Universal
FeynRules Output~\cite{Alloul:2013bka} implementation of the SM Higgs
effective couplings to gluons in \textsc{MadGraph
  v.1.5.14}~\cite{Alwall:2011uj}, and is included only as an example
of a heavy real scalar that couples dominantly to longitudinal vector
bosons.  The spin-1 $W'$ and spin-1 $Z'$ possibilities are based on
the Heavy Vector Triplet model~\cite{Pappadopulo:2014qza,
  Thamm:2015csa}, whose phenomenology related to the ATLAS 2 TeV
diboson excess was described in detail in Ref.~\cite{Thamm:2015csa}.
The spin-1 $W_R$ explanation is taken from the UFO model files that
accompany Ref.~\cite{Dobrescu:2015jvn}.  The spin-2 heavy graviton
resonance is adapted from a Randall-Sundrum
scenario~\cite{Randall:1999ee, Randall:1999vf} as a \textsc{MadGraph}
model file implementation~\cite{Hagiwara:2008jb}.

Each of these signal possibilities is generated as an on-shell
resonance in \textsc{MadGraph} with subsequent decays to massive
electroweak diboson and then final state SM fermions.  These parton
level events are then showered and hadronized with \textsc{Pythia
  v.8.2}~\cite{Sjostrand:2014zea}, processed through \textsc{Delphes
  v.3.1}~\cite{deFavereau:2013fsa} for detector simulation, and
clustered into jets using the \textsc{FastJet
  v.3.1.0}~\cite{Cacciari:2011ma} as each ATLAS or CMS analysis
requires.  Because \textsc{Delphes} does not include parametrized
detector simulation of jet constituents, which are the basis for
studying jet substructure and angular correlations between subjets, we
also post-process the jet constituents to smear their $p_T$, $\phi$,
and $\eta$ to mimic detector resolution effects: the constituent
smearing parameters are rescaled by the respective energy fraction of
the constituent compared to the full jet.

We simulate QCD dijet background with \textsc{Pythia
  v.8.2}~\cite{Sjostrand:2014zea}.  The subsequent event evolution is
the same as described above.

\subsection{ATLAS and CMS analysis cuts at 8 TeV and 13 TeV}
We recast the ATLAS and CMS searches for heavy resonances with
hadronic diboson decays at 8 TeV~\cite{Aad:2015owa,
  Khachatryan:2014hpa} and 13 TeV~\cite{ATLAS-CONF-2015-073,
  CMS-PAS-EXO-15-002}. As the angular correlations in term of the
parameterization of Sec.~\ref{sec:framework} are skewed by the actual
analyses, we briefly summarize the basic selection criteria for the
different searches.
\newline

\paragraph*{\bf $4q$ Final State by ATLAS at 8 TeV\\}
In the fully hadronic ATLAS search for diboson resonances at 8 TeV,
jets are clustered with the C/A algorithm with radius $R = 1.2$, and
events must have two jets with $p_T > 20$ GeV and $|\eta| < 2.0$.  If
there are electrons with $E_T > 20$ GeV and $|\eta| < 1.37$ or $1.52 <
|\eta| < 2.47$, or if there are muons with $p_T > 20$ GeV and $|\eta|
< 2.5$, the event is vetoed.  Events must also have $\slashed{E}_T <
350$ GeV.  

The two fat jets are then filtered with $y_{\text{min}} = 0.04$.  The
constituents of the two subjets of the groomed jet are then
reclustered again with the C/A algorithm but with a smaller cone size
of $R = 0.3$. The up to three highest-$p_T$ jets, which we will call
{\it filtered jets}, are used to reconstruct the $W$ or $Z$ boson
candidate.  Having reconstructed the ungroomed, groomed, and filtered
jets, further event selection cuts are applied.  The rapidity
difference between the ungroomed jets must satisfy $|y_{J_1}-y_{J_2}|
< 1.2$.  Additionally, the $p_T$ asymmetry of ungroomed jets must be
small, $\left( p_{T,\ J_1} - p_{T,\ J_2} \right) / \left( p_{T,\ J_1}
+ p_{T,\ J_2} \right) < 0.15$.  The ungroomed and corresponding
groomed and filtered jets are tagged as a $W$ or $Z$ boson if they
fulfill the following three criteria:
\begin{itemize}
 \item The pair of subjets of the groomed jet must satisfy a stronger
   transverse momentum balance requirement, $y \geq y_\text{min} =
   0.2025$.
 \item The number of charged tracks associated to the ungroomed jet
   has to be less than $n_\text{trk} < 30$.  Only well-reconstructed
   tracks with $p_T \geq 500$~MeV are used.
 \item The $W$ or $Z$ boson candidates, reconstructed from the
   filtered jets, are finally tagged as a $W$ and/or $Z$, if their
   invariant mass fulfills $|m_J - m_V| < 13$~GeV. Here, $m_V$ is
   either 82.4~GeV for a $W$ boson or 92.8~GeV for a $Z$ boson, as
   determined ATLAS full simulation.
\end{itemize}
Finally, the event is required to have the two highest-$p_T$ jets be
boson-tagged and $m_{JJ} > 1.05$~TeV.
\newline

\paragraph*{\bf $4q$ Final State by CMS at 8 TeV\\}
The CMS 8~TeV analysis uses jet pruning to reconstruct a diboson
resonance.  Jets are reconstructed with the C/A algorithm using $R =
0.8$, and events must have at least two jets with $p_T > 30$~GeV and
$|\eta| < 2.5$, where the two leading jets must satisfy $|\Delta \eta|
< 1.3$ and $m_{JJ} > 890$~GeV.  The two jets are pruned with
$z_\text{min} = 0.1$ (roughly equivalent to $y_\text{min} = 0.11$) and
the corresponding $W$/$Z$ candidate must satisfy $70$~GeV $< m_J <$
100~GeV.  Jets are further categorized according to their purity using
the $N$-subjettiness ratio $\tau_{21}$, where high-purity $W$/$Z$
candidates have $\tau_{21} < 0.5$ and low-purity $W$/$Z$ candidates
have $0.5 < \tau_{21} < 0.75$.  The diboson resonance search requires
at least one high-purity $W$/$Z$ jet, and the second $W$/$Z$ can be
either high- or low-purity.
\newline

\paragraph*{\bf $4q$ Final State by ATLAS at 13 TeV\\}
In ATLAS 13 TeV analysis, events are again vetoed if they contain
electrons or muons with $p_T > 25$~GeV and $|\eta| < 2.5$, and events
must have $\slashed{E}_T < 250$ GeV.  In contrast to earlier, though,
jets are now clustered using the anti-$k_T$ cluster
algorithm~\cite{Cacciari:2008gp} with $R = 1.0$, and events must have
two fat jets with $p_T > 200$~GeV, $|\eta| < 2.0$ and $m_J > 50$~GeV.
The leading jet must have $p_T > 450$ GeV, the invariant mass of the
two fat jets must lie between 1~TeV and 2.5~TeV, and the rapidity
separation must be small, $|y_{J_1} - y_{J_2}| < 1.2$.  Furthermore,
the leading two jets must also have a small $p_T$ asymmetry, $\left(
p_{T,\ J_1} - p_{T,\ J_2} \right) / \left( p_{T,\ J_1} + p_{T,\ J_2}
\right) < 0.15$.

Jets are then trimmed, instead of filtered, by reclustering with the
$k_T$ algorithm using $R = 0.2$ and using hardness parameter
$z_\text{min} = 0.05$, and the energy correlation functions for
$D_2^{(\beta = 1)}$ are then calculated on the trimmed jets to help
distinguish $W$ bosons, $Z$ bosons, and multijet background.  The
upper limit on $D_2$ varies for $W$ and $Z$ candidates as well as the
$p_T$ of the trimmed jet: to implement this $D_2$ cut, we linearly
interpolate between the two cut values, $D_2 = 1.0$ at $p_T = 250$ GeV
and $D_2 = 1.8$ at $p_T = 1500$ GeV, quoted in their analysis.  The
trimmed jets are tagged as bosons if they fulfill two final criteria:
$N_\text{trk} < 30$ for charged-particle tracks associated with the
ungroomed jet and $|m_{J}-m_V| < 15$~GeV, where $m_V = 84$~GeV for a
$W$ boson and $m_V = 96$~GeV for a $Z$ boson.
\newline

\paragraph*{\bf $4q$ Final State by CMS at 13 TeV\\}
The CMS 13 TeV analysis shares many of the same selection criteria as
their 8 TeV analysis, with the following adjustments.  The two
anti-$k_T$, $R = 0.8$, $p_T > 30$ GeV jets must now lie within $|\eta|
< 2.4$.  The pseudorapidity separation between the two jets must again
satisfy $|\Delta \eta| < 1.3$, and the minimum invariant mass cut on
$m_{JJ}$ is raised to 1~TeV.  The two jets are again pruned with
$z_{\text{min}} = 0.1$ and the pruned jet mass window is widened,
allowing 65~GeV$< m_{W/Z} <$ 105~GeV.  Finally, the $N$-subjettiness
ratio $\tau_{21}$ is again calculated, where high-purity $W/Z$ jets
must have a slightly harder requirement, $\tau_{21} \leq 0.45$, and
low-purity $W/Z$ jets satisfy $0.45 < \tau_{21} < 0.75$.  The event
must have at least one high-purity $W/Z$ jet and is classified as
high-purity or low-purity according to the second jet.

\subsection{Analysis effects and reconstruction}
We implement the fully hadronic ATLAS and CMS diboson searches on the
signal samples presented in Sec.~\ref{subsec:signals}, and we extract
the angular observables reviewed in
Sec.~\ref{subsec:angles_definition} from the subjets of the
reconstructed $W/Z$-tagged boson.  Since $W/Z$ discrimination is very
difficult in this final state, we merge the $\cos \theta_{p_1}$ and
$\cos \theta_{p_3}$ distributions into a single differential
distribution labeled $\cos \theta_q$ and do not differentiate between
$W$ and $Z$ candidates.  We also recognize that these analyses do not
attempt to distinguish quarks from anti-quarks, hence we randomly
assign the $p_1$ and $p_2$ labels (or $p_3$ and $p_4$ labels) to
subjets of a given $W/Z$ candidate, which renders the signs of
different angles ambiguous.  Finally, we merge the high-purity and
low-purity tagged events in the CMS analyses to ensure our angular
sensitivity analysis has reasonable statistics.

We find that of the angles defined in Sec.~\ref{sec:framework}, the
main discrimination power between different spin scearios comes from
$\cos{\theta^*}$, $\cos{\theta_{q}}$ and $\Psi$.  In the remainder of
this Section, we will present the individual differential shapes for
the different Monte Carlo samples and experimental studies, and
explain how they are skewed by the respective event selection and jet
substructure cuts.  All of our figures show both parton and
reconstruction level unit-normalized distributions for the different
signal samples and QCD multijet background, where all showering,
hadronization, detector resolution, jet reconstruction, and
substructure analysis effects have been included in the reconstructed
differential distributions.
\newline

\begin{figure}[tb]
\begin{center}
\includegraphics[scale=0.36]{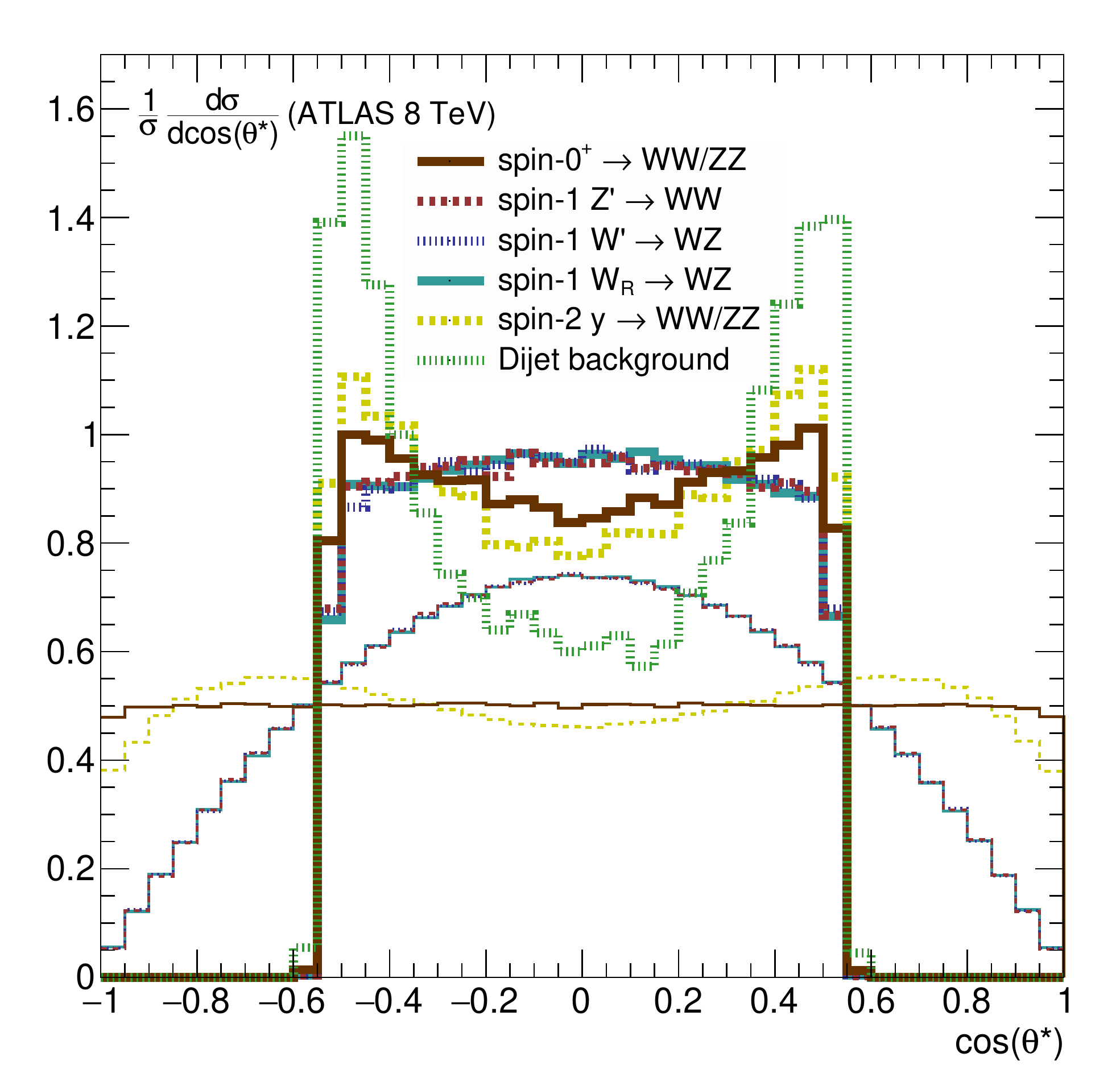}
\includegraphics[scale=0.36]{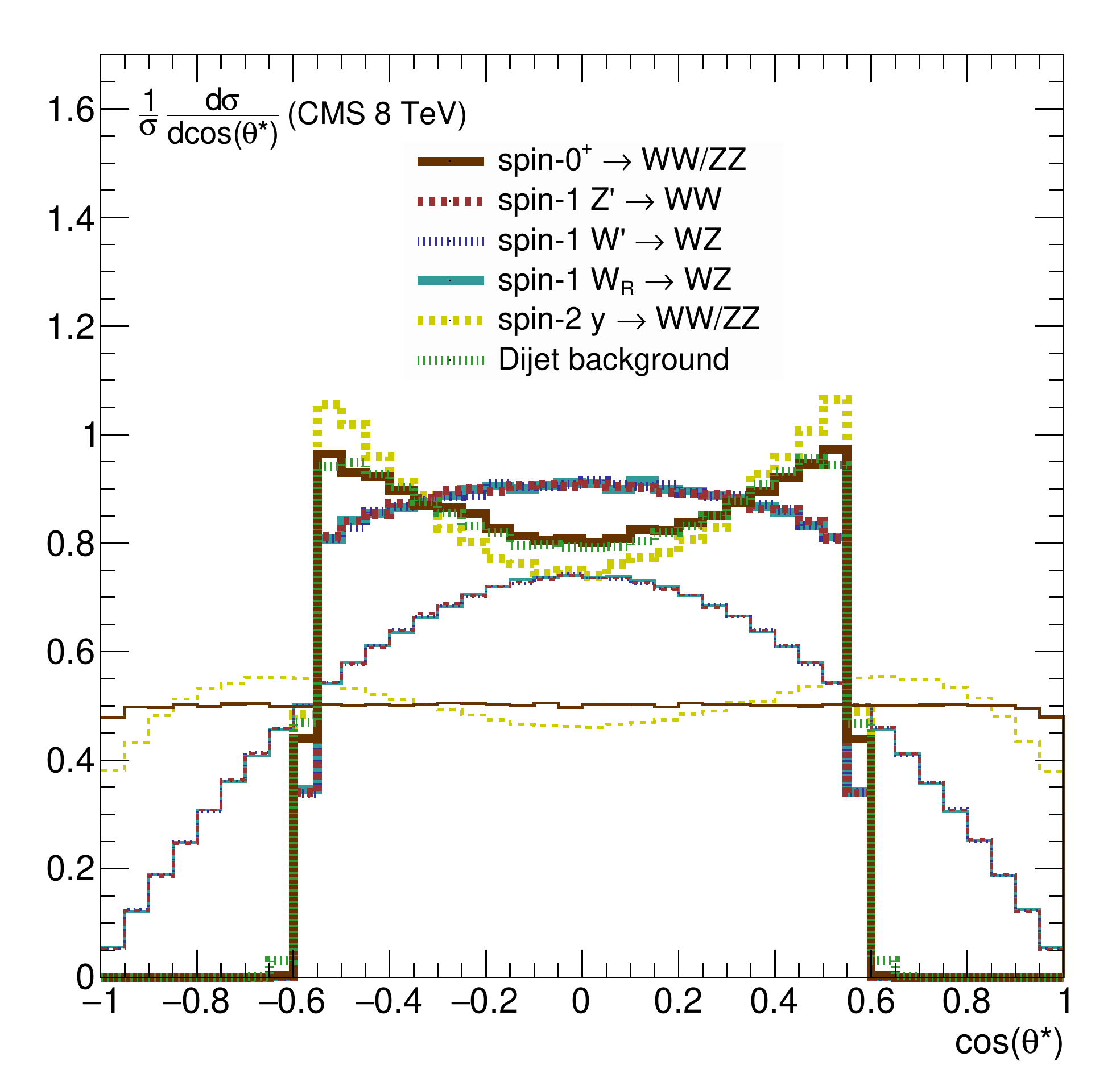}\\
\includegraphics[scale=0.36]{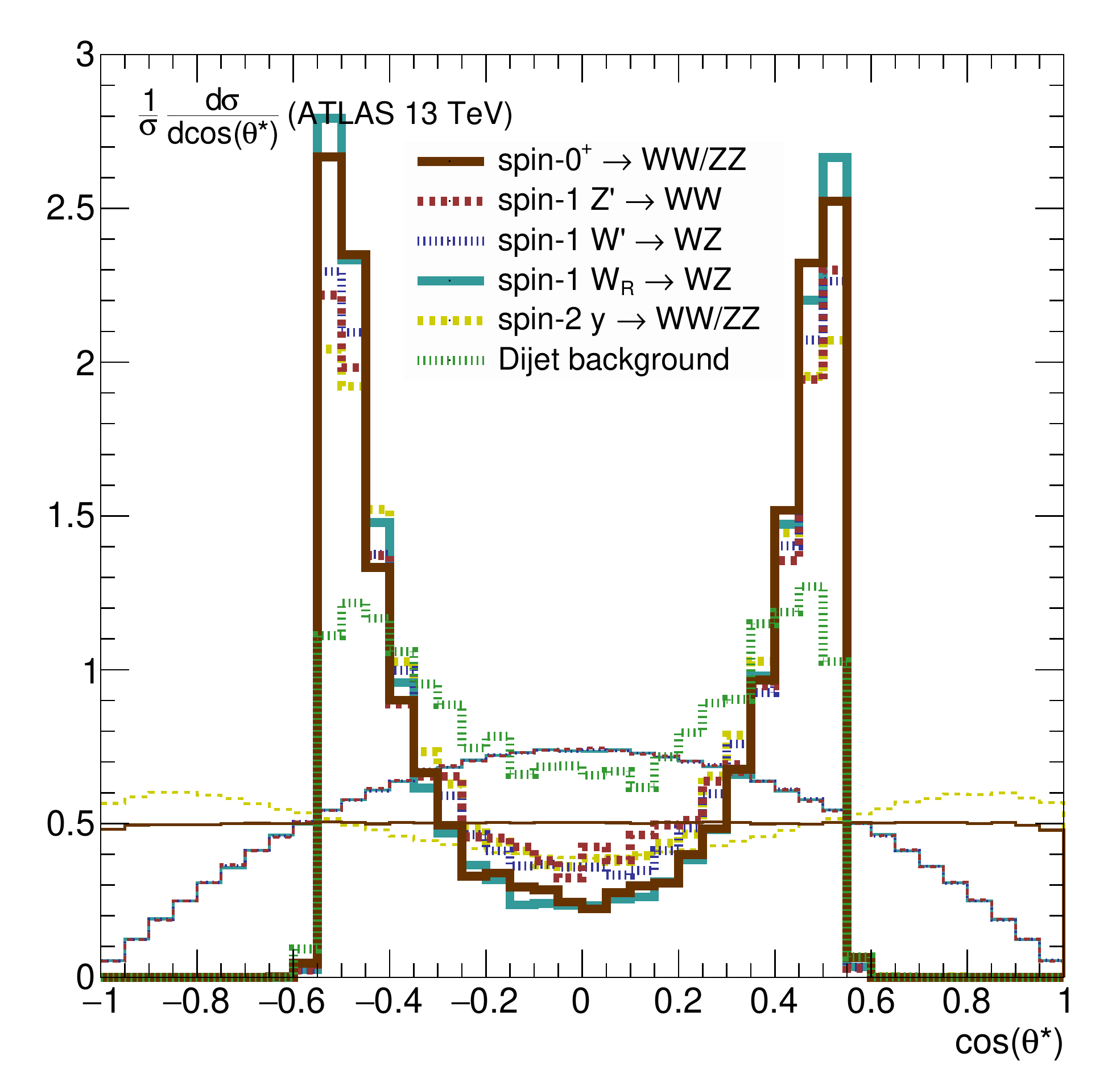}
\includegraphics[scale=0.36]{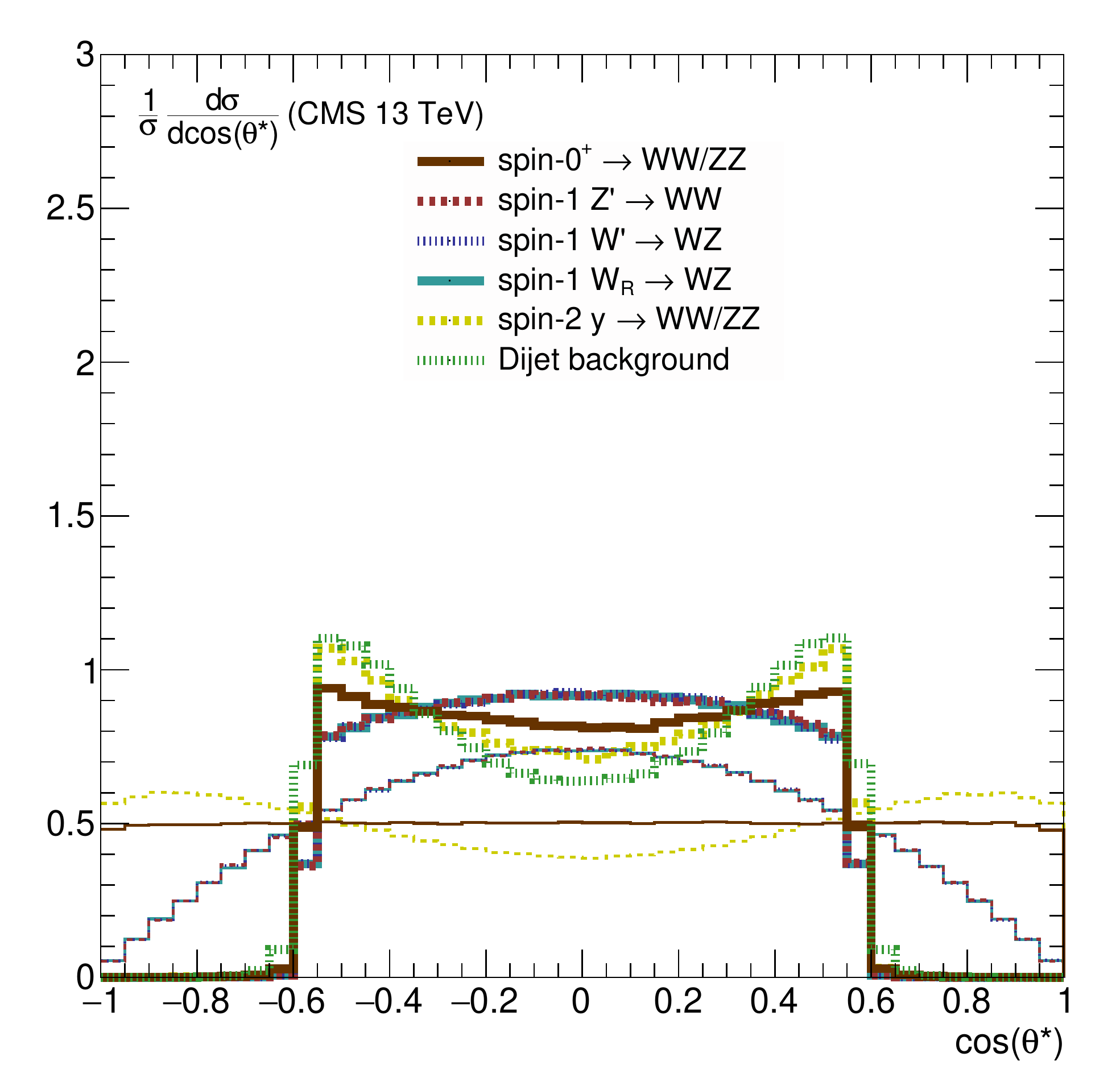}
\caption{Comparison of the $\cos{\theta^*}$ angle between MC parton
  level results (thin lines) and reconstruction of showered events via
  jet substructure (thick lines) for the ATLAS (left) and CMS (right)
  hadronic diboson search at 8 TeV (top) and 13 TeV (bottom).  Each
  distribution is unit-normalized.}
\label{fig:MCRecon_CosStar}
\end{center}
\end{figure}

\paragraph*{\bf Differential Shape of $\cos{\theta^*}$\\}
We first show the angular observable $\cos{\theta^*}$ in
Fig.~\ref{fig:MCRecon_CosStar} for various spin-0, spin-1, and spin-2
signal benchmarks and the QCD background after implementing the ATLAS
8~TeV (upper left), CMS 8~TeV (upper right), ATLAS 13~TeV (lower
left), and CMS 13~TeV (lower right) analyses.  This angle measures the
alignment of the vector bosons from $X$ decay with the beam axis, if
we use the threshold approximation to identify the $X$ rest frame with
the lab frame.  We see significant discrimination power at parton
level (thin lines) between the different signal benchmarks, especially
between the spin-0 and spin-2 signals compared to the spin-1
benchmark.  The extra oscillations in the spin-2 signal, however, are
lost when comparing the reconstruction level (thick lines)
distributions, leaving only the overal concavity of the spin-1
distribution the main discriminant from the spin-0, spin-2, and QCD
background shapes.  Comparing parton level to reconstruction level
results for each signal sample, we see the experimental analyses cause
significant hard cuts in $\cos{\theta^*}$, effectively requiring
$|\cos{\theta^*}| \lesssim 0.55$ for ATLAS and $|\cos{\theta^*}|
\lesssim 0.6$ for CMS, and we also see a deficit of events around
$\cos{\theta^*} \approx 0$ is induced by each analysis, most notably
in the ATLAS 13~TeV analysis.

We can identify the sharp cliffs in the $|\cos{\theta^*}|$
distribution with the cut on the maximum pseudorapidity difference
$|\Delta \eta|$ between the two fat jets, since the $\theta^*$ angle
is directly related to the pseudorapidity via $\eta = -\log \tan(
\theta/2 )$.  Therefore $\cos{\theta^*}$ can be rewritten in the $X$
rest frame as
\begin{align}
|\cos{\theta^*}| = \cos \left( 2\arctan
      e^{-\frac{|\Delta\eta|}{2}} \right)
      = \tanh{ \frac{|\Delta\eta|}{2} }
      \leq \tanh{ \frac{|\Delta\eta_\text{max}|}{2} } \ .
\label{eq:cosstar}
\end{align}
Given $|\Delta y_\text{max}| = 1.2$ at ATLAS and
$|\Delta\eta_\text{max}| = 1.3$ at CMS, and since differences in
pseudorapidity are invariant under longitudinal boosts, we therefore
expect sharp cuts at $|\cos{\theta^*}| \approx 0.54$ and $0.57$,
respectively, where the steepness of the cliff is only spoiled by the
net transverse momentum of the $X$ resonance in the lab frame.

\begin{figure}[tb]
\begin{center}
\includegraphics[scale=0.4]{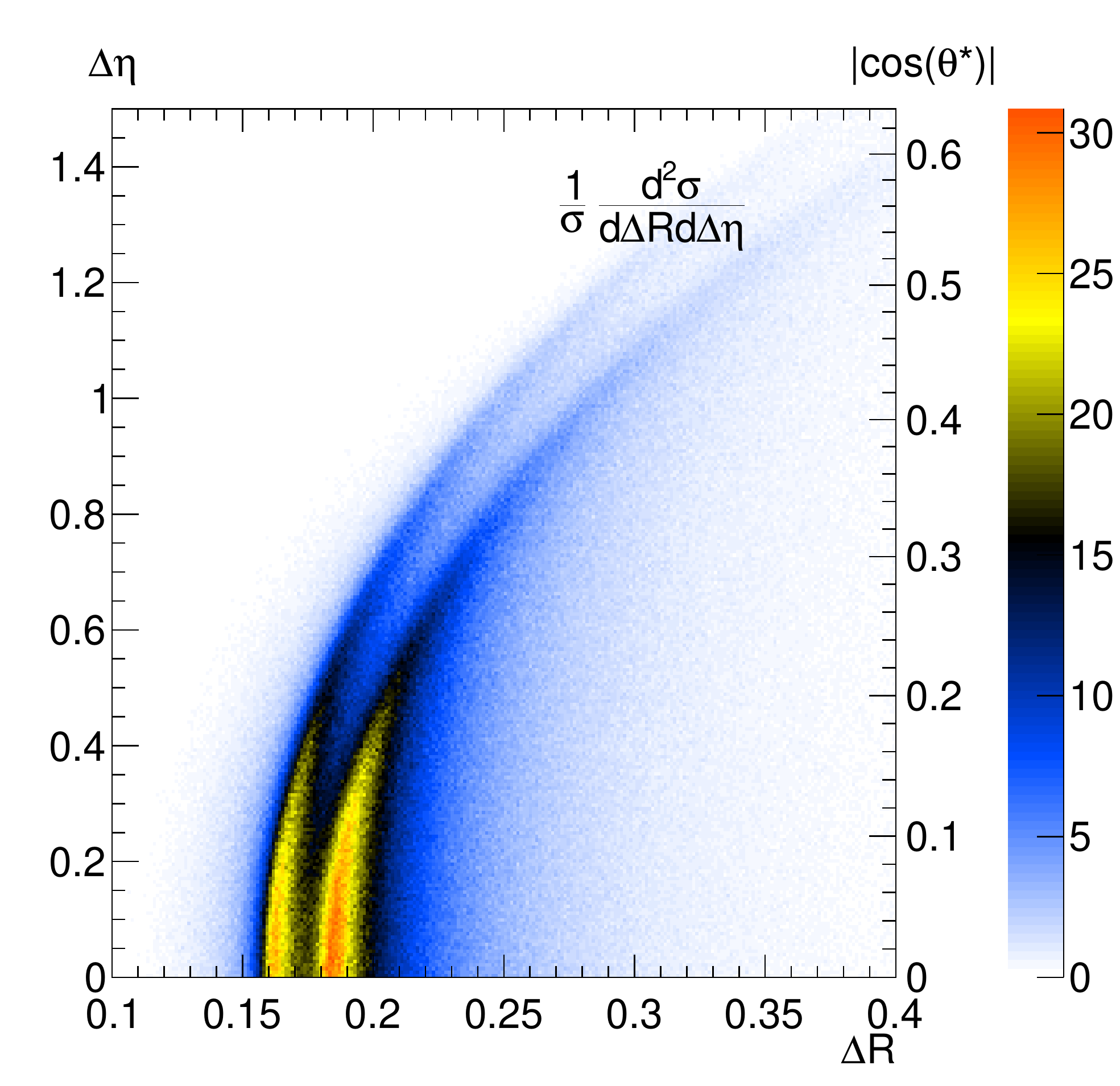}
\caption{Spin-1 $W'$ parton level correlation of the angular
  separation $\Delta R$ between the $W/Z$ decay products and the
  rapidity difference $\Delta \eta$ of the two $W' \to WZ$ fat jets,
  where the left band shows the $W$ decay products and the right band
  shows the $Z$ decay products, and the shading shows the relative
  event weight.  This correlation holds also for other spin
  scenarios. The $\Delta\eta$ axis is translated to a
  $|\cos{\theta^*}|$ axis according to Eq.~\ref{eq:cosstar}.
}
\label{fig:Correlations}
\end{center}
\end{figure}

The deficit of events around $\cos{\theta^*} \approx 0$ is a direct
result of the angular scale chosen for the jet substructure analysis,
where a larger angular scale causes a stronger sculpting behavior
around $\cos{\theta^*} \approx 0$.  We know from Eq.~\ref{eq:cosstar}
that small $\cos{\theta^*}$ is identified with small $\Delta \eta$
between the two fat jets, and we also show in
Fig.~\ref{fig:Correlations} the parton level correlation between
$\Delta \eta$ for the two fat jets and $\Delta R$ of the resulting $W$
and $Z$ decay products for a spin-1 $W'$ example.  Other signal
samples would show a similar correlation, albeit with only one $W$
(left color band) or $Z$ (right color band) as appropriate.  The bulk
of the $W/Z$ subjets lie at $\Delta R \approx 2 m_{W/Z} /
(1~\text{TeV})$ as expected, where 1~TeV is a rough estimate of the
$W$ and $Z$ transverse momenta when the vector bosons are central, but
we also see a clear correlation between larger $\Delta \eta$
separation between the fat jets and the corresponding $\Delta R$ of
the resulting subjets.  As $\Delta \eta$ grows, the vector bosons from
the $W'$ decay become more forward, and thus the corresponding $p_T$
of each vector boson decreases, leading to larger $\Delta R$
separation of their subjets.

As a result of this correlation, using a large fixed angular scale
during jet substructure reclustering leads to a deficit of events with
small $\Delta \eta$ separation between fat jets and hence leads to the
sculpting effect around $\cos \theta^* \approx 0$ observed in
Fig.~\ref{fig:MCRecon_CosStar}.  A relatively large angular scale for
subjet clustering will merge nearby partons together, and the
resulting event will not have the requisite subjets to define the
$\cos \theta^*$ angle and fail the reconstruction of angular
observables.  The ATLAS 13~TeV analysis has the most pronounced
deficit of events around $\cos \theta^* \approx 0$, since this
analysis uses a fixed radius of $R = 0.2$ during trimming.  Most
notably, using an angular scale of $R = 0.2$ during subjet clustering
causes most of the quarks to merge into a single subjet, which
severely limits the viability of such a subjet identification
technique for a post-discovery study of angular correlations.  We
remark that the $D_2^{(\beta=1)}$ discriminant also used in the ATLAS
13 TeV analysis to identify a prevalence of two-prong energy
correlations compared to one-prong and three-prong energy correlations
fails to ameliorate the situation, as the events with the strongest
two-prong behavior would still need to be reclustered to identify the
appropriate subjets for angular observable studies.
\newline

\begin{figure}[tb]
\begin{center}
\includegraphics[scale=0.36]{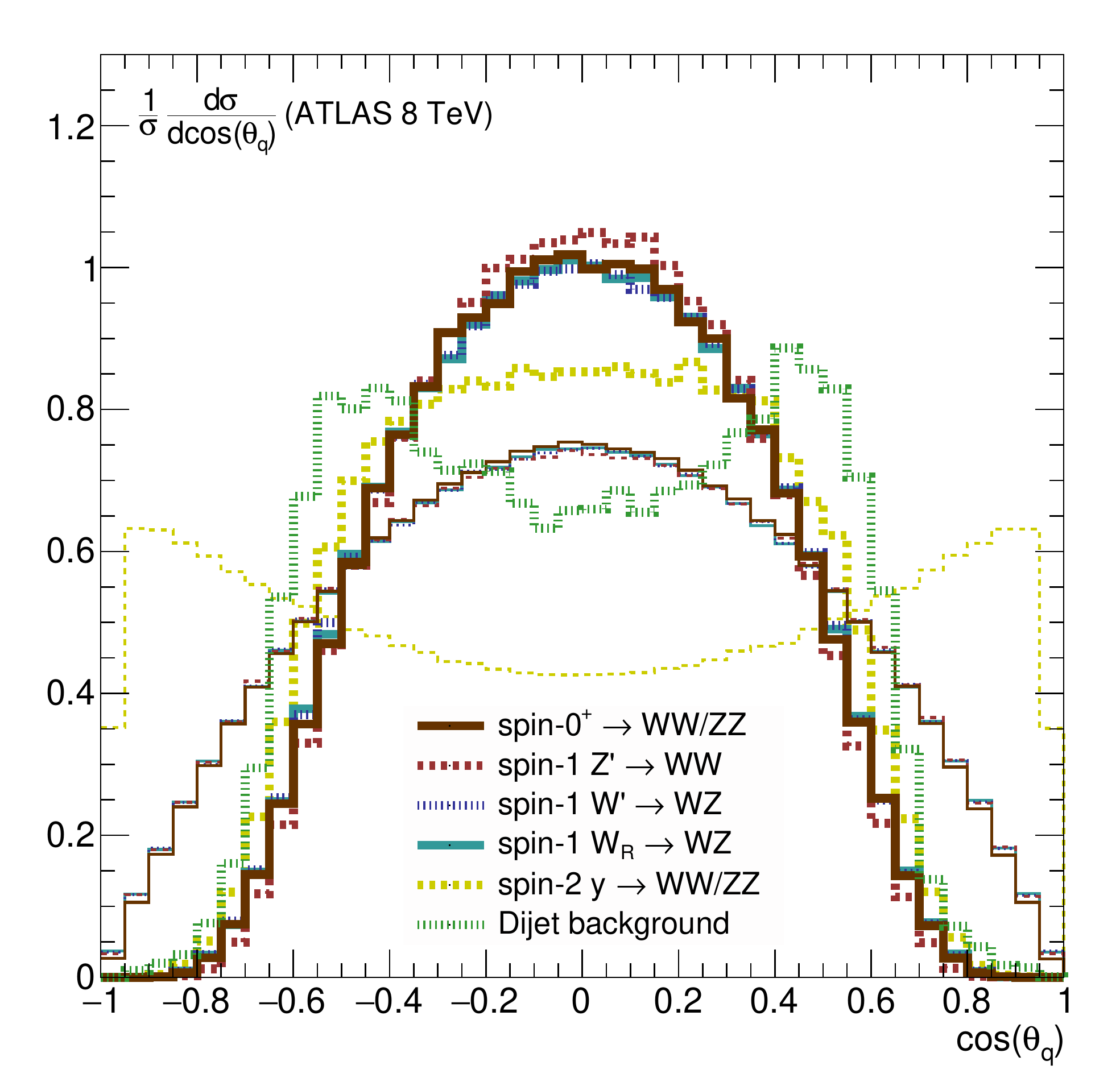}
\includegraphics[scale=0.36]{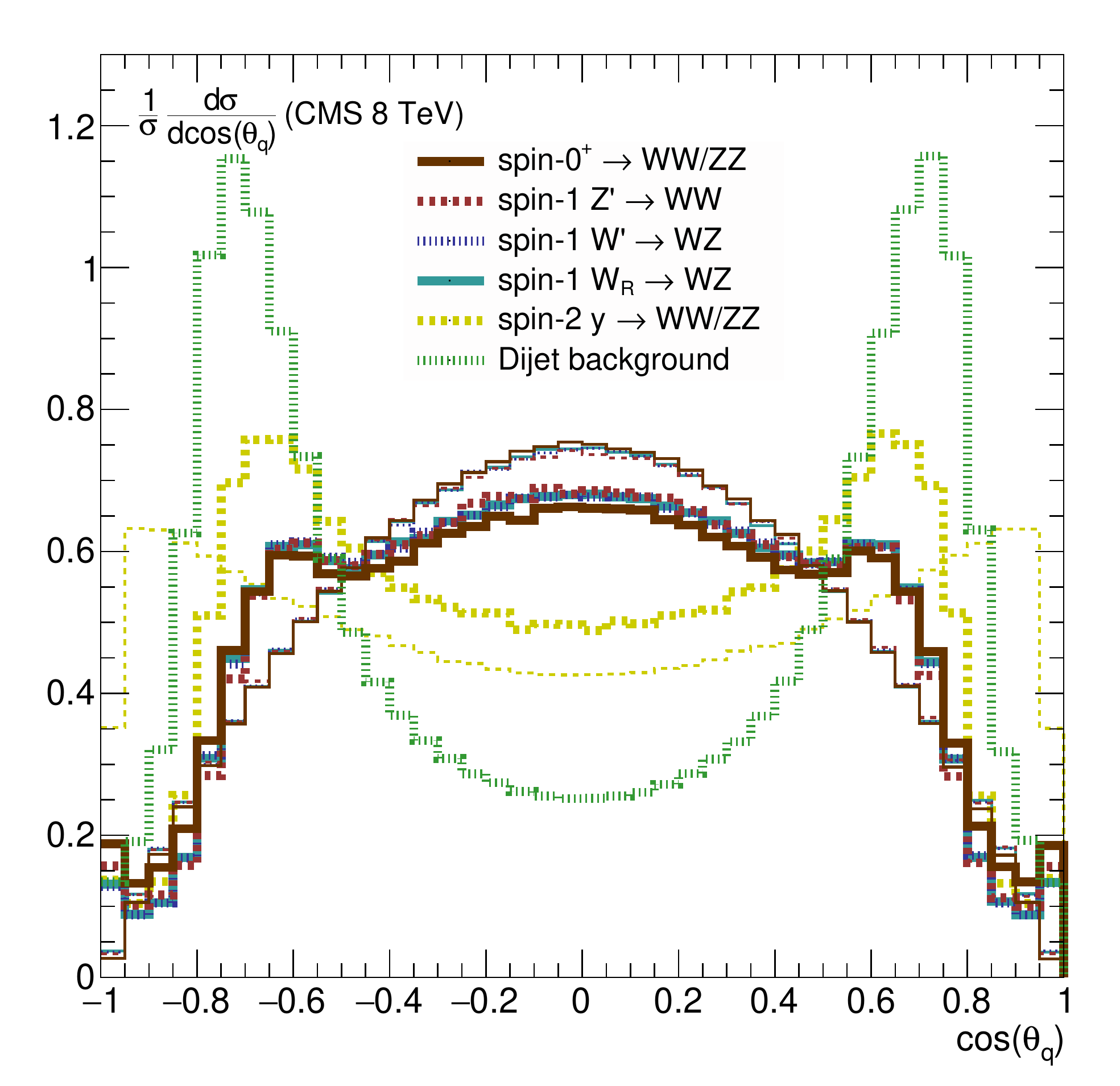}\\
\includegraphics[scale=0.36]{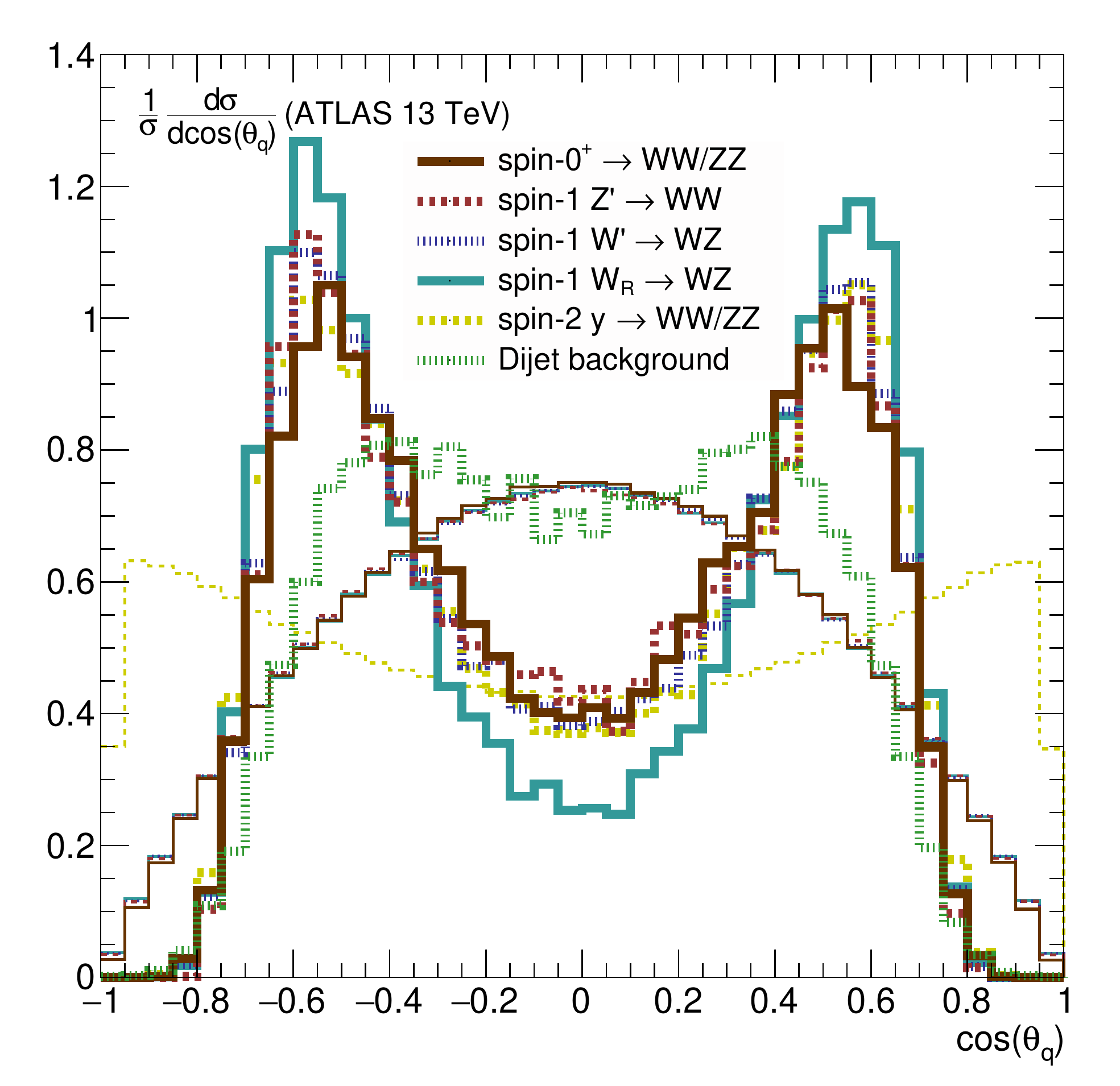}
\includegraphics[scale=0.36]{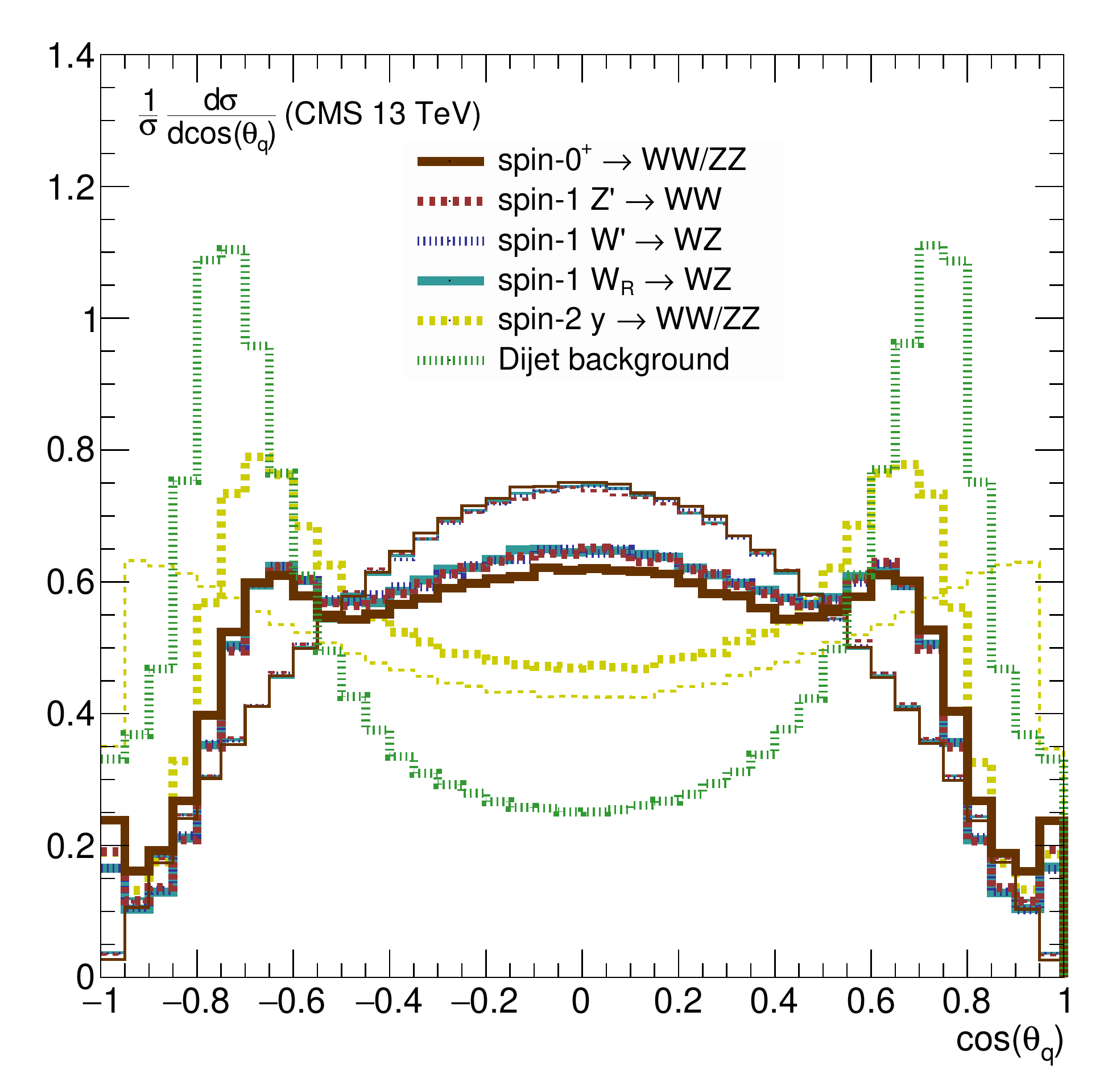}
\caption{\label{fig:MCRecon_Cosq} Comparison of the $\cos{\theta_q}$
  angle between MC parton level results (thin lines) and reconstruction of
  showered events via jet substructure (thick lines) for the ATLAS (left)
  and CMS (right) hadronic diboson search at 8 TeV (top) and 13 TeV
  (bottom).}
\end{center}
\end{figure}

\paragraph*{\bf Differential Shape of $\cos{\theta_q}$\\}
The second main discriminant between different spin signal hypotheses
is the $\cos{\theta_q}$ angle, shown in Fig.~\ref{fig:MCRecon_Cosq},
which combines the $\cos{\theta_{p_1}}$ and $\cos{\theta_{p_3}}$
angles defined in Sec.~\ref{sec:framework}.  This angle measures the
alignment of the outgoing quark with the boost vector of its parent
vector in the parent rest frame, and since each event has two vector
candidates, each event contribues twice to the distribution.  Again we
first focus on the parton level results (thin lines), which show that the
spin-2 RS graviton hypothesis has the opposite concavity to the spin-0
and spin-1 signals.  We note that the spin-2 resonance dominantly
couples to tranversely polarized electroweak bosons, while the spin-0
and spin-1 resonances dominantly couple to longitudinal bosons.
Hence, the pronounced difference in shape between the signals is a
realistic proxy for studying the sensitivity of different jet
substructure analyses to the polarization of $W$ and $Z$ bosons.  For
longitudinal bosons, the expected analytic shape of the $\cos
\theta_q$ distribution is $\dfrac{3}{4} \left(1 - \cos^2 \theta_q
\right)$, while the shape is $\dfrac{3}{8} \left( 1 + \cos^2 \theta_q
\right)$ for transverse bosons~\cite{Bolognesi:2012mm}.  We remark
that enhancing sensitivity to either the center or edges of the $\cos
\theta_q$ distribution will emphasize sensitivity to longitudinal or
transverse gauge bosons, respectively.  These results also agree with
an earlier analysis by CMS~\cite{Khachatryan:2014vla}, but we carry
the analysis further by studying multiple state-of-the-art jet
substructure techniques to understand the impact of vector boson
polarization on the resulting reconstruction efficiency.

Turning to the reconstructed angular distributions (thick lines) in
Fig.~\ref{fig:MCRecon_Cosq}, we again see the full phase space of the
parton decays gets significantly molded by the experimental analyses,
where events close to $\cos{\theta_q} \approx \pm 1$ are cut away.  In
contrast to the sharp cliffs in $\cos{\theta^*}$, though, the
$\cos{\theta_q}$ distribution exhibits a milder transformation, and
start and strength of the deviations depend strongly on the individual
analysis.  At 8 TeV, ATLAS shows a reversal point at $\cos{\theta_q}
\approx \pm 0.6$, whereas the CMS reversal point is $\cos{\theta_q}
\approx \pm 0.8$.  We also observe a deficit of events with
$\cos{\theta_q} \approx 0$, most notably in the ATLAS 13 TeV analysis.

In order to understand the behavior around $\cos{\theta_q} \approx \pm
1$, we derive an approximate relation between $\cos{\theta_q}$ and the
subjet $p_T$ ratio $y$.  Identifying $\cos{\theta_q}$ with
$\cos{\theta_{p_1}}$ for the moment, we write $\cos{\theta_q} \equiv
\hat{p}_{p_1} \cdot \hat{p}_{V_2}$ from Eq.~\ref{eq:Nangles}, where
the $p_1$ and $V_2$ four-momenta are boosted to the $V_1$ rest frame.
If we assume threshold production of $X$, then the $X$ rest frame is
identified with the lab frame, and the two vectors $V_1$ and $V_2$ are
completely back-to-back in both frames.  Hence, $\hat{p}_{V_2}$ in the
$V_1$ rest frame can be replaced by the (negative) boost direction
$-\hat{p}_{V_1}$ going from the lab frame to the $V_1$ rest frame.  If
we now take the limiting case that $V_1$ and $V_2$ have no
longitudinal momentum, then we are left with six four-momentum
components of $p_1$ and $p_2$, which are the decay products of $V_1$,
subject to four constraints: $(p_1 + p_2)^2 = m_{V_1}^2$, $p_1^2 =
p_2^2 = 0$, and $y = p_{T_2} / p_{T_1}$ given by the $y_{\text{min}}$
cut parameter.  We choose the two remaining free parameters to be the
transverse momentum of the boson, $p_{T,\ V_1}$ and the angle between
the decay plane spanned by $p_1$ and $p_2$ relative to the transverse
plane.  We have three planes: the plane spanned by the beam axis and
the $V_1$ boson, the transverse plane, and the decay plane spanned by
$p_1$ and $p_2$, where the common axis of intersection is the $V_1$
transverse momentum vector.

For the limiting case that the decay plane spanned by $p_1$ and $p_2$
aligns with the transverse plane, the cut on $y$ provides a lower
bound on $|\cos \theta_q|$, while the case when the decay plane aligns
with the plane spanned by the beam axis and the $V_1$ boson provides
an upper bound on $|\cos \theta_q|$, where we can only bound $|\cos
\theta_q|$ because we order the two subjets in $p_T$.  These lower and
upper limits are\footnote{It is easiest to derive these limits by
  performing an azimuthal rotation of the system to fix the $V_1$
  transverse momentum in the $\hat{y}$ direction.}
\begin{align}
\frac{p_{T,\ V}}{\sqrt{m_V^2+p_{T,\ V}^2}} \frac{1-y}{1+y} \leq
|\cos{\theta_{q}}| \leq 
\frac{\sqrt{m_V^2+p_{T,\ V}^2}}{p_{T,\ V}} \frac{1-y}{1+y} \ .
\label{eq:yrelation}
\end{align}
Note that the upper bound can in principle exceed $1$, and at this
point, for a given $p_{T,\ V}$ and $y$, the solution with the decay
plane aligned with the beam axis becomes unphysical and a rotation of
the decay plane away from the beam axis is needed to obtain a physical
solution.  If we relax the initial conditions and allow longitudinal
boosts of the system, the resulting $y$ cut will, by construction,
project out only the transverse components of the boost needed to
transform the lab frame into the rest frame of $V_1$.  This smears the
expression in Eq.~\ref{eq:yrelation} for both the upper and lower
limits.

Nevertheless, we can see that in the limit $p_{T,\ V} \gg m_V$,
\begin{align}
|\cos{\theta_{q}}| \approx \frac{1-y}{1+y} \leq 
\frac{1-y_\text{min}}{1+y_\text{min}} \ .
\label{eq:cosq}
\end{align}
For $y_\text{min} = 0.20$, $0.11$, or $0.05$ for the ATLAS 8~TeV, CMS,
and ATLAS 13~TeV analyses, respectively, we expect edges in the
$|\cos{\theta_q}|$ distribution at approximately $0.66$, $0.80$, and
$0.90$.  As mentioned before, the analytic calculation above requires
assumptions about the necessary boost to move from the lab frame to
the $V_1$ rest frame and taking $p_{T,\ V} \gg m_V$, and if these
assumptions are violated, the upper limit on $|\cos{\theta_q}|$ can be
exceeded.
\newline

\begin{figure}[tb]
\begin{center}
\includegraphics[scale=0.4]{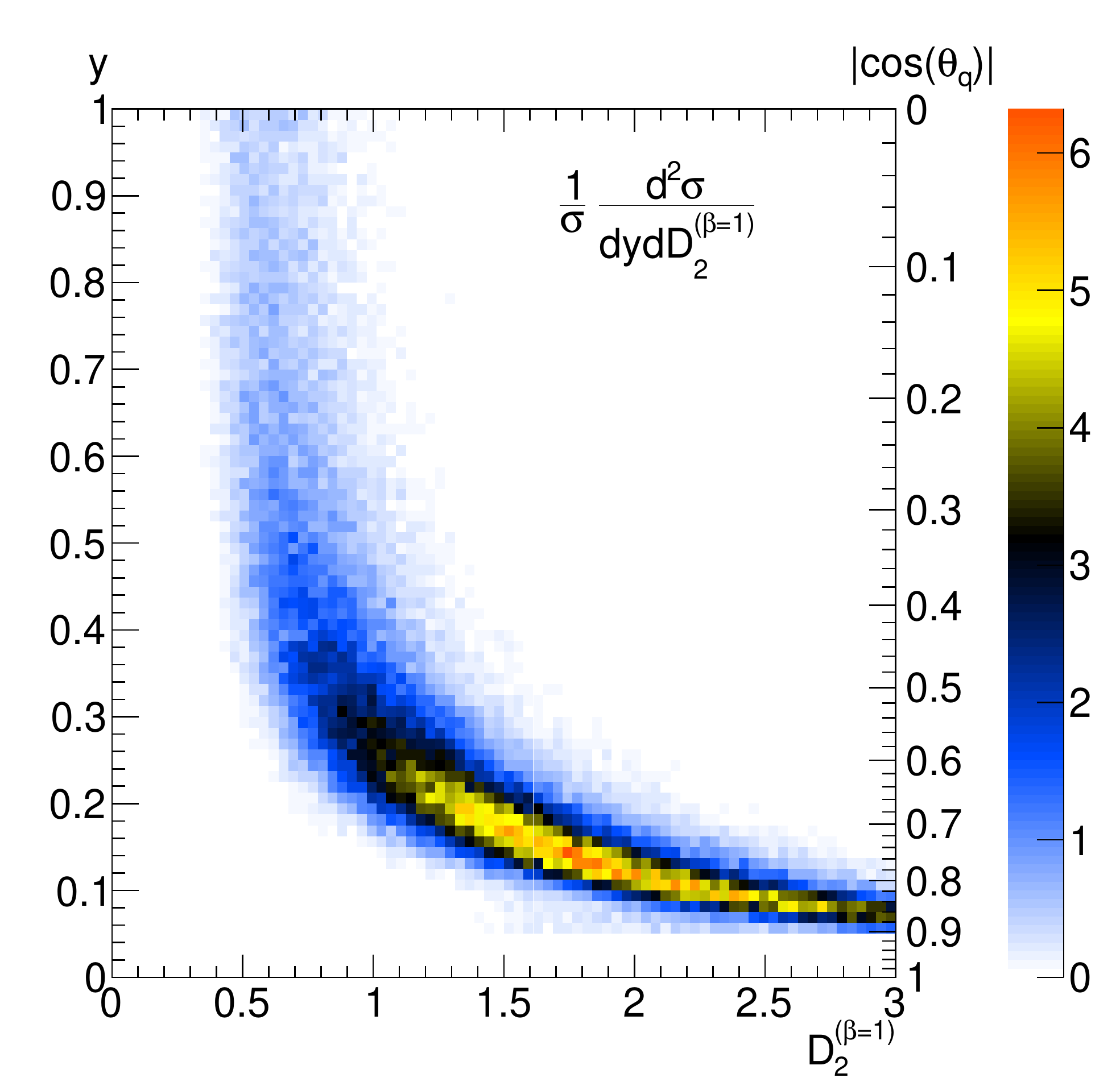}
\caption{\label{fig:Correlations2} Correlation between the energy
  correlation function $D_2^{(\beta=1)}$ and the ratio of transverse
  momentum $y$ of the two leading subjets, where the shading shows the
  relative event weight.  All analysis cuts of the ATLAS 13 TeV
  analysis are applied, except the cut on $D_2^{(\beta=1)}$
  itself. This particular plot is based on the spin-1 $W'$ model, but
  the correlation seen holds also for other spin scenarios.  The $y$
  axis is translated to a $|\cos{\theta_q}|$ axis according to
  Eq.~\ref{eq:cosq}.}
\end{center}
\end{figure}

This discussion explains the results in Fig.~\ref{fig:MCRecon_Cosq},
except for the ATLAS 13~TeV analysis, where many more events are lost
then simply those beyond the derived edge at $|\cos{\theta_{q}}| =
0.9$.  This is because the ATLAS 13~TeV imposes an effectively tighter
$y_\text{min}$ criteria via the $D_2^{(\beta=1)}$ discriminant, which
we demonstrate in Fig.~\ref{fig:Correlations2}.  We see that an event
with a low subjet $p_T$ ratio would generally have a large value of
$D_2^{(\beta=1)}$ and thus be removed given the $D_2$ cut.  As a
reminder, the $D_2$ cut parameter varies from $D_2 = 1.0$ for a
trimmed jet of $p_T = 250$ GeV to $D_2 = 1.8$ for $p_T = 1500$ GeV,
which corresponds to $y_\text{min} \approx 0.1$--$0.2$, in agreement
with the resulting sculpting seen in Fig.~\ref{fig:MCRecon_Cosq}.
\newline

\begin{figure}[tb]
\begin{center}
\includegraphics[scale=0.4]{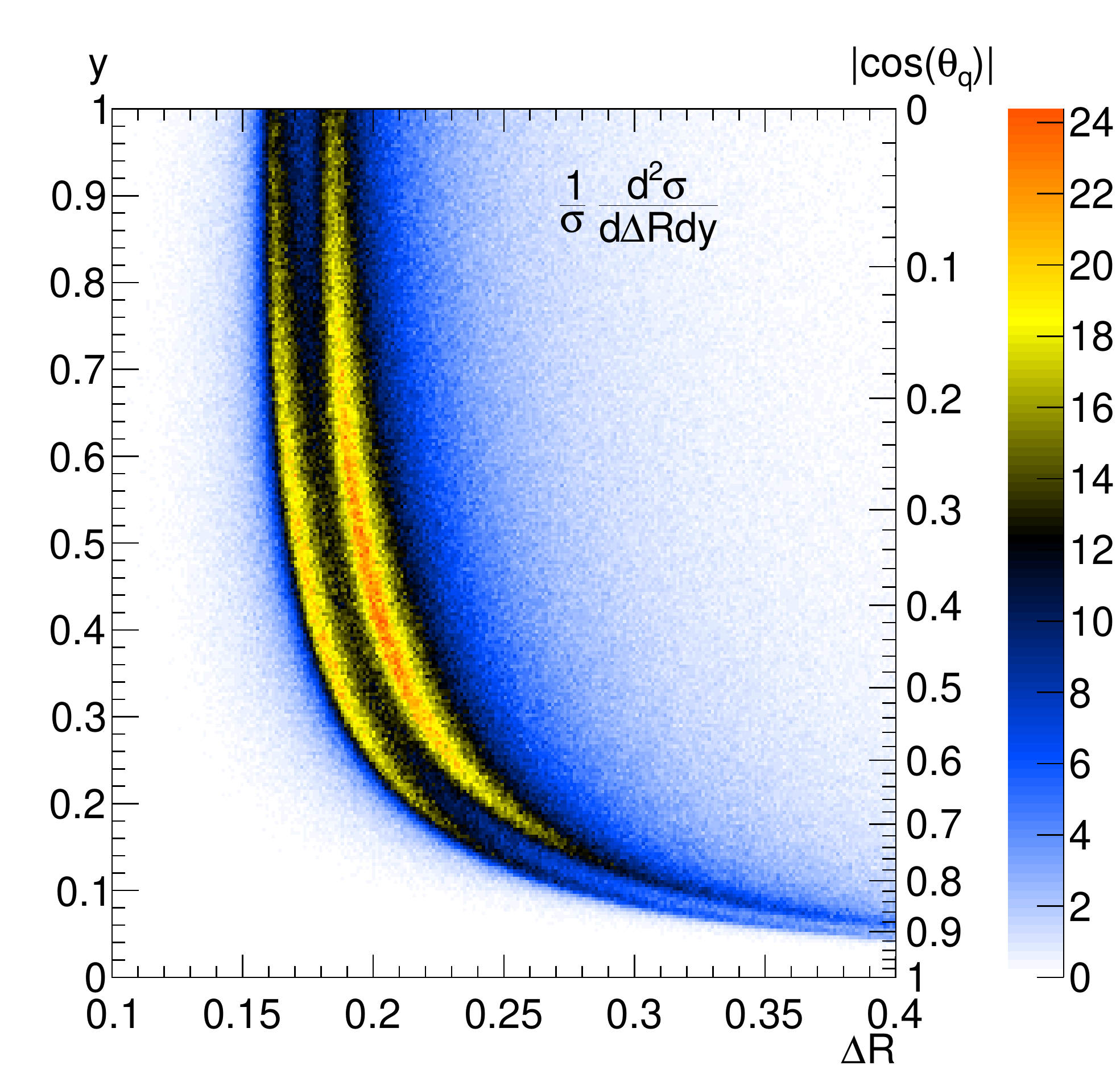}
\caption{\label{fig:Correlations3} Spin-1 $W'$ parton level
  correlations of the angular separation $\Delta R$ between the
  $W$/$Z$ decay products and their ratio in transverse momentum $y$,
  where the shading shows the relative event rate.  This basic
  correlation holds also for other spin scenarios.  The $y$ axis is
  translated to an approximate $|\cos{\theta_q}|$ axis according to
  Eq.~\ref{eq:cosq}.}
\end{center}
\end{figure}

Finally, the deficit of events with $\cos{\theta_q} \approx 0$ is the
same sculpting effect as seen before around $\cos{\theta^*} \approx
0$.  In Fig.~\ref{fig:Correlations3}, we show the correlation between
$\Delta R$ of the $W$/$Z$ decays and the ratio of quark transverse
momentum $y$ for parton-level $W' \to WZ$ events.  As before, the left
band shows the $W^\pm$ daughter partons and the right band shows the
$Z$ daughter quarks.  Since using a large $\Delta R$ during subjet
finding causes the $W$/$Z$ decay partons to be merged, events with
large $y$ are more likely to be removed from the event sample by
subsequent kinematic cuts.  Using Eq.~\ref{eq:cosq}, we can relate $y$
to an effective cut on $\cos{\theta_q}$, which explains the deficit of
events seen around $\cos{\theta_q} \approx 0$ in
Fig.~\ref{fig:MCRecon_Cosq}, most notably in the lower left panel for
the ATLAS 13~TeV analysis.
\newline

\paragraph*{\bf Differential Shape of $\Psi$\\}

\begin{figure}[tb]
\begin{center}
\includegraphics[scale=0.35]{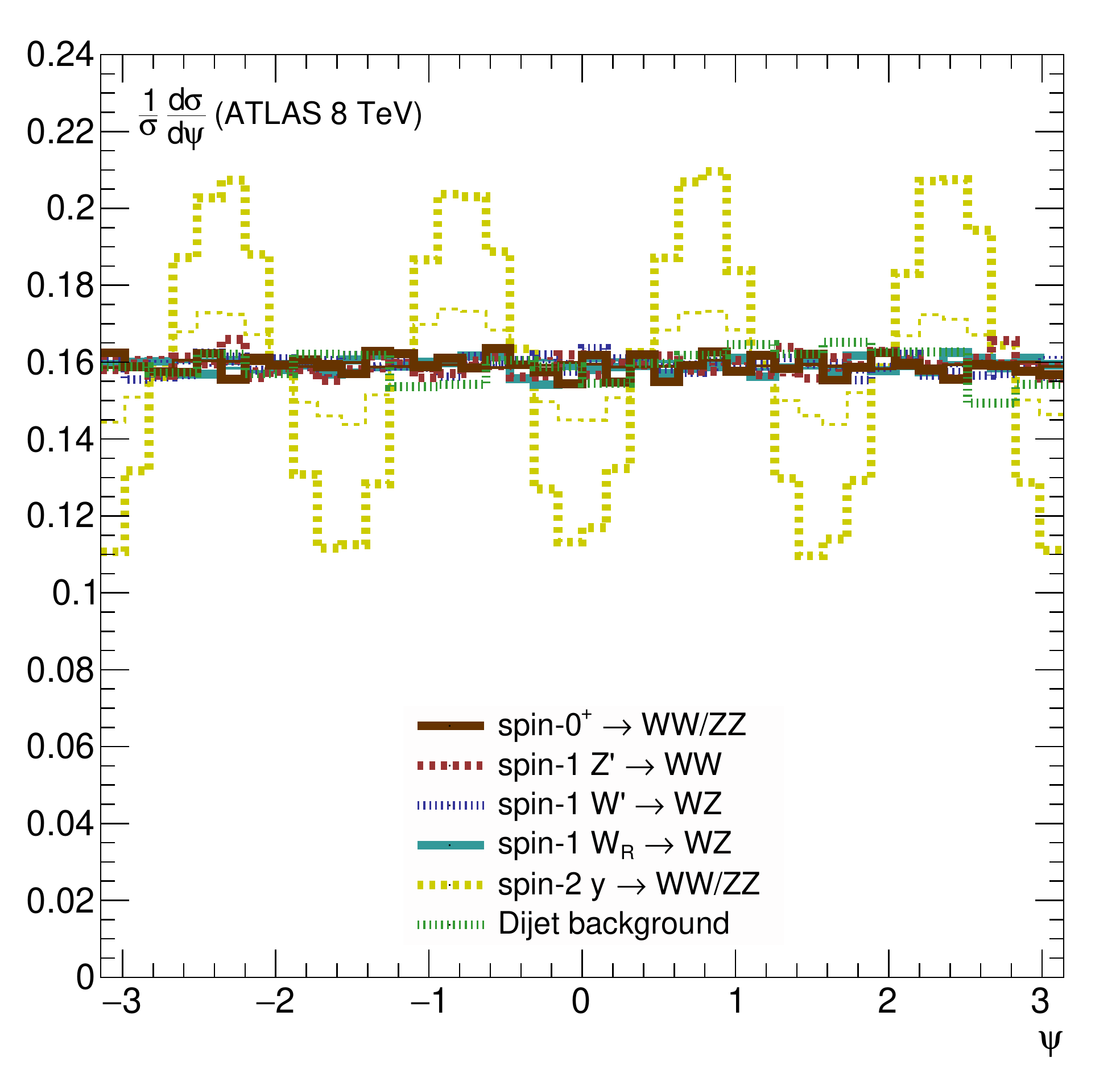}
\includegraphics[scale=0.35]{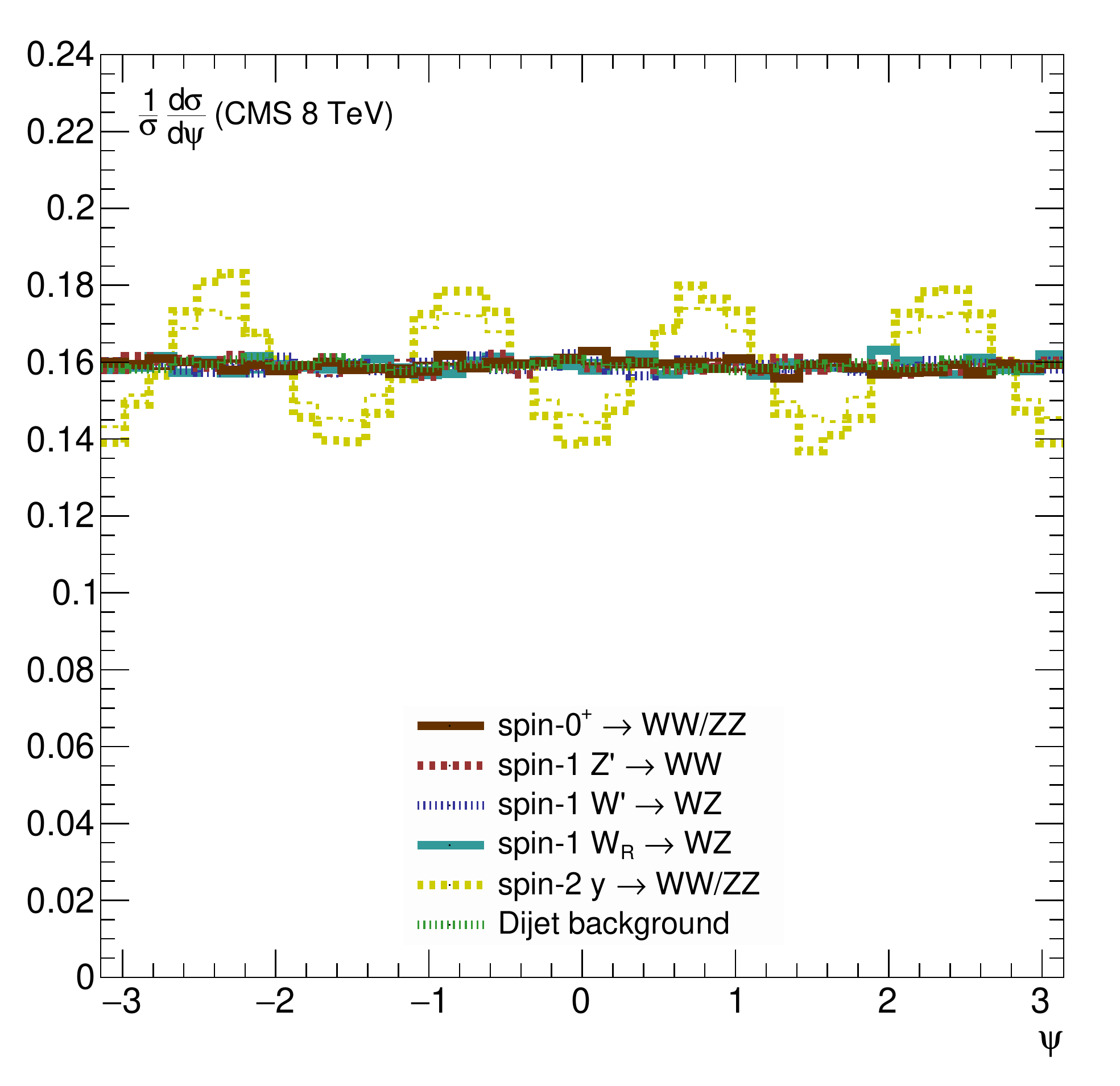}\\
\includegraphics[scale=0.35]{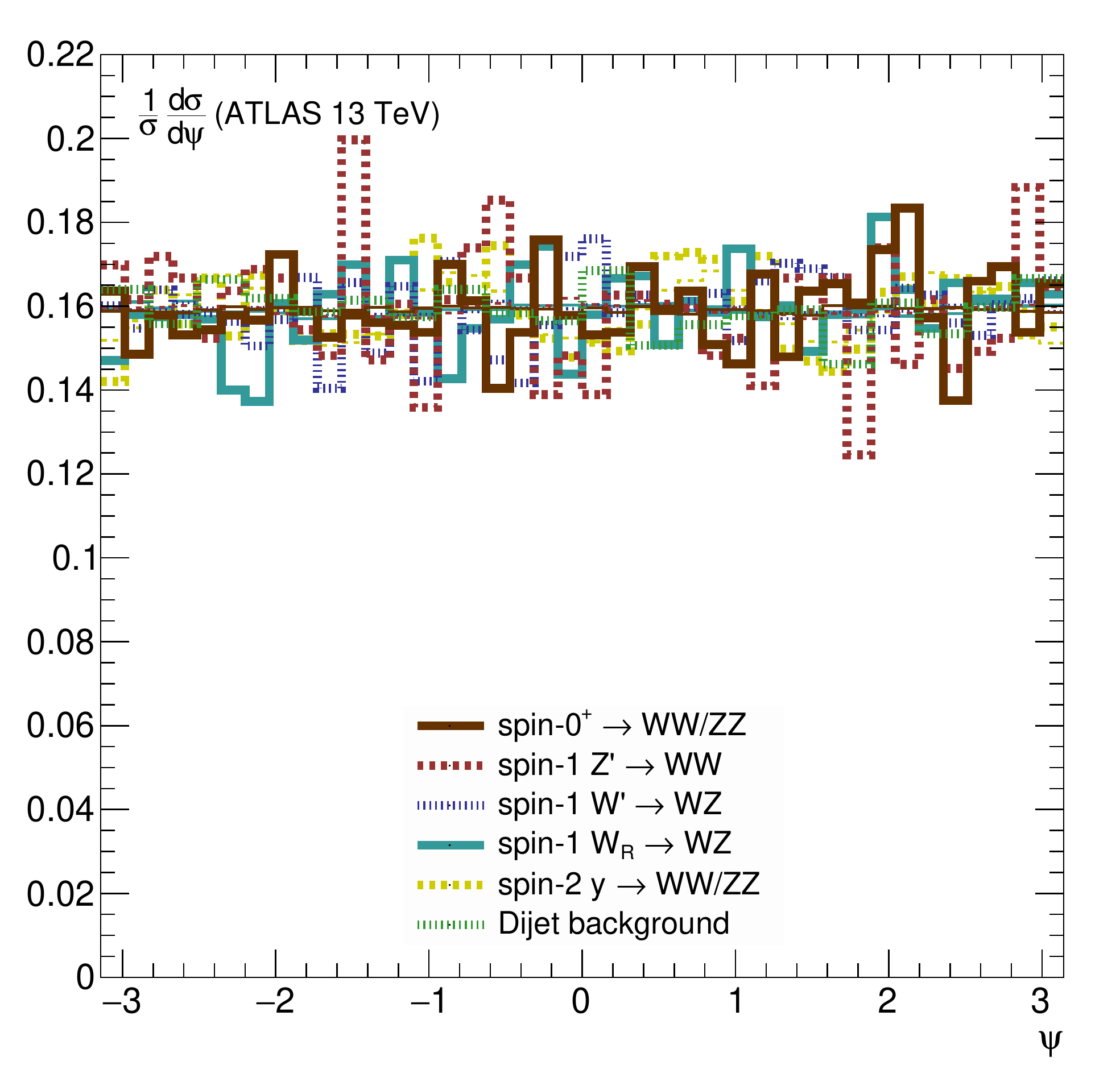}
\includegraphics[scale=0.35]{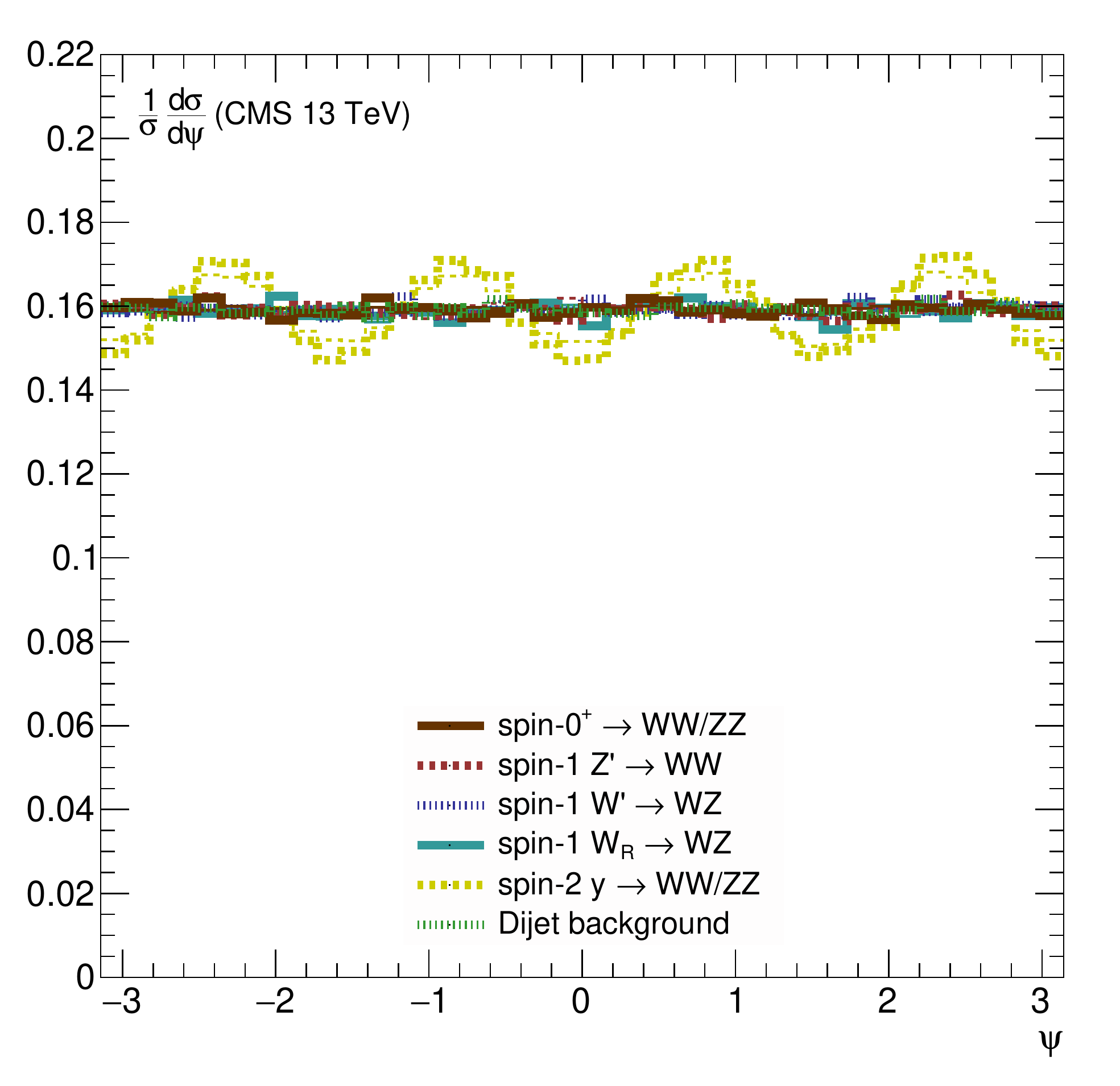}
\caption{\label{fig:MCRecon_Psi} Comparison of the $\Psi$ angle
  between MC parton level results (thin lines) and reconstruction of
  showered events via jet substructure (thick lines) for the ATLAS (left)
  and CMS (right) hadronic diboson search at 8 TeV (top) and 13 TeV
  (bottom).}
\end{center}
\end{figure}

As shown in Fig.~\ref{fig:MCRecon_Psi}, the differential distribution
in the angle $\Psi$ is flat for all spin hypotheses except for the
spin-2 resonance.\footnote{Recall $\Psi = \Phi_{V_1} + \Phi/2$ is the
  average azimuthal angle of the two decay planes formed by the vector
  boson decay products.  Also, note that the lower left panel showing
  the ATLAS 13~TeV analysis is dominated by statistical fluctuations,
  which occurs because the $R = 0.2$ substructure angular scale has
  poor efficiency at finding four distinct subjets needed to reconstruct
  the two decay planes.}  We will thus focus on explaining the
behavior of the spin-2 scenario. In this distribution, we expect
amplitudes proportional to $1$ and $\cos{(4\Psi)}$, where the
respective amplitudes at parton level depend on the helicity states of
the vector bosons and the production level partons~\cite{Gao:2010qx,
  Bolognesi:2012mm}. A $\cos{(2\Psi)}$ contribution would only appear
when particles and anti-particles of the $V$ decay can be
distinguished.  Curiously, the differential distribution of $\Psi$
after cuts causes the $\cos{(4\Psi)}$ amplitude to increase. This is
related to the same two cuts on $\Delta \eta_{JJ, \text{ max}}$ and
subjet $p_T$ ratio $y_\text{min}$, which already skewed the
$\cos{\theta^*}$ and $\cos{\theta_q}$ angle.

\begin{figure}[tb]
\begin{center}
\includegraphics[scale=0.4]{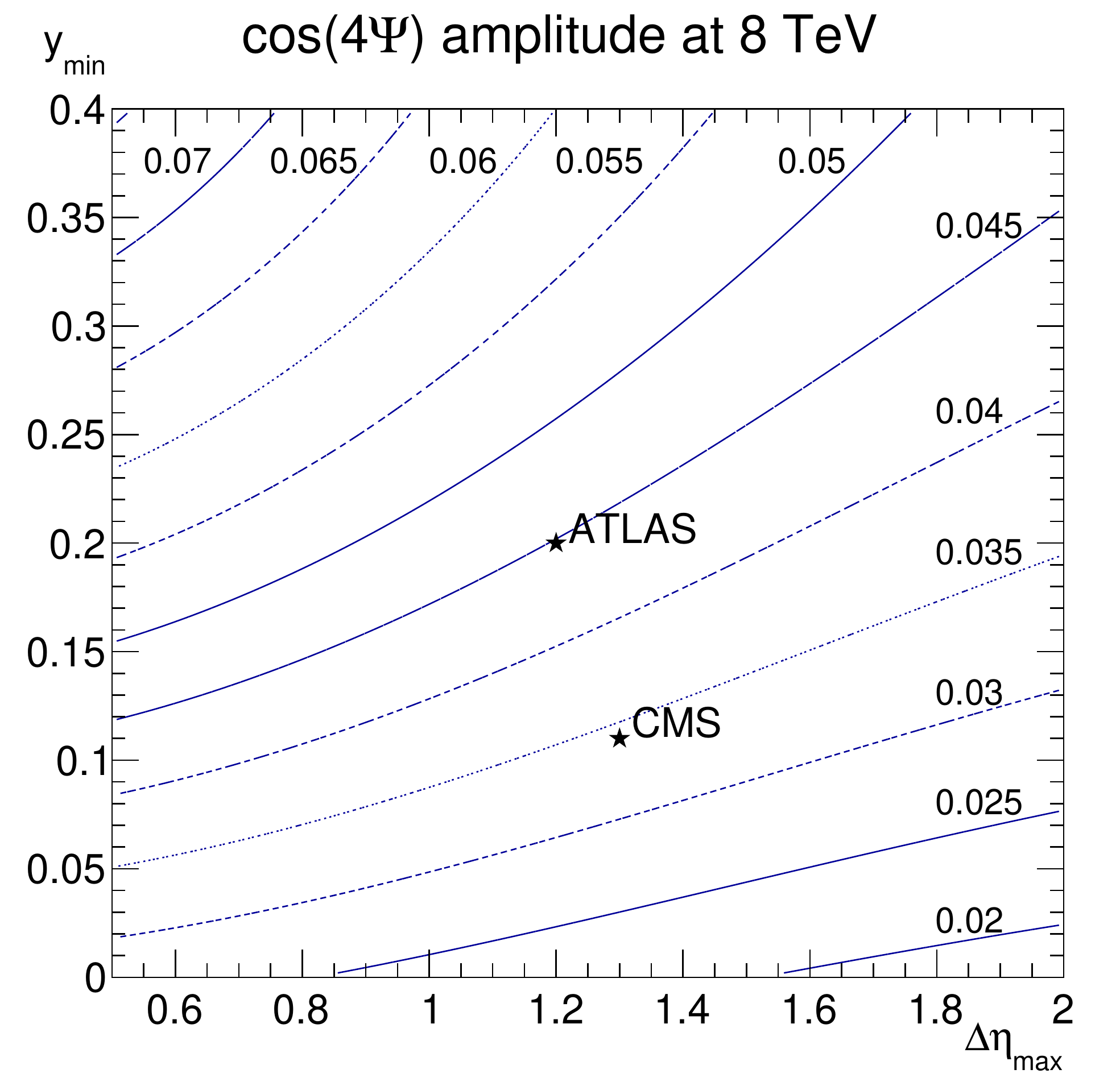}
\includegraphics[scale=0.4]{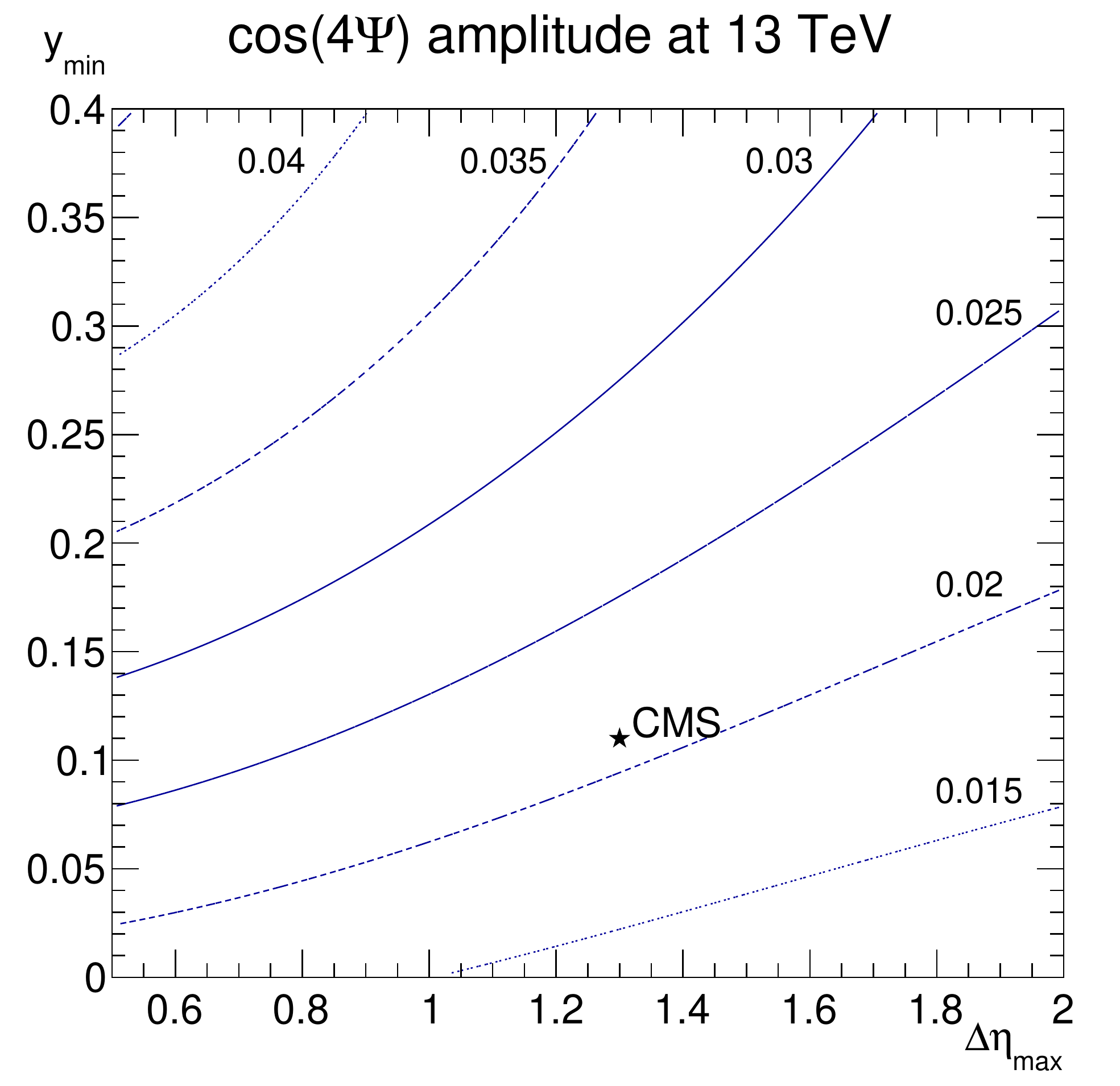}
\caption{\label{fig:Amplitudes} Expected $\cos(4\Psi)$ amplitude
  $\mathcal{A}$ (contours) for a spin-2 resonance at 8 TeV (left) and 
  13 TeV (right) as
  function of the cut parameter $y_\text{min}$ and $\Delta \eta_{JJ,
    \text{ max}}$, as shown in Eq.~\ref{eq:amplitude}.  We use $F_{+-}
  = F_{-+} = 45.8\%$ and $F_{00} = 7.8\%$, as determined from our
  underlying Monte Carlo simulation, and $f_{q\bar{q}} = 65.5\%$ and
  $f_{q\bar{q}} = 45.0\%$ for 8 TeV and 13 TeV, respectively.  We also
  show the respective working points of ATLAS and CMS, except for
  ATLAS at 13 TeV, where the effective $y_\text{min}$ is not a fixed
  parameter.}
\end{center}
\end{figure}

We can analytically determine the differential shape of the $\Psi$
distribution as a function of the cut values on $\Delta\eta_{JJ,
  \text{ max}}$ and $y_\text{min}$, using the fully differential
results in Ref.~\cite{Bolognesi:2012mm}. The normalized shape can be
expressed as
\begin{align}
\frac{1}{\sigma^\text{(spin-2)}} \frac{d\sigma^\text{(spin-2)}}{d\Psi}=\frac{1}{2\pi}-\mathcal{A}
	(y_\text{min},\Delta\eta_\text{max})\cos(4\Psi) \ ,
\end{align}
with
\begin{align}
\label{eq:amplitude}
 \mathcal{A}
    =\frac{1}{24\pi}
    &F_{+-}\left(1+4y_\text{min}+y_\text{min}^2\right)^2
	        (5f_{q\bar{q}}-1)
		(8+6\cosh{\Delta\eta_\text{max}}
		  +\cosh{2\Delta\eta_\text{max}})\Big/\\\nonumber
	\biggl[&F_{+-}\left(1+y_\text{min}+y_\text{min}^2\right)^2
	     \Bigl((5f_{q\bar{q}}+1)(1+2\cosh{\Delta\eta_\text{max}})
		+2\cosh{2\Delta\eta_\text{max}}\Bigr)\ +\\\nonumber
	    &F_{00}\left(1+4y_\text{min}+y_\text{min}^2\right)^2
	    (-15f_{q\bar{q}}+8+6\cosh{\Delta\eta_\text{max}}
	       +\cosh{2\Delta\eta_\text{max}})	     
	 \biggr] \ .	 
\end{align}
Here, $F_{\lambda_1 \lambda_2}$ is the fraction of events with two
gauge bosons having a helicity $\lambda_1$ and $\lambda_2$
respectively, and $f_{q\bar{q}}$ is the production fraction from
$q\bar{q}$ initial state quarks.  From our Monte Carlo simulation at
8~TeV, we find $F_{+-} = F_{-+} = 45.8\%$, $F_{00} = 7.8\%$ and
$0.6\%$ others, and thus we neglected the subleading helicity
components, which are suppressed by powers of $m_{W/Z} / m_X$.
Furthermore, we find $f_{q\bar{q}} \approx 65.5\%$ at 8~TeV LHC, while
it drops to $f_{q\bar{q}} \approx 45.0\%$ at 13~TeV LHC.  We show the
scaling behaviour of $\mathcal{A}$ in Fig.~\ref{fig:Amplitudes}.

From Fig.~\ref{fig:Amplitudes}, we can directly read off the expected
$\cos(4\Psi)$ amplitude for our 8~TeV and 13~TeV signal sample. Using
$y_\text{min} \rightarrow 0$ and $\Delta \eta_{JJ, \text{
    max}}\rightarrow \infty$ the predicted amplitudes at parton level
match with $\mathcal{A} \approx 0.014$ at 8 TeV and $\mathcal{A}
\approx 0.0077$ at 13 TeV very well our Monte Carlo
simulation. Including cuts we expect $\mathcal{A}\approx0.045$ and
$\mathcal{A} \approx 0.034$ for ATLAS and CMS at 8 TeV, respectively,
and $\mathcal{A} \approx 0.021$ for CMS at 13 TeV. For CMS the
expected amplitude is slightly larger than that seen in
Fig.~\ref{fig:MCRecon_Psi}, which can be explained by the
approximation of Eq.~\ref{eq:cosq} used to relate $\cos\theta_q$ with
$y_\text{min}$.


\section{Angular observables in semi-leptonic final states}
\label{sec:semileptonic}
We now turn to the semi-leptonic analyses, $X \to \ell \ell qq$ and $X
\to \ell \nu qq$, which provide important cross-channels for a future
discovery of a diboson resonance.  To reiterate, the relative rates of
the $4q$, $\ell \ell qq$, and $\ell \nu qq$ final states will
disentangle the intermediate $W^+ W^-$, $W^\pm Z$, and $ZZ$ nature of
the resonance, which is very difficult to do using only the $4q$
analysis.  Moreover, the semileptonic channels enjoy cleaner
reconstruction of angular observables, larger signal efficiencies, and
better control of systematic uncertainties, counterbalanced by lower
overall statistical power.  The importance of the semileptonic
channel, especially compared to the fully leptonic channel, was
emphasized, for example, in Refs.~\cite{Hackstein:2010wk,
  Englert:2010ud}.  In particular, the angular observables $\cos
\theta_{p_1}$ and $\cos \theta_{p_3}$, which were previously combined
into $\cos \theta_q$ because we could not trace a given parent from
one event to the next, are now assigned as $\cos \theta_q$ and $\cos
\theta_l$.  In addition, for the $\ell \nu qq$ analysis, the $\cos
\theta_l$ distribution is asymmetric because the charge of the lepton
distinguishes leptons from the anti-lepton, in constrast to the $4q$
case.  We begin again by summarizing the semi-leptonic analyses by
ATLAS and CMS~\cite{CMS-PAS-EXO-15-002, Aad:2014xka, Aad:2015ufa,
  Khachatryan:2014gha, ATLAS-CONF-2015-075, ATLAS-CONF-2015-071} and
then present the corresponding angular distributions.

\subsection{ATLAS and CMS semi-leptonic analyses at 8 TeV and 13 TeV}

\paragraph*{\bf $\ell\ell qq$ Final State by ATLAS at 8 TeV\\}
In the $\ell \ell qq$ ATLAS analysis at 8~TeV, events are required to
have exactly two muons of opposite charge or two electrons, where
muons must have $p_T > 25$~GeV and $|\eta| < 2.4$ and electrons must
have $p_T > 25$~GeV and $|\eta| < 2.47$, excluding $1.37 < |\eta| <
1.52$.  In addition, all leptons must pass a track isolation
(calorimeter isolation) requirement (see Ref.~\cite{Aad:2014xka} for
details).  The lepton pair must have 66 GeV$ < m_{\ell\ell} <$ 116 GeV
and $p_T^{\ell\ell} > 400$~GeV.

Jets are clustered using the C/A algorithm with $R = 1.2$ and need to
have $p_T > 100$~GeV and $|\eta| < 1.2$.  One jet needs to survive the
grooming procedure with $y_{\text{min}} = 0.2025$ and fulfill $p_T >
400$~GeV and 70~GeV$ < m < $110~GeV.
\newline

\paragraph*{\bf $\ell\ell qq$ Final State by CMS at 8 TeV\\}
In the CMS 8~TeV $\ell \ell qq$ analysis, electrons with $p_T >
40$~GeV and $|\eta| < 2.5$, excluding $1.44 < |\eta| < 1.56$, muons
with $p_T > 20$~GeV and $|\eta| < 2.1$ are selected, and all leptons
must be isolated from other tracks as well as in the calorimeter.  Two
same flavor, opposite charge, leptons are required, and for dimuon
events, the leading muon must have $p_T > 40$~GeV.  The lepton pair
must have 70~GeV $< m_{\ell \ell} <$ 110~GeV.

Jets are reconstructed with the C/A algorithm using $R = 0.8$ and must
have $p_T > 30$~GeV and $|\eta| < 2.4$.  They are pruned with
$z_\text{min} = 0.1$ and are categorized by purity according the
$N$-subjettiness variable $\tau_{21}$, analogous to the CMS $4q$
search.  The pruned jet mass must lie within 65~GeV $< m_J <$ 110~GeV.
Both the leptonic and hadronic vector boson candidates must have
$p_T^V > 80$~GeV and satisfy $m_{VV} > 500$~GeV.  If there are
multiple hadronic $V$ candidates, the hardest $p_T$ candidate in the
higher purity category is used.
\newline

\paragraph*{\bf $\ell \ell qq$ Final State by ATLAS at 13 TeV\\}
ATLAS uses the same kinematic acceptance cuts on electrons and muons
in the 13~TeV analysis as the 8~TeV analysis, and track isolation
requirements are imposed.  Two muons of opposite charge or two
electrons are required, where the lepton pair must have 66~GeV $<
m_{\mu^+ \mu^-} <$ 116~GeV or 83~GeV$< m_{e^+e^-} <$ 99~GeV,
respectively.

Jets are clustered using the anti-$k_T$ algorithm with $R = 1.0$ and
are required to have $p_T > 200$~GeV and $|\eta| < 2.0$.  The leading
jet must satisfy the trimming procedure with $z_\text{min} = 0.05$,
and fulfill $p_T^J > 0.4 m_{\ell\ell J}$ and $68.2\text{ GeV} < m_J <
108.4$~GeV.  Additionally, the jet needs to statisfy an upper bound on
the $D_2^{(\beta=1)}$ energy correlator function.  For simplicity, we
linearly interpolate the $D_2$ cut between the two points quoted, $D_2
< 1.0$ at $p_T^J = 250$ GeV and $D_2 < 1.8$ at $p_T^J = 1500$ GeV.
Finally, the dilepton system must have $p_T^{\ell \ell} > 0.4 m_{\ell
  \ell J}$.
\newline

\paragraph*{\bf $\ell\nu qq$ Final State by ATLAS at 8 TeV\\}
In the 8~TeV ATLAS $\ell \nu qq$ analsis, the lepton kinematic
criteria are the same as their 8~TeV $\ell \ell qq$ search, and a
similar isolation criteria is used.  Missing transverse energy
$\slashed{E}_T$ (MET) must exceed $30$~GeV and is used to calculate
the corresponding neutrino four-momentum assuming no other source of
MET and $m_W^2 = (p_\ell + p_\nu)^2$:
\begin{align}
  p_z^\nu = \frac{1}{2 p_T^\ell} \left[ (m_W^2 + 
2 \vec{p}_T^{~\ell} \cdot \vec{p}_T^{~\nu}) p_z^{~\ell} \pm 
E^\ell \sqrt{(m_W^2 + 2 \vec{p}_T^{~\ell} \cdot \slashed{\vec{E}}_T)^2 -
      4(p_T^{~\ell})^2 \slashed{E}_T^2} \right] \ .
\end{align}
In the case of two complex solutions for $p_Z^\nu$, the real part is
used, otherwise the smaller solution in absolute value is used.
Events are required to have $p_T^{\ell \nu} > 400$~GeV.

Jets are clustered using the C/A algorithm with $R = 1.2$.  One jet
must survive the grooming procedure with $y_\text{min} = 0.2025$ and
fulfill $p_T^J > 400$~GeV, $|\eta| < 2.0$ and $65\text{ GeV} < m_J <
105$~GeV, and the $\Delta \phi$ between this jet and the MET vector
must exceed 1.  Events with at least one $b$-tagged jet are vetoed
(see Ref.~\cite{Aad:2015ufa} for details).
\newline

\paragraph*{\bf $\ell\nu qq$ Final State by CMS at 8 TeV\\}
At CMS, electrons with $p_T > 90$~GeV and $|\eta| < 2.5$, excluding
$1.44 < |\eta| < 1.56$, and muons with $p_T > 50$~GeV and $|\eta| <
2.1$ are selected.  The same isolation criteria from the CMS $\ell\ell
qq$ search are applied.  A single muon or electron is required and MET
must exceed 40~GeV or 80~GeV, respectively.  The corresponding
neutrino four-momentum is reconstructed as in the ATLAS $\ell\nu qq$
search, and $p_T^{\ell\nu} > 200$~GeV is required.

Jets are reconstructed with the C/A algorithm using $R = 0.8$, $p_T >
30$ and $|\eta| < 2.4$.  They are pruned with $z_\text{min} = 0.1$ and
categorized by purity using $\tau_{21}$, as in the CMS $4q$ and $\ell
\ell qq$ searches.  The pruned jet mass must again lie within 65~GeV
$< m_J <$ 110~GeV and have $p_T^J > 200$~GeV, and if there are
multiple hadronic $V$ candidates, the hardest $p_T$ candidate in the
higher purity category is used.  Furthermore, $\Delta R_{J,\ (\ell \nu)}
> \pi/2$, $\Delta \phi_{J,\ \slashed{E}_T} > 2.0$, $\Delta
\phi_{J,\ (\ell\nu)} > 2.0$ and $m_{J \ell \nu} > 700$~GeV are required.
Events with one $b$-tagged jet are vetoed.
\newline

\paragraph*{\bf $\ell\nu qq$ Final State by ATLAS at 13 TeV\\}
For the ATLAS 13~TeV $\ell \nu qq$ search, leptons are identified as
in the ATLAS $\ell\ell qq$ final state search at 8 TeV.  Events must
have one lepton and $\slashed{E}_T > 100$~GeV, and the neutrino
four-momentum is reconstructed as in the 8~TeV analysis.

Jets are clustered using the anti-$k_T$ algorithm with $R = 1.0$. The
leading jet must survive the trimming procedure with $z_\text{min} =
0.05$ and fulfill $p_T > 200$~GeV, $|\eta| < 2.0$, $70.2\text{ GeV} <
m_J < 106.4$~GeV, and $p_T^J > 0.4 m_{\ell \nu J}$.  The same
$D_2^{(\beta=1)}$ energy correlator cut as the 13~TeV $\ell \ell qq$
ATLAS search is imposed.  Finally, events must have $p_T^{\ell \nu} >
0.4 m_{\ell \nu J}$ and $p_T^{\ell \nu} > 200$~GeV, and events with
$b$-tagged jets are vetoed.
\newline

\paragraph*{\bf $\ell\nu qq$ Final State by CMS at 13 TeV\\}
Lastly, for the CMS $\ell \nu qq$ search at 13~TeV, events must have a
single electron or muon, where electron candidates must have $p_T >
120$~GeV and $|\eta| < 2.5$, excluding $1.44 < |\eta| < 1.56$, and
muon candidates must have $p_T > 53$~GeV and $|\eta| < 2.1$.  The same
lepton isolation criteria as the CMS $\ell\ell qq$ search are applied.
Electron (muon) events must have at least 80~GeV (40~GeV) of MET.  The
neutrino four-momentum is reconstructed as in the ATLAS $\ell\nu qq$
final state search, and the lepton-neutrino system must have
$p_T^{\ell \nu} > 200$~GeV.

Jets are reconstructed with the anti-$k_T$ algorithm using $R = 0.8$,
$p_T > 30$~GeV and $|\eta| < 2.4$.  They are pruned with $z_\text{min}
= 0.1$ and categorized by purity using the same criteria as the 13~TeV
CMS $4q$ search.  To satisfy the boson tagging requirements, a pruned
jet $J$ has to fulfill $65\text{ GeV} < m_J < 105$~GeV and $p_T^J >
200$~GeV, and for events with multiple hadronic boson candidates, the
highest $p_T$ jet with the higher purity category is used.  Events
must also pass $\Delta R_{J,\ (\ell \nu)} > \pi/2$, $\Delta
\phi_{J,\ \slashed{E}_T} > 2.0$, $\Delta \phi_{J,\ (\ell \nu)} > 2.0$,
and $m_{J\ell\nu}>700$~GeV cuts, and events with $b$-tagged jets are
vetoed.
\newline


\subsection{Angular observables in semi-leptonic final states and 
comparison with fully hadronic final states}

\begin{figure}[tb!]
\begin{center}
\includegraphics[scale=0.27]{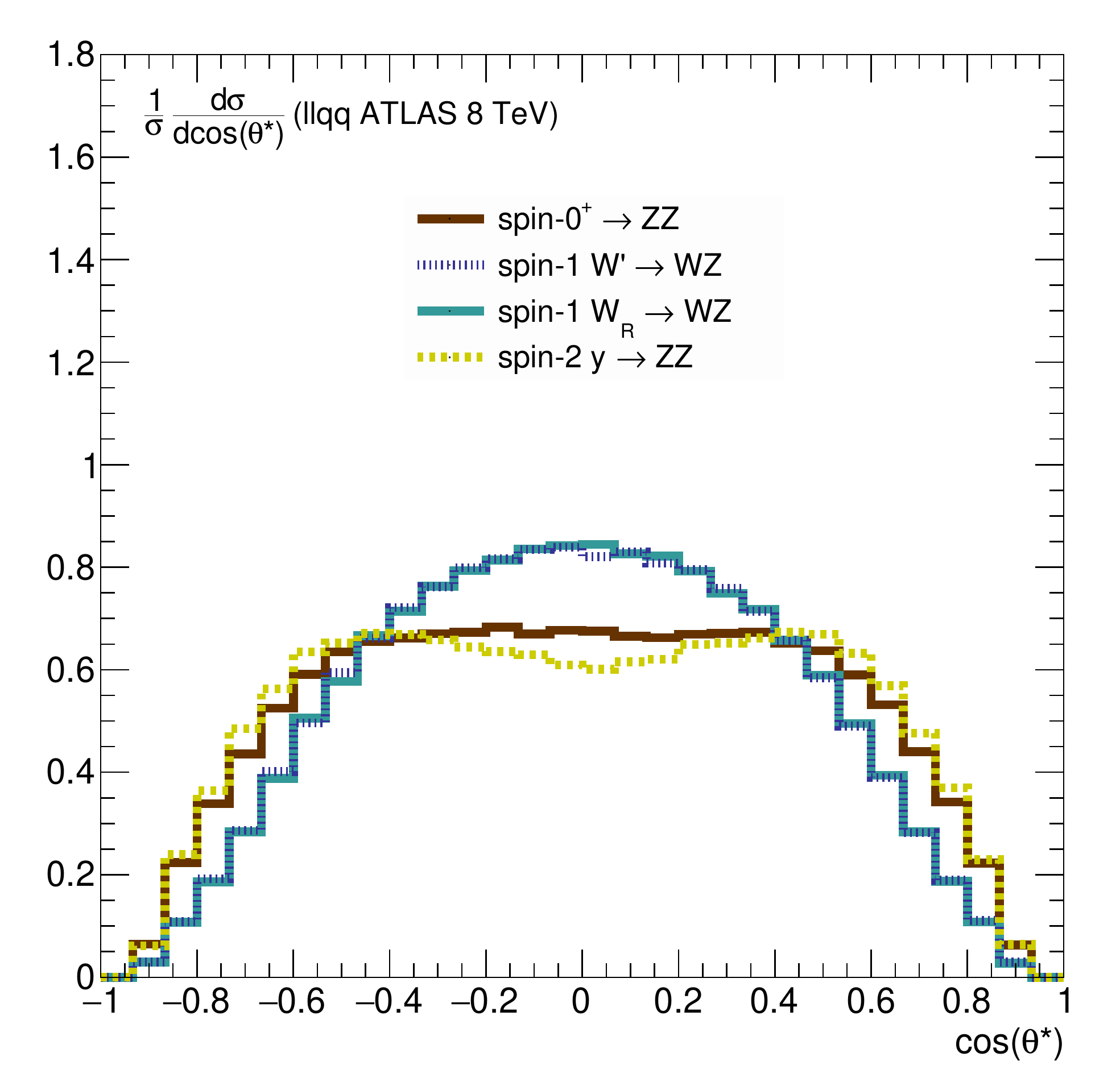}
\includegraphics[scale=0.27]{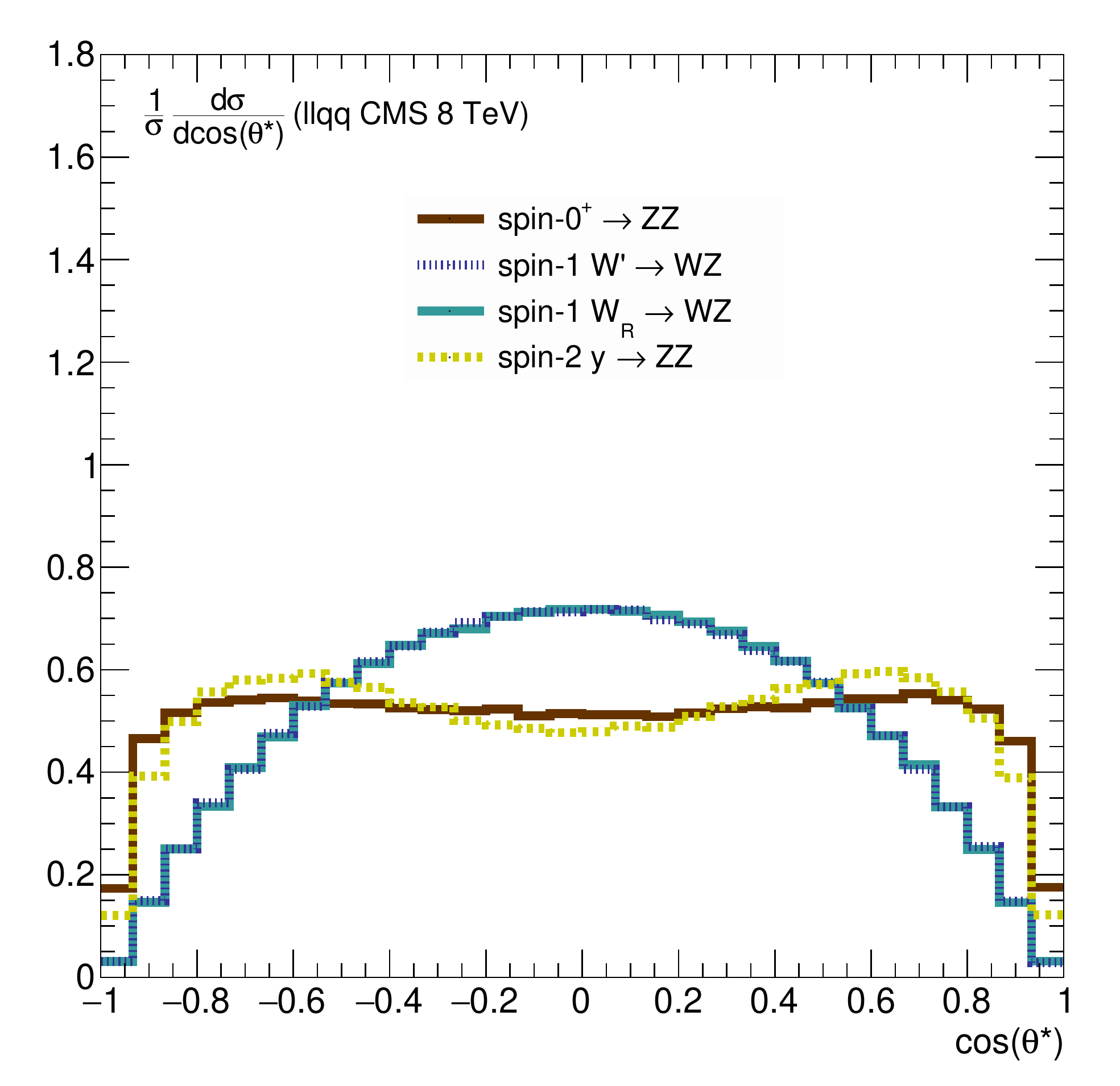}
\includegraphics[scale=0.27]{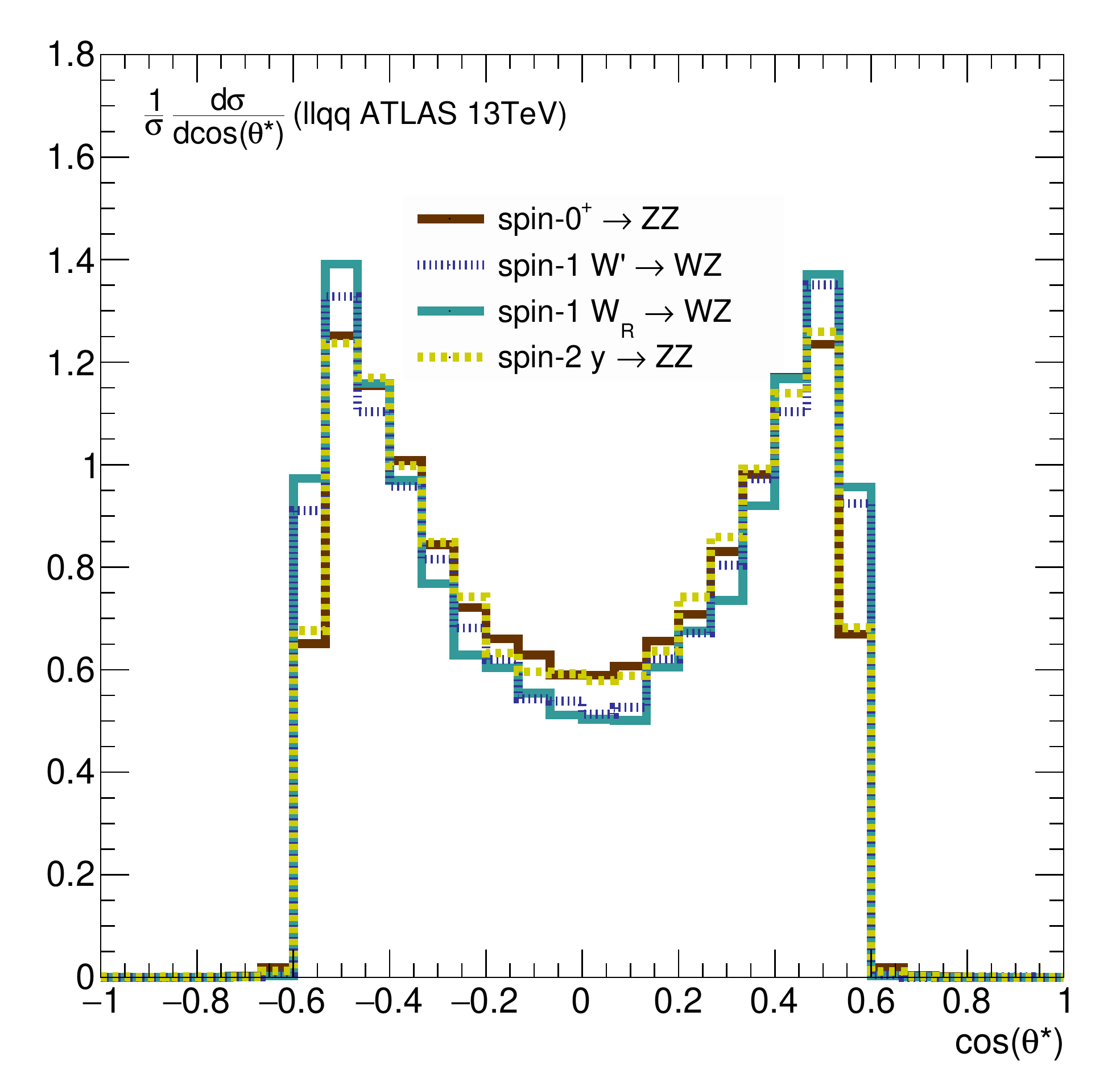} \\
\includegraphics[scale=0.27]{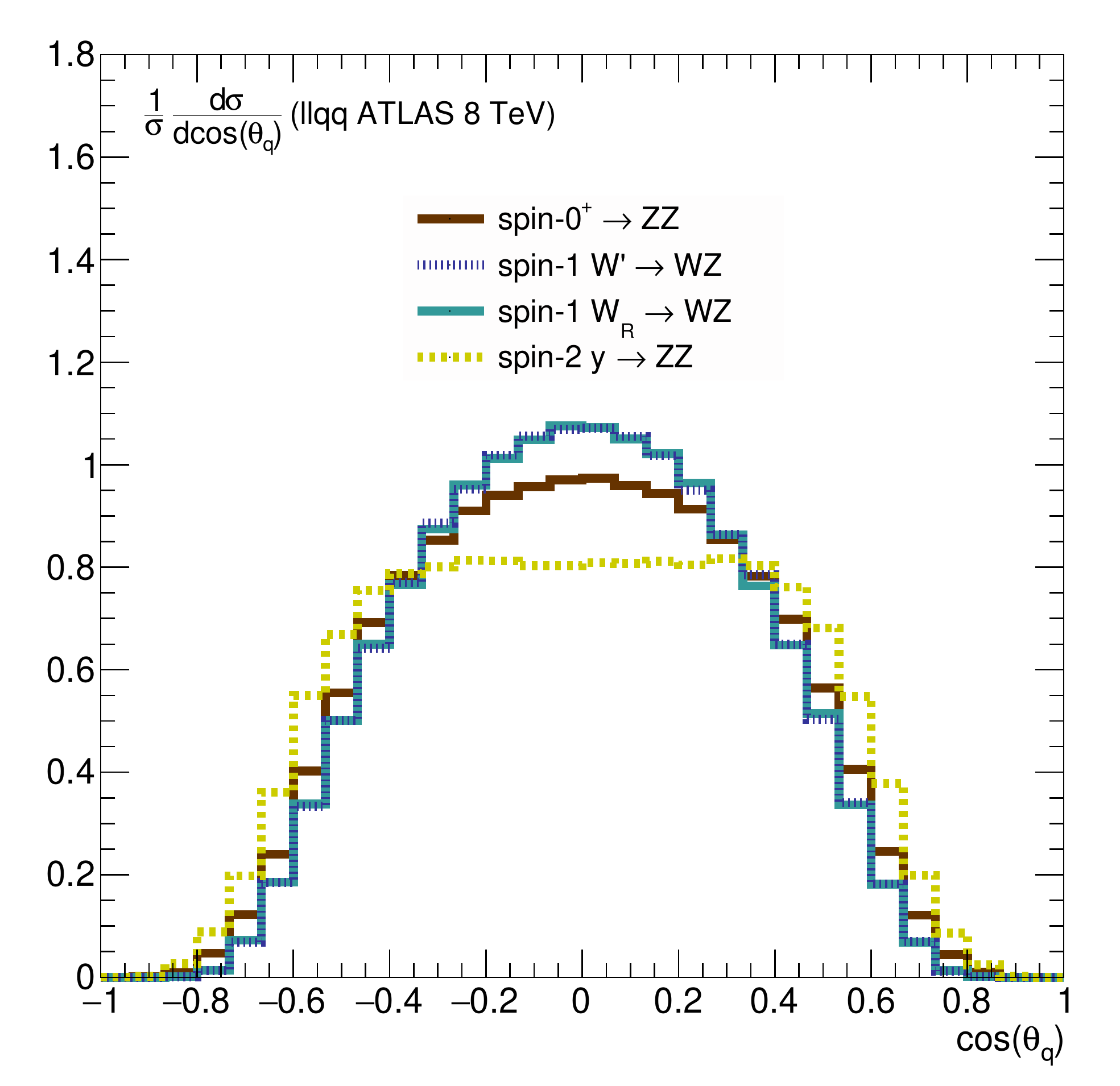}
\includegraphics[scale=0.27]{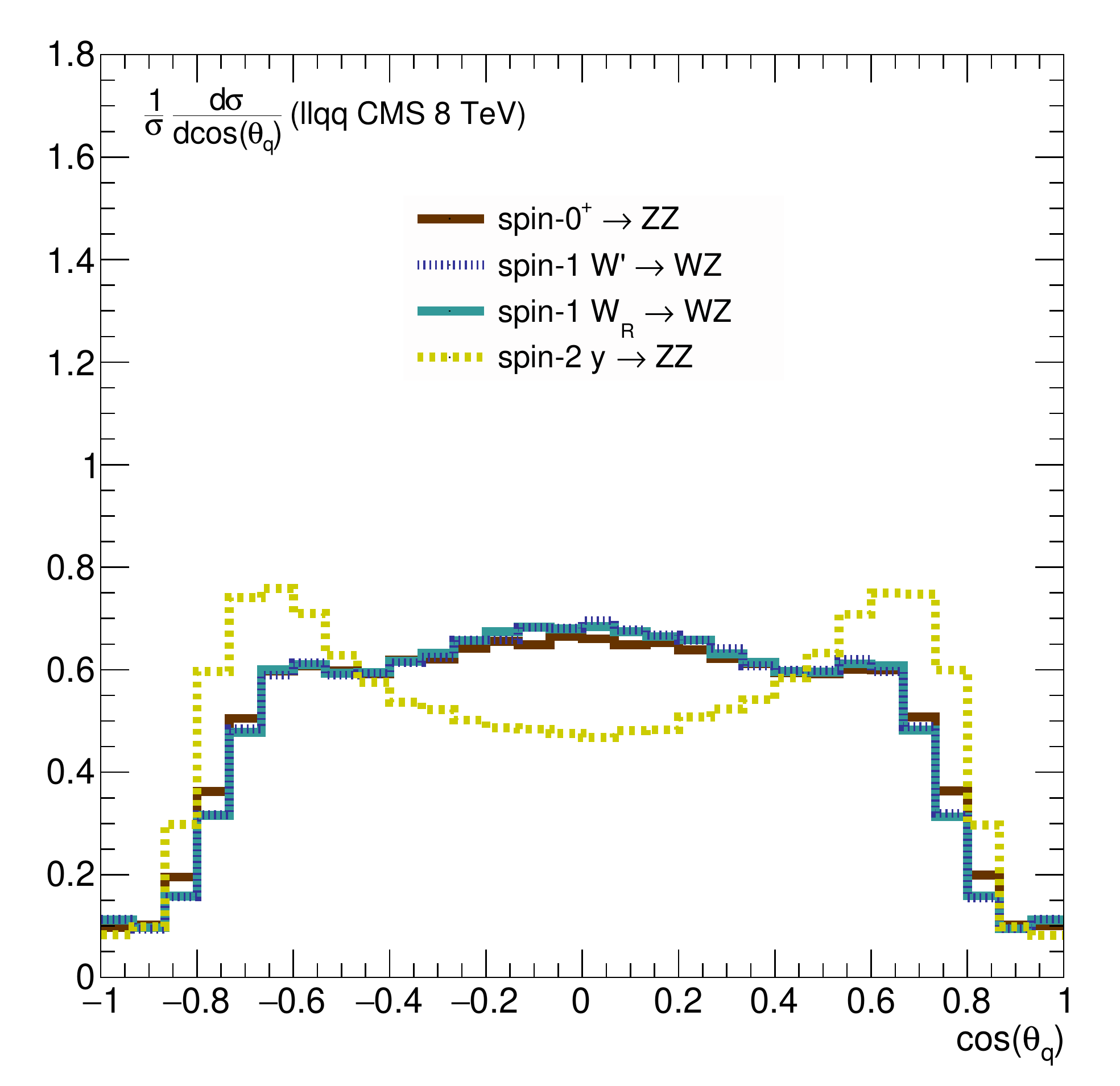}
\includegraphics[scale=0.27]{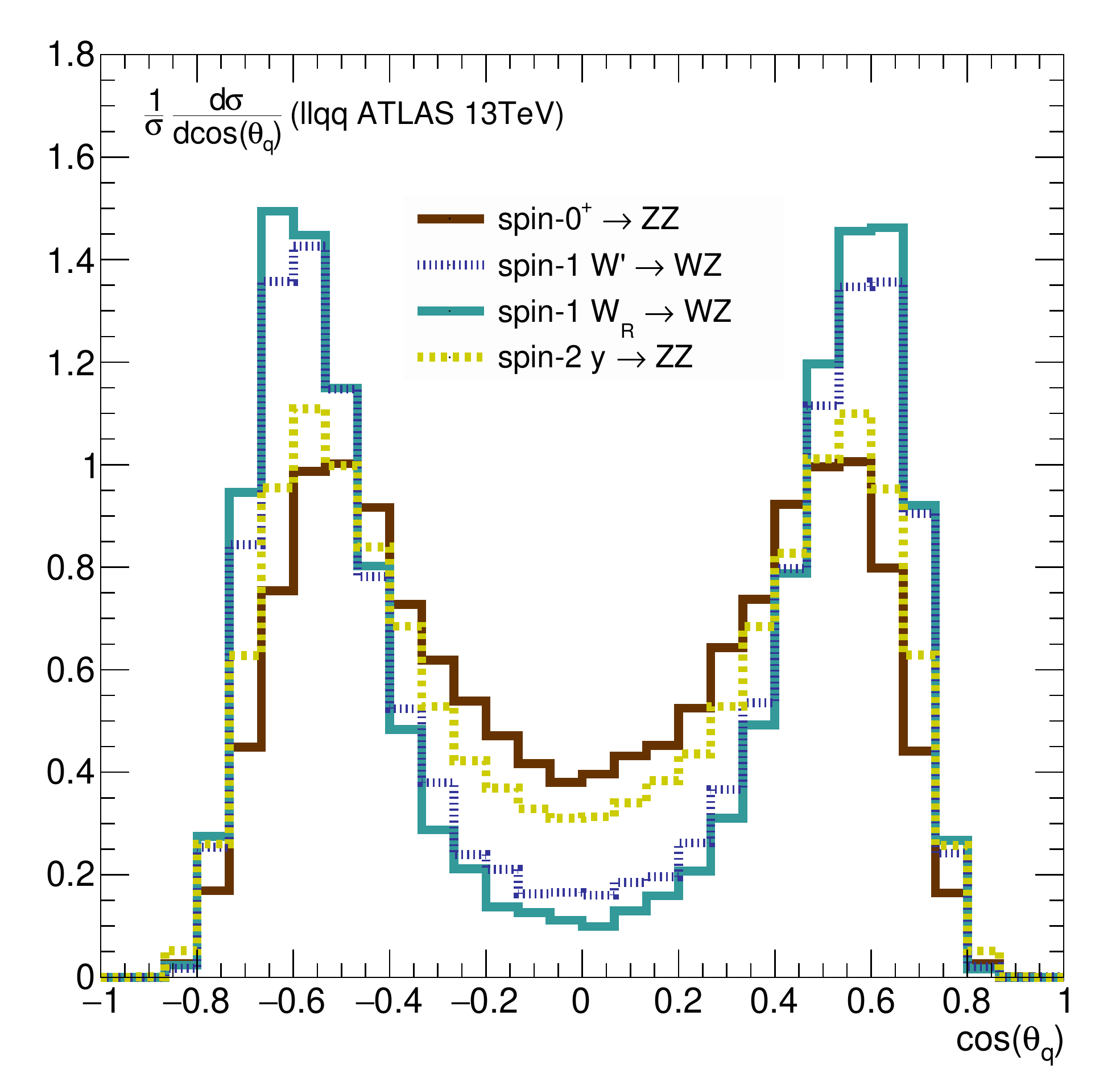} \\
\includegraphics[scale=0.27]{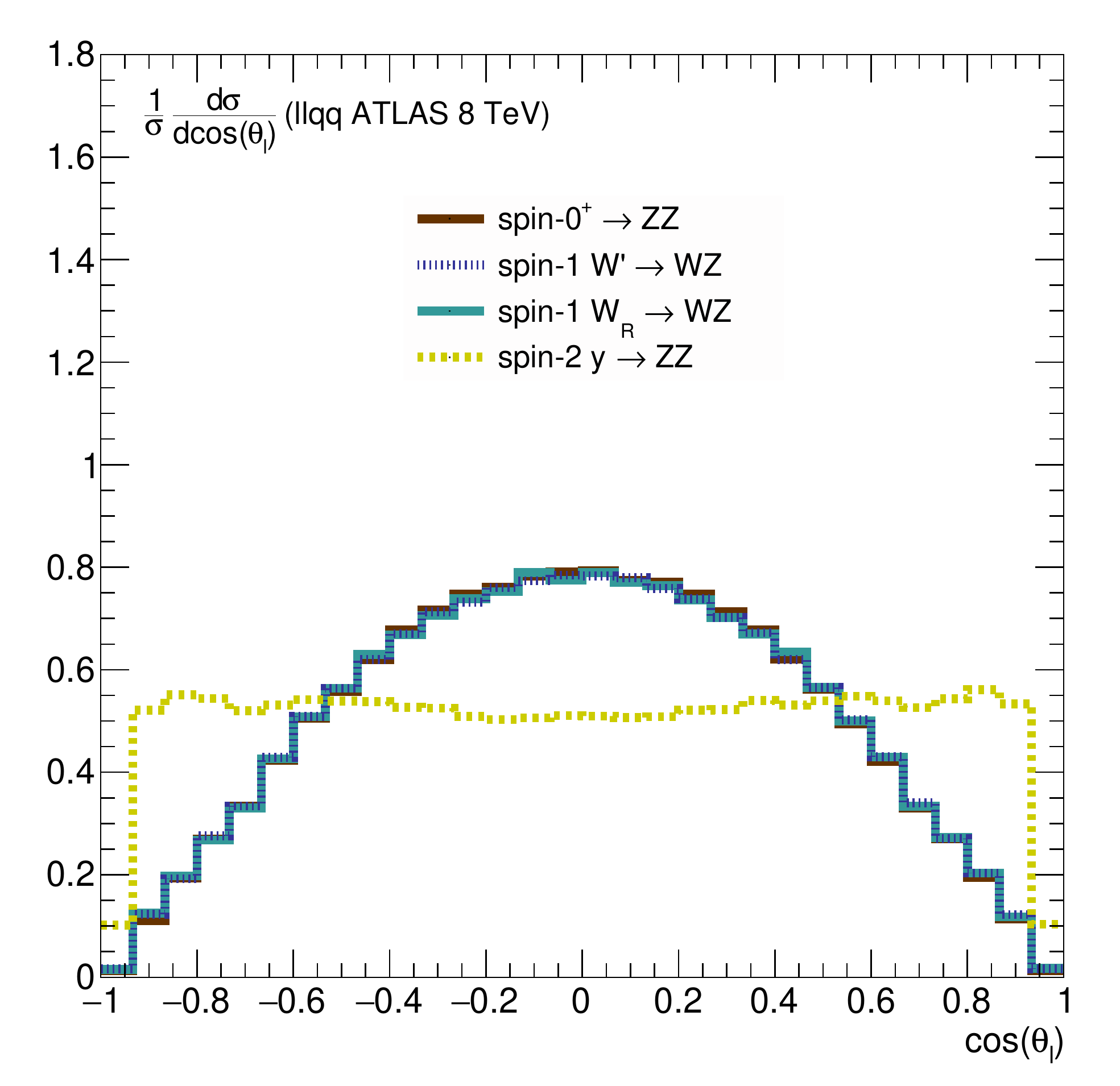}
\includegraphics[scale=0.27]{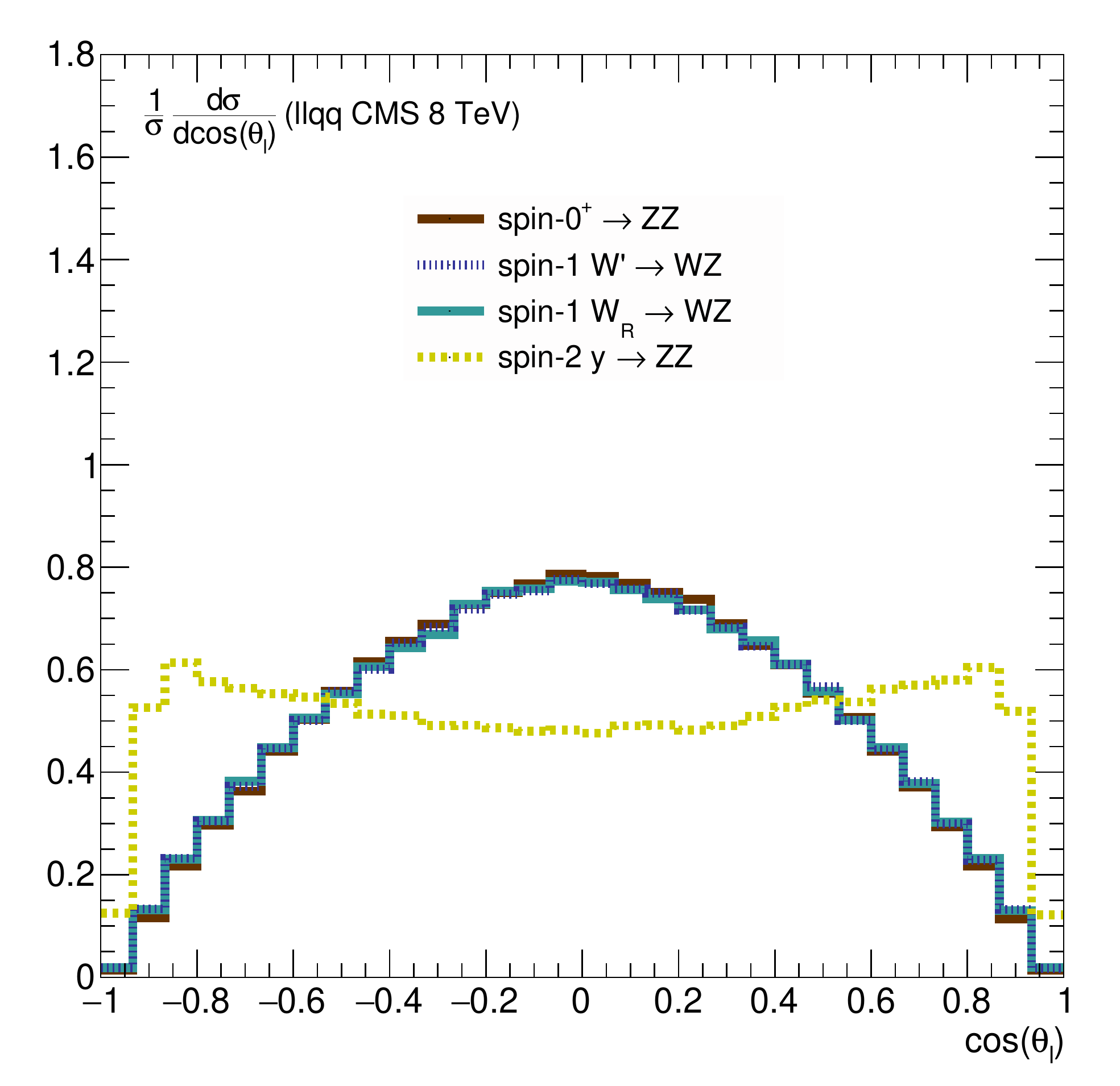}
\includegraphics[scale=0.27]{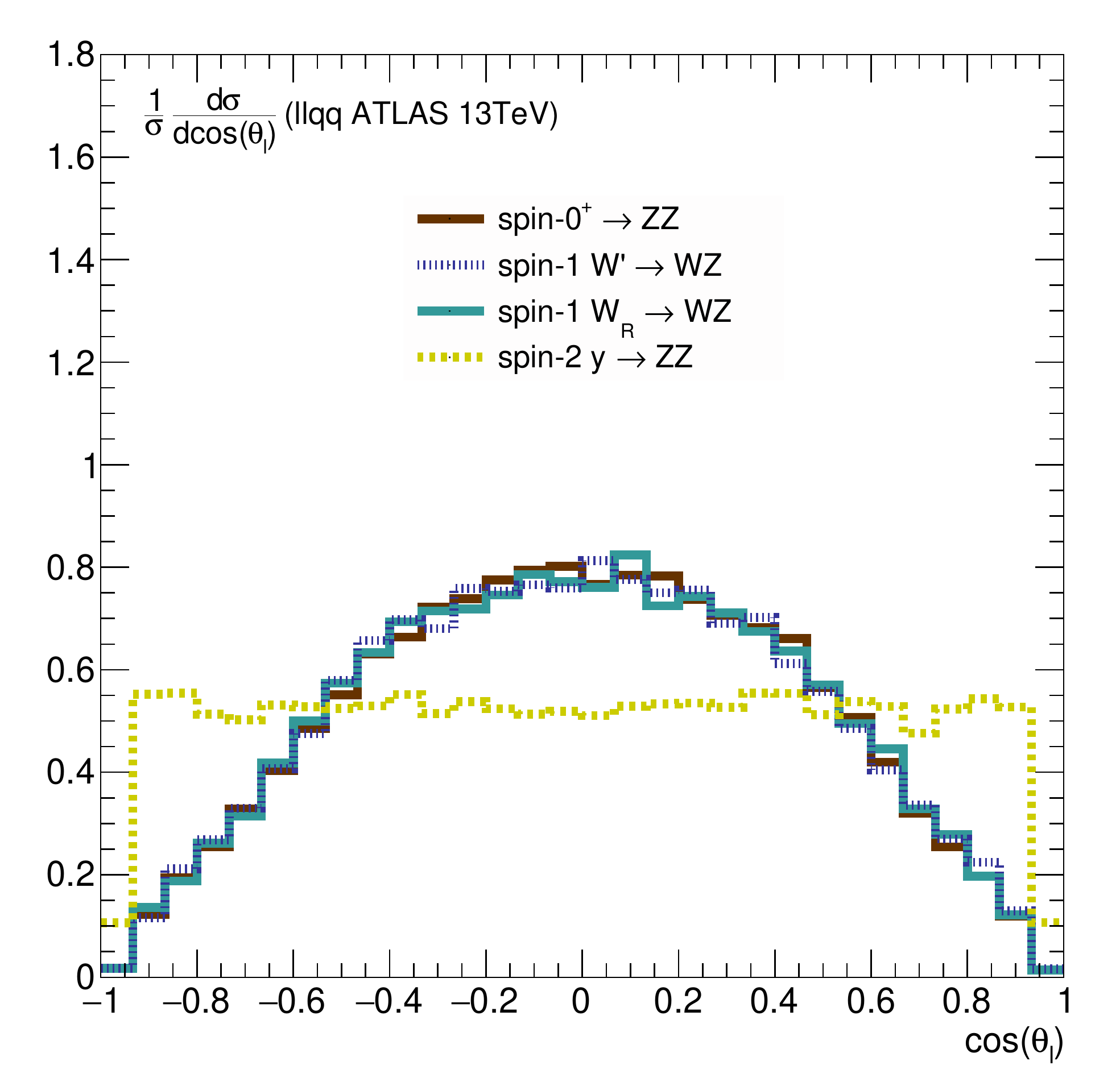} \\
\includegraphics[scale=0.27]{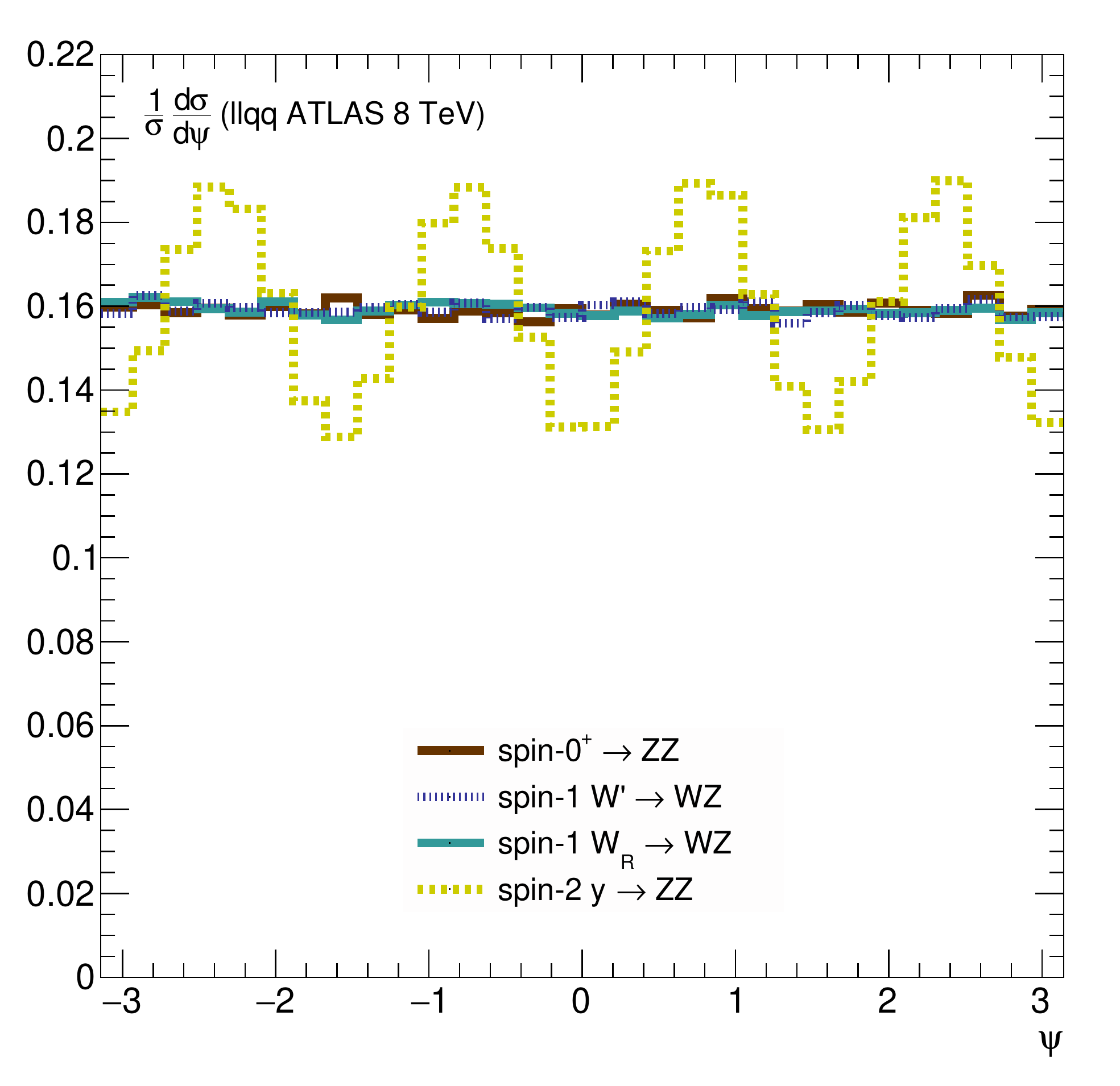}
\includegraphics[scale=0.27]{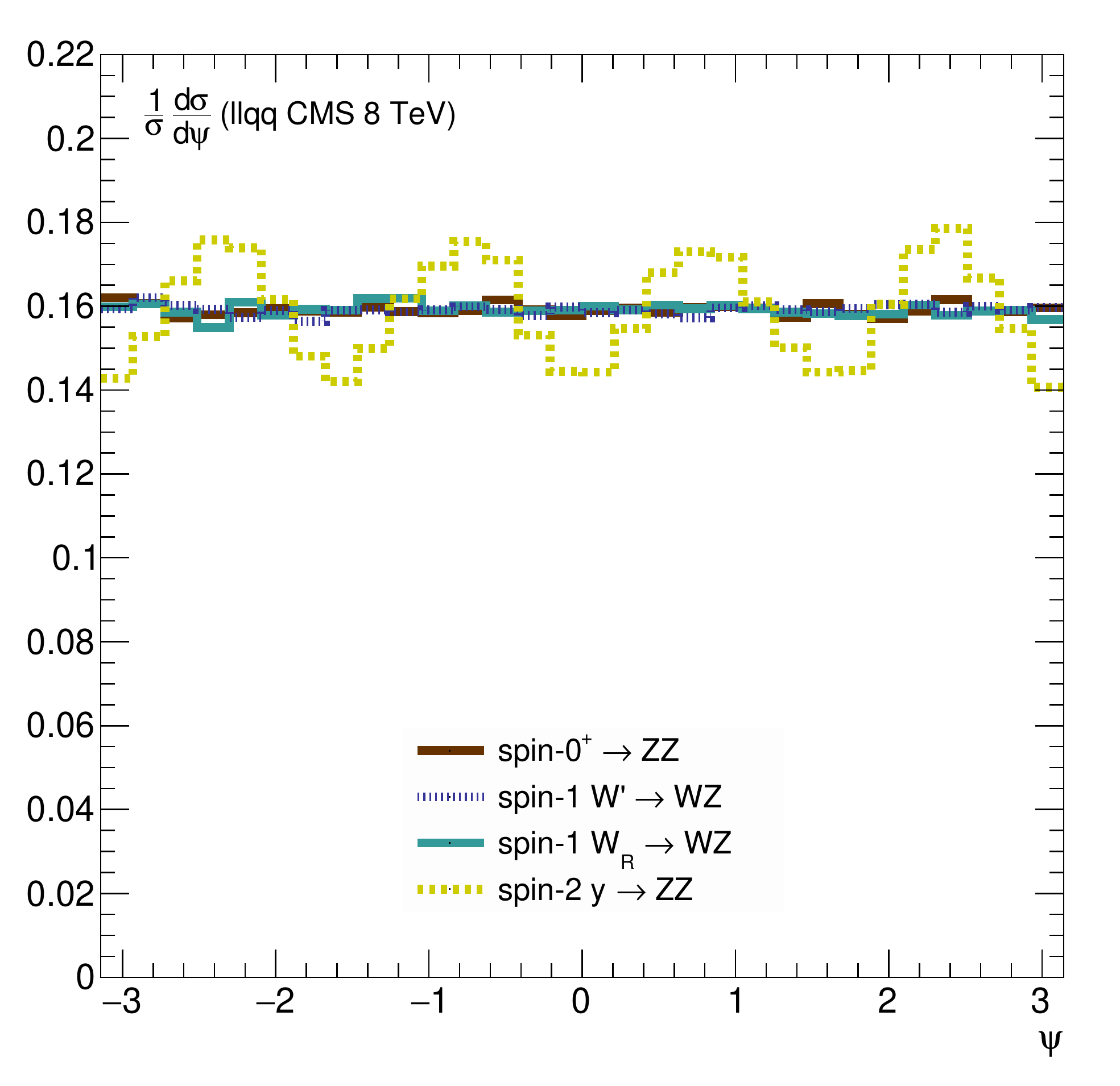}
\includegraphics[scale=0.27]{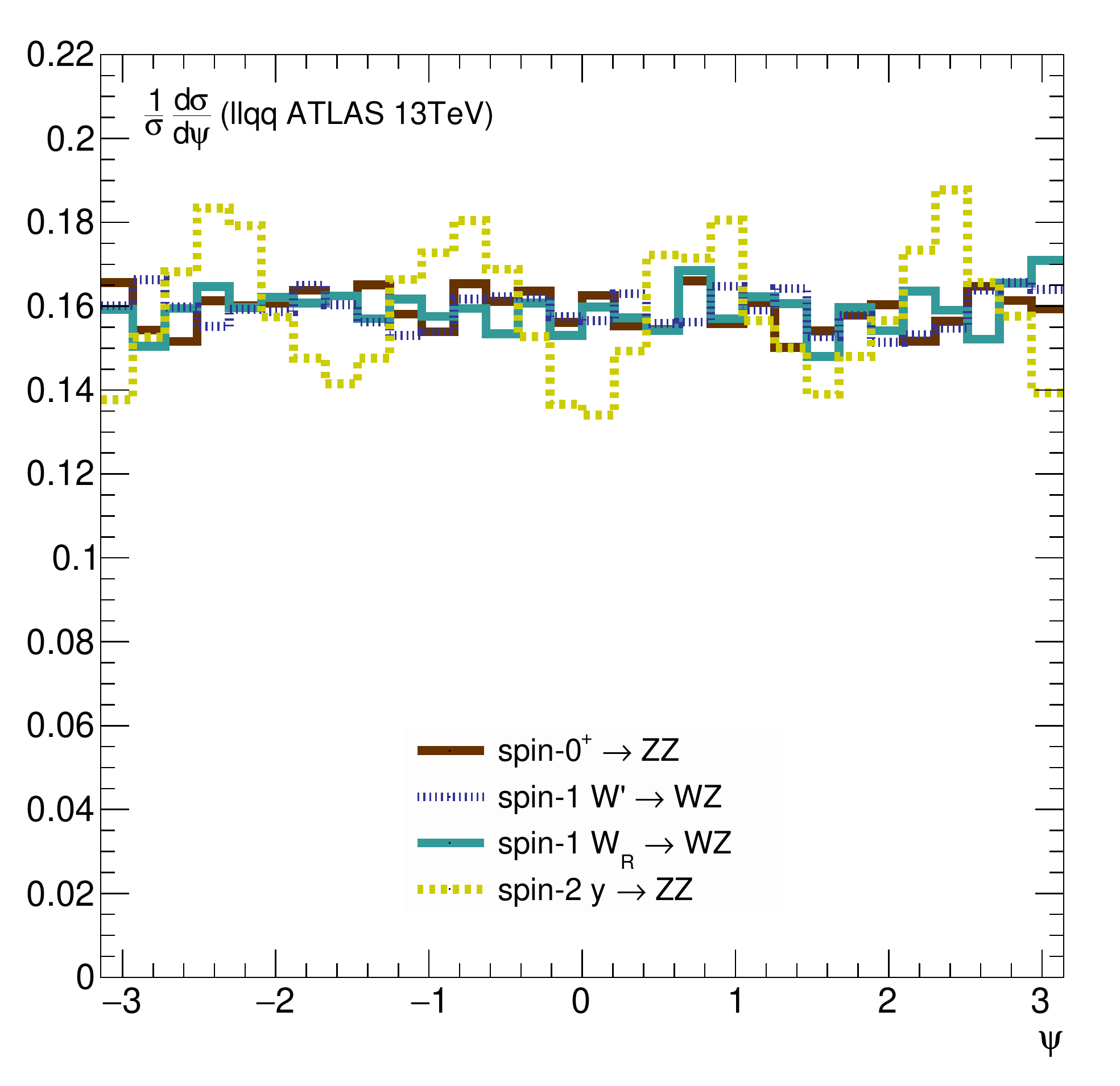} \\
\caption{\label{fig:MCLepto_LLQQ} Normalized differential
  distributions for $\cos \theta^*$ (top row), $\cos \theta_q$ (second
  row), $\cos \theta_l$ (third row), and $\Psi$ (bottom row) angles in
  the semi-leptonic final state $\ell \ell qq$, after imposing the
  ATLAS 8~TeV (left column), CMS 8~TeV (middle), and ATLAS 13~TeV
  (right) analysis cuts.}
\end{center}
\end{figure}

In Fig.~\ref{fig:MCLepto_LLQQ}, we show the normalized distributions
for the $\cos \theta^*$, $\cos \theta_q$, $\cos \theta_l$, and $\Psi$
angles for the relevant ATLAS and CMS $\ell \ell qq$ analyses.  Note
that we do not show the $\ell \ell qq$ background or the parton-level
results in this plots.  The $\ell \ell qq$ final state mimics the $4q$
final state, since the entire $X \to \ell \ell qq$ system is in
principle reconstructible.  Moreover, as mentioned before, the $\cos
\theta_q$ distribution for the $4q$ final state splits into the new
$\cos \theta_q$ and $\cos \theta_l$ angles, because the final state
partons are distinguishable.  On the other hand, the $\ell \ell qq$
final state pays an intrinsic penalty in statistical power, since the
branching ratio Br$(W^\pm Z \to \ell \ell qq)$ / Br$(W^\pm Z \to 4q)
\approx 0.094$, for $\ell = e$, $\mu$, is only partially mitigated by
an improved semileptonic signal efficiency.  Thus, the $4q$ and
semileptonic channels play important complementary roles both in the
discovery of a new resonance but also give significant cross-checks
for spin discrimination.

From Fig.~\ref{fig:MCLepto_LLQQ}, we see that angular observables
again provide important discrimination power between spin-2 and the
other spin hypotheses, while the main sensitivity to distinguish
spin-0 from spin-1 resonances comes from the $\cos \theta^*$ angle.
The sculpting effects we identified earlier are still evident in $\cos
\theta_q$ as a result of the jet substructure cuts, but on the other
hand, most of the phase space is preserved for the $\cos \theta_l$
distribution. Note that there is no $p_T$ 
requirement on the individual subjets in contrast to the hard cut on the 
lepton $p_T$. This effectively flattens the $\cos \theta_l$ shape
for the spin-2 resonance compared to $\cos \theta_q$, 
as events with large lepton $p_T$ imbalance 
near $\cos \theta_l=\pm 1$ tends to miss one of the leptons.

One interesting feature is the sharp cliff in $\cos \theta^*$ for the
ATLAS 13~TeV analysis, shown in the top row, rightmost panel of
Fig.~\ref{fig:MCLepto_LLQQ}.  This is directly connected to the
$p_T^{\ell \ell} > 0.4 m_{\ell \ell J}$ and $p_T^J > 0.4 m_{\ell \ell
  J}$ cuts, because from Eq.~\ref{eqn:m0pTpTdR}, we see that the
corresponding maximum pseudorapidity gap between the vector boson
candidates is $\Delta \eta_{\text{max}} \sim 2.1$, which leads to a
maximum of $|\cos \theta^*| = 0.6$.  We also note the ATLAS 8~TeV
analysis has cliffs at $|\cos (\theta^*)| \lesssim 0.92$ in the ATLAS
8 TeV analysis, driven by their milder cuts on $p_T^{\ell \ell} > 400$
GeV and $p_T^J > 400$ GeV.  

In this regard, the most discrimination
power between the various spin scenarios follows from the CMS 8~TeV
analysis, where the spin-0 and spin-2 curves are readily distinguished
from the spin-1 shapes.  In contrast, the ATLAS 13~TeV analysis molds
the $\cos \theta^*$ distribution to eliminate any possibility of
distinguishing these different spins.

\begin{figure}[tb!]
\begin{center}
\includegraphics[scale=0.21]{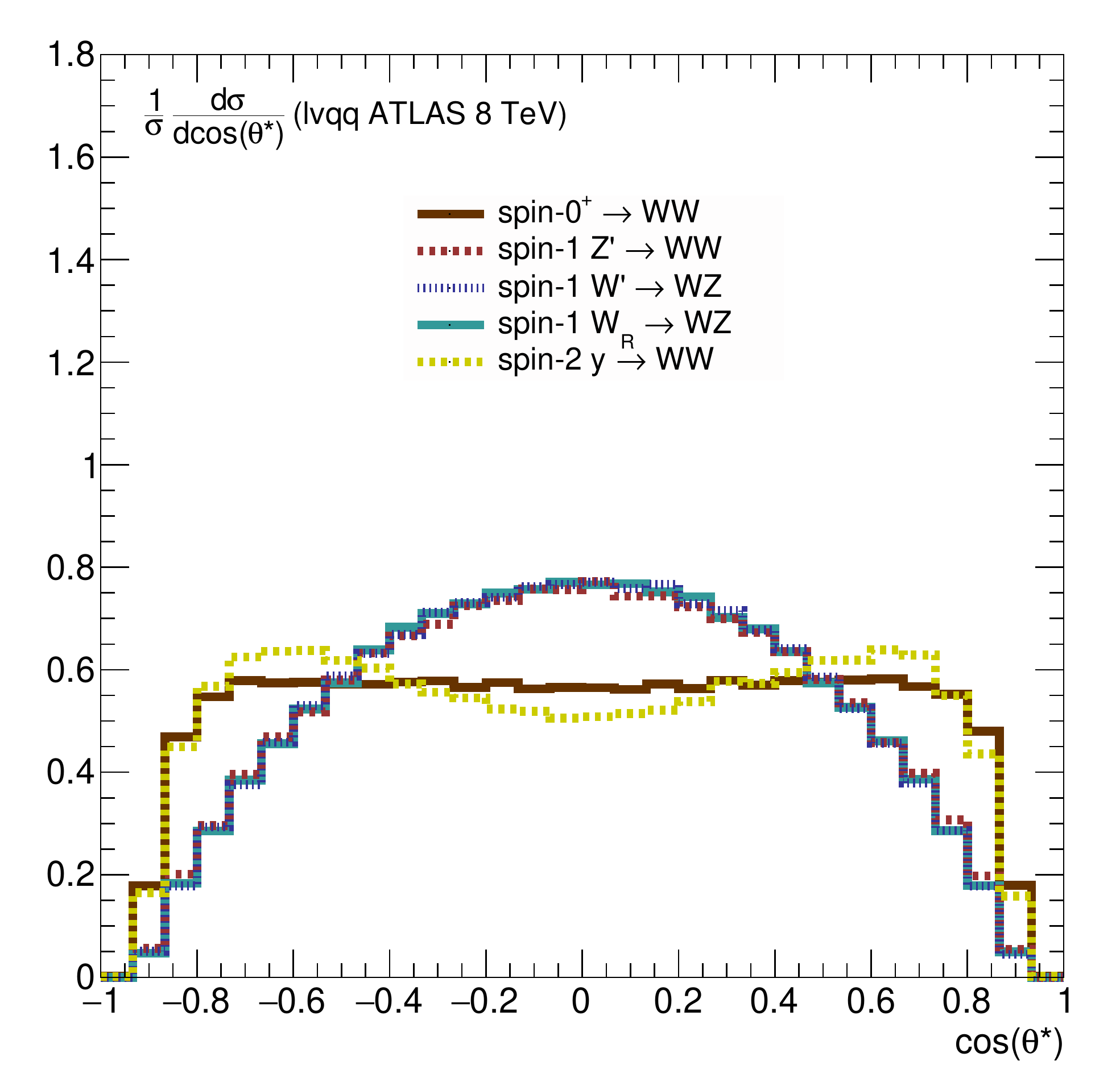} \hspace{-10px}
\includegraphics[scale=0.21]{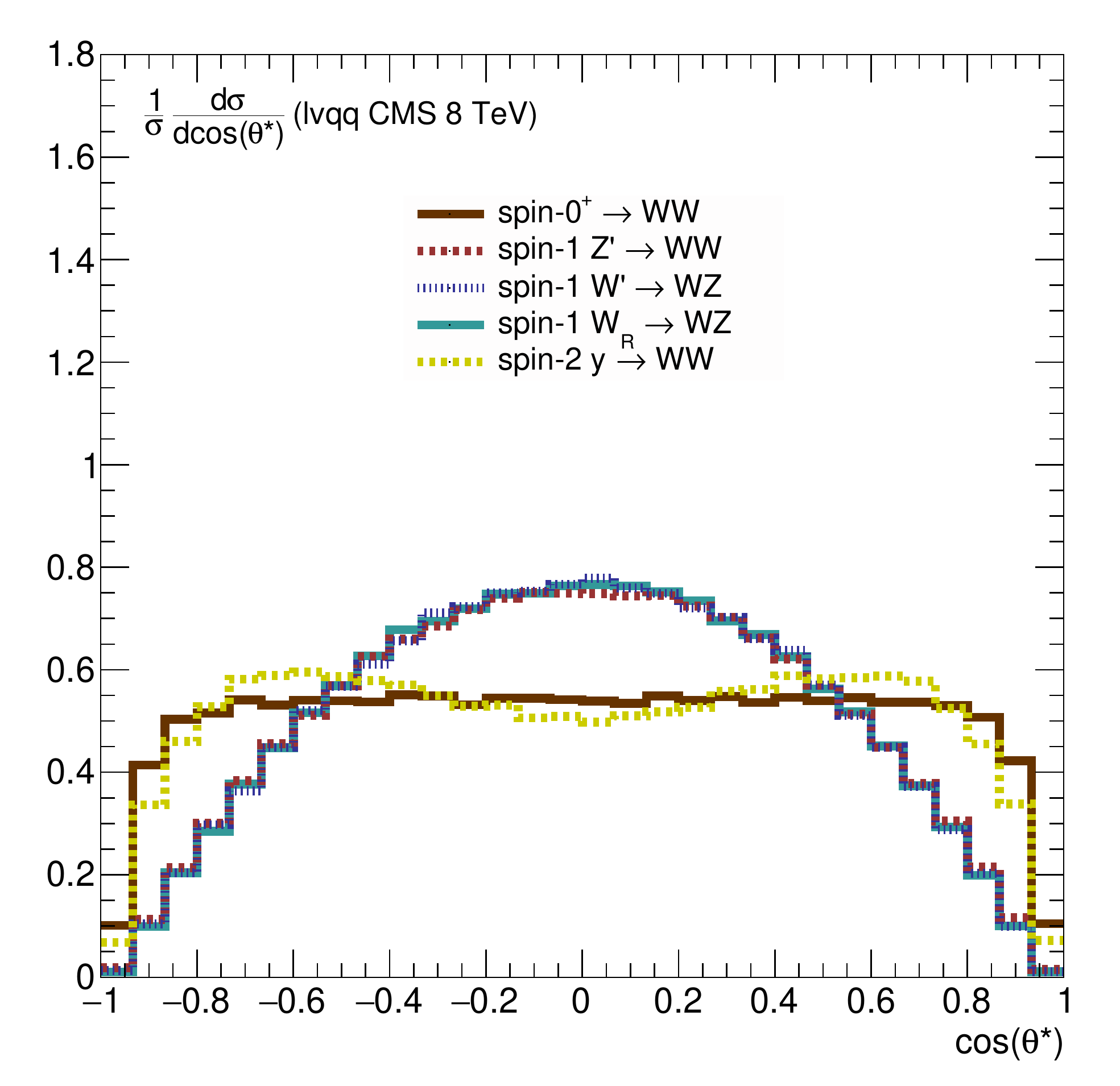} \hspace{-10px}
\includegraphics[scale=0.21]{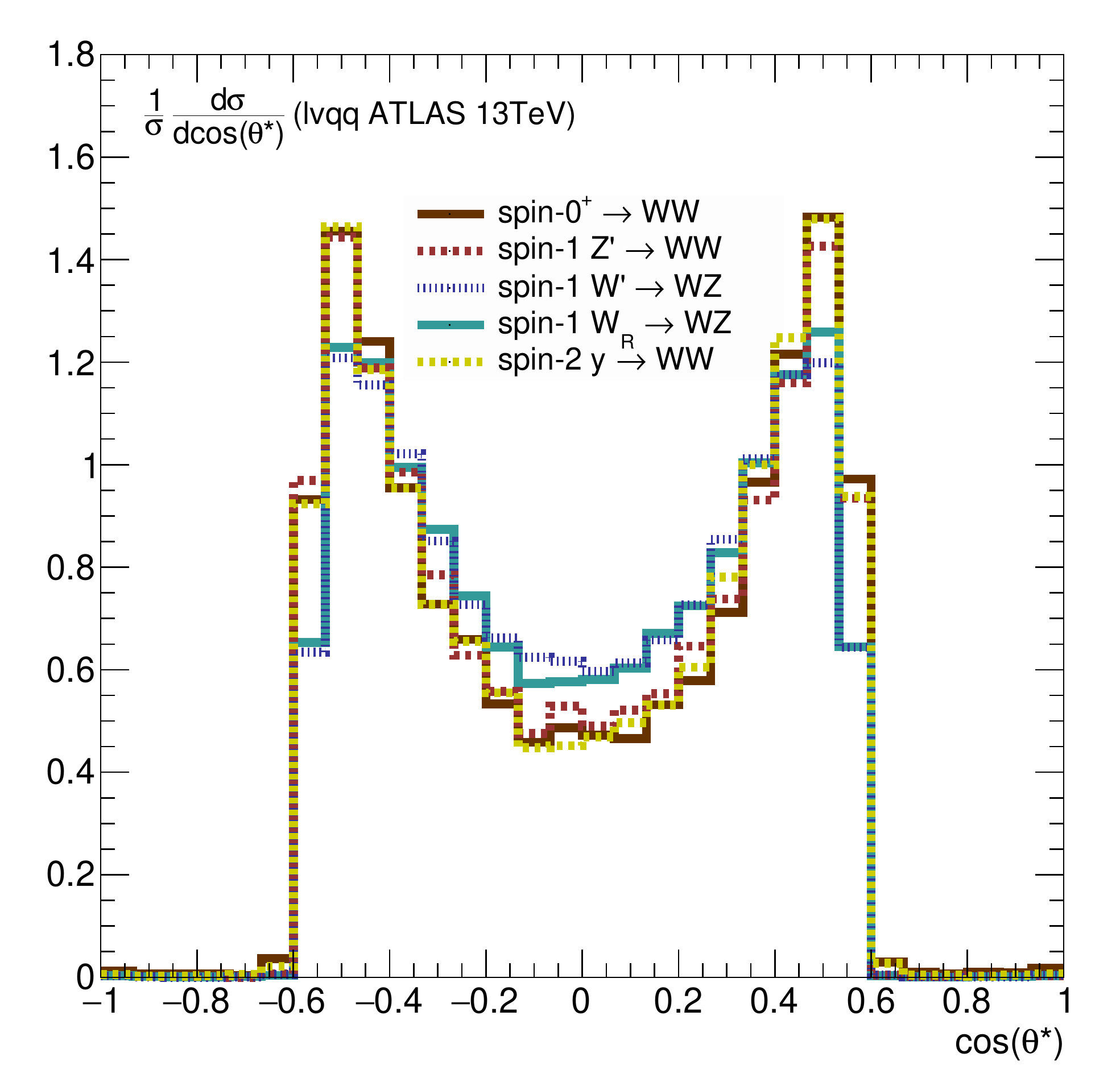} \hspace{-10px}
\includegraphics[scale=0.21]{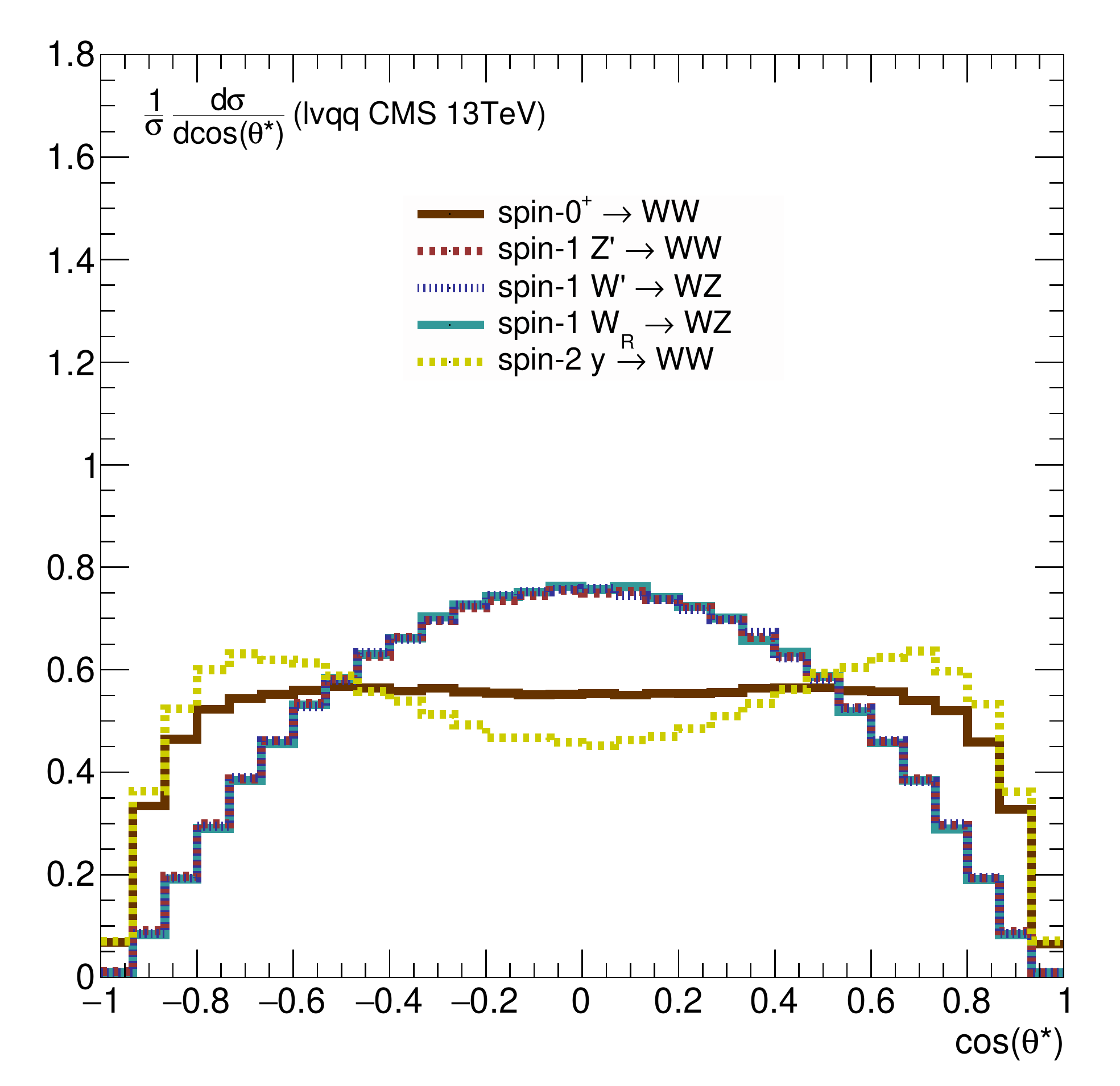} \\
\includegraphics[scale=0.21]{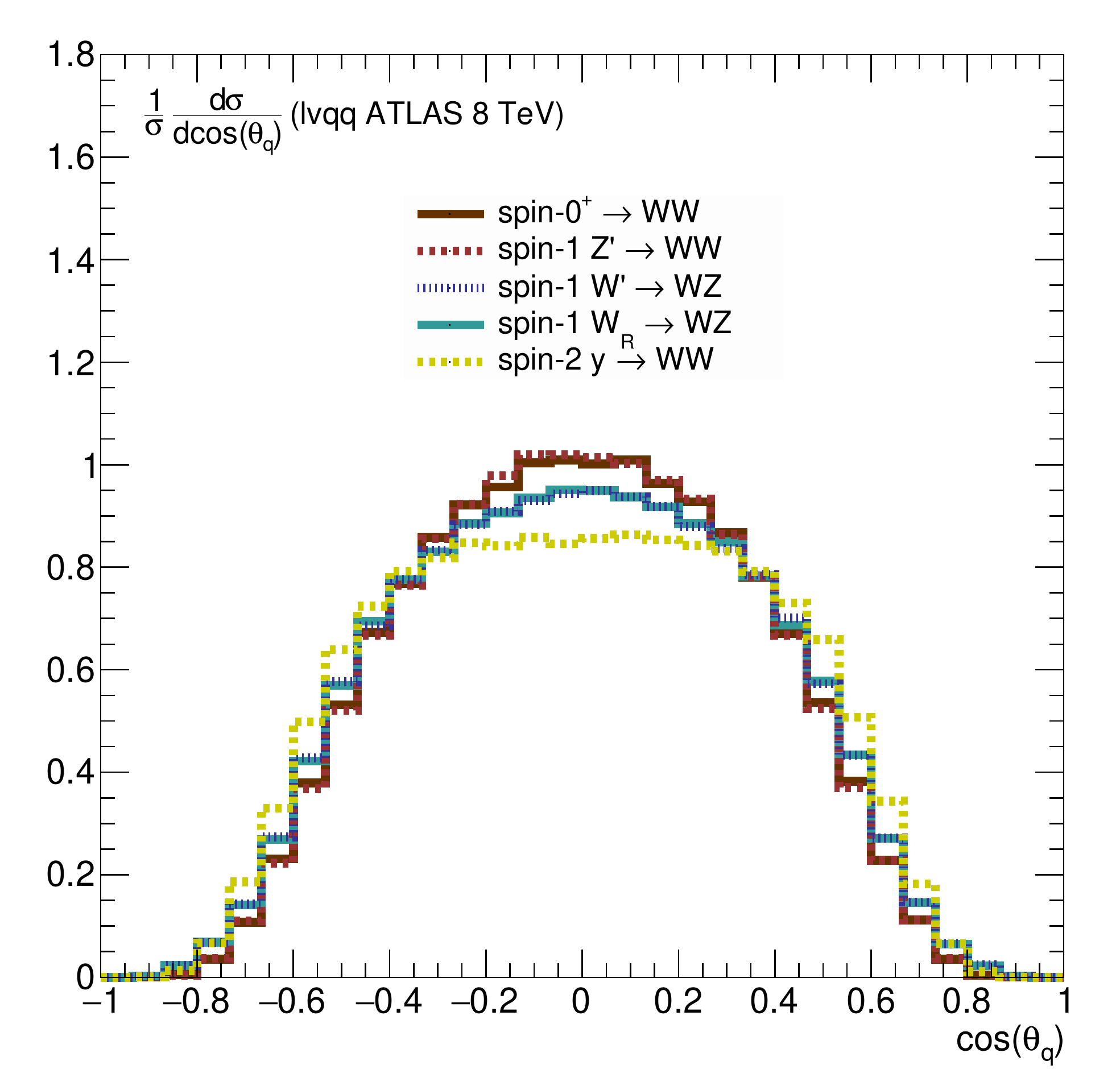} \hspace{-10px}
\includegraphics[scale=0.21]{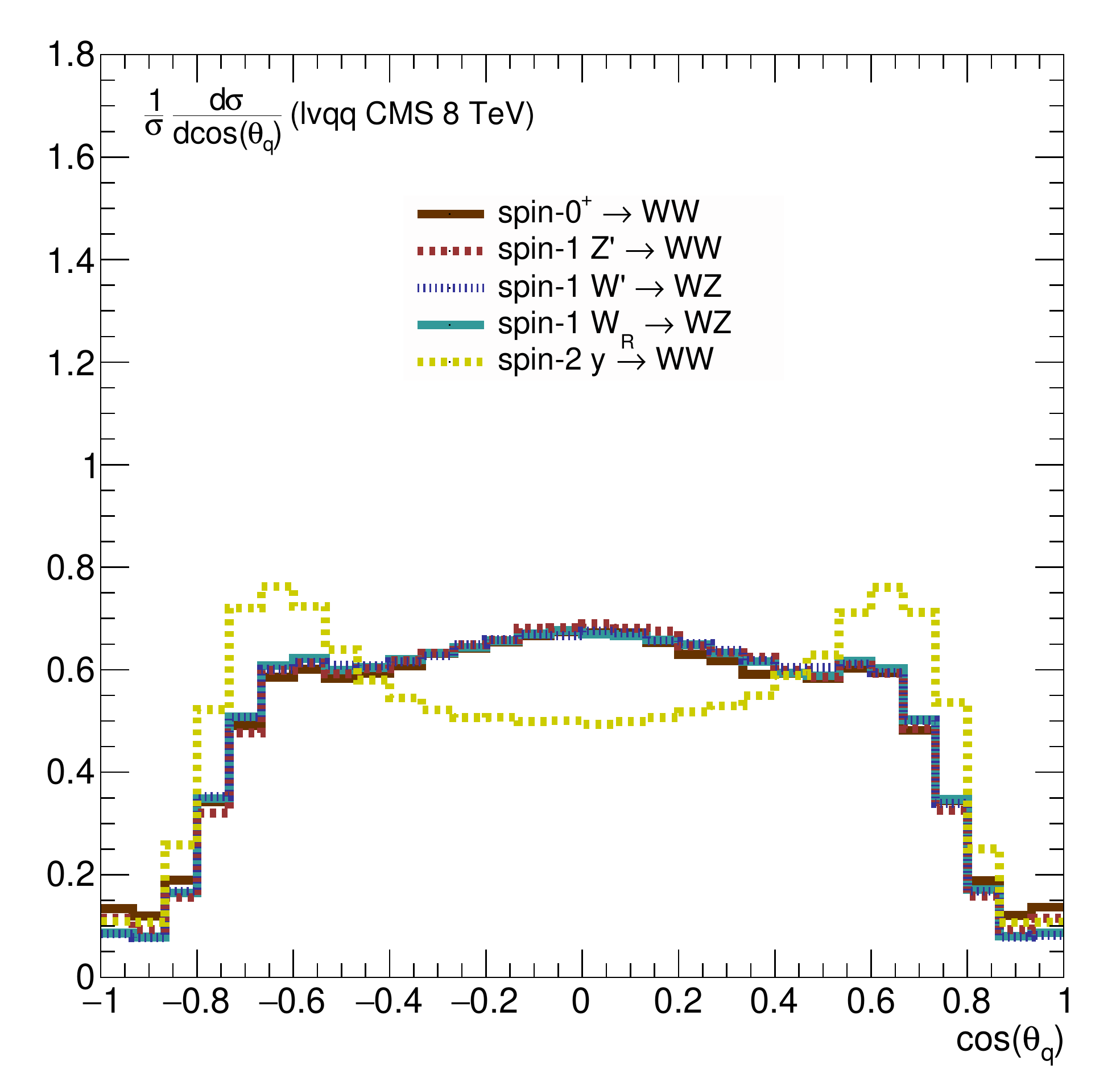} \hspace{-10px}
\includegraphics[scale=0.21]{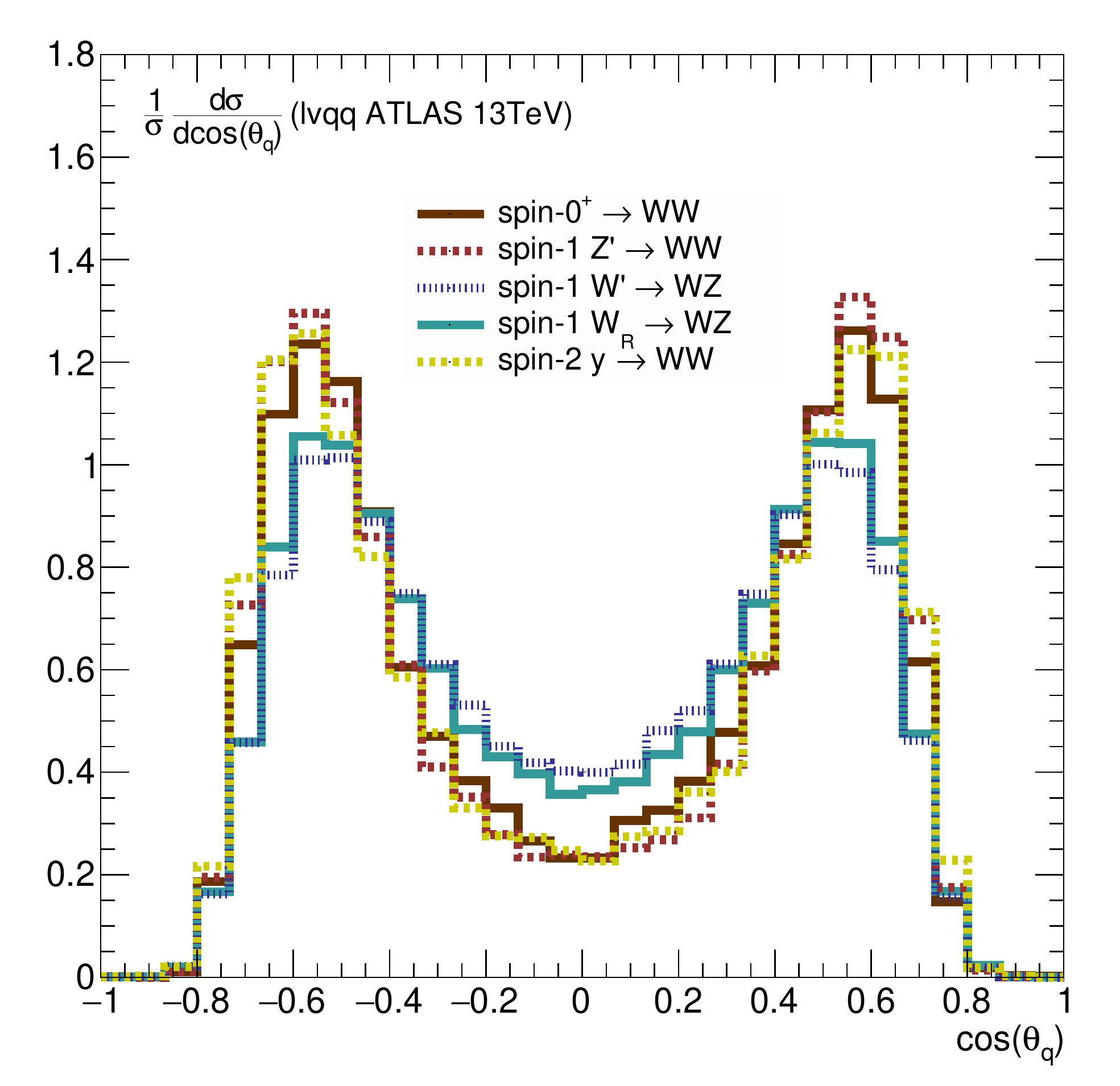} \hspace{-10px}
\includegraphics[scale=0.21]{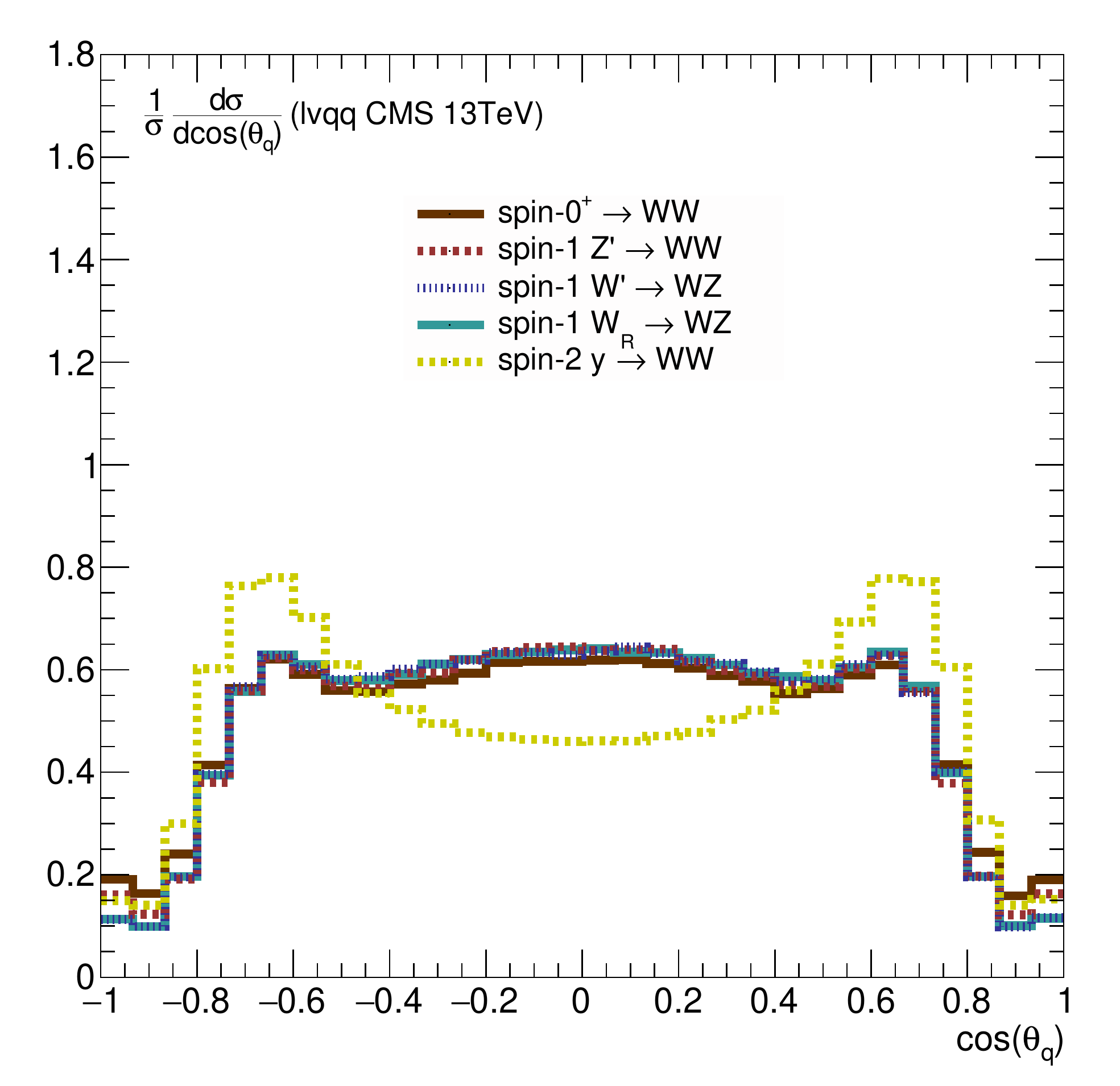} \\
\includegraphics[scale=0.21]{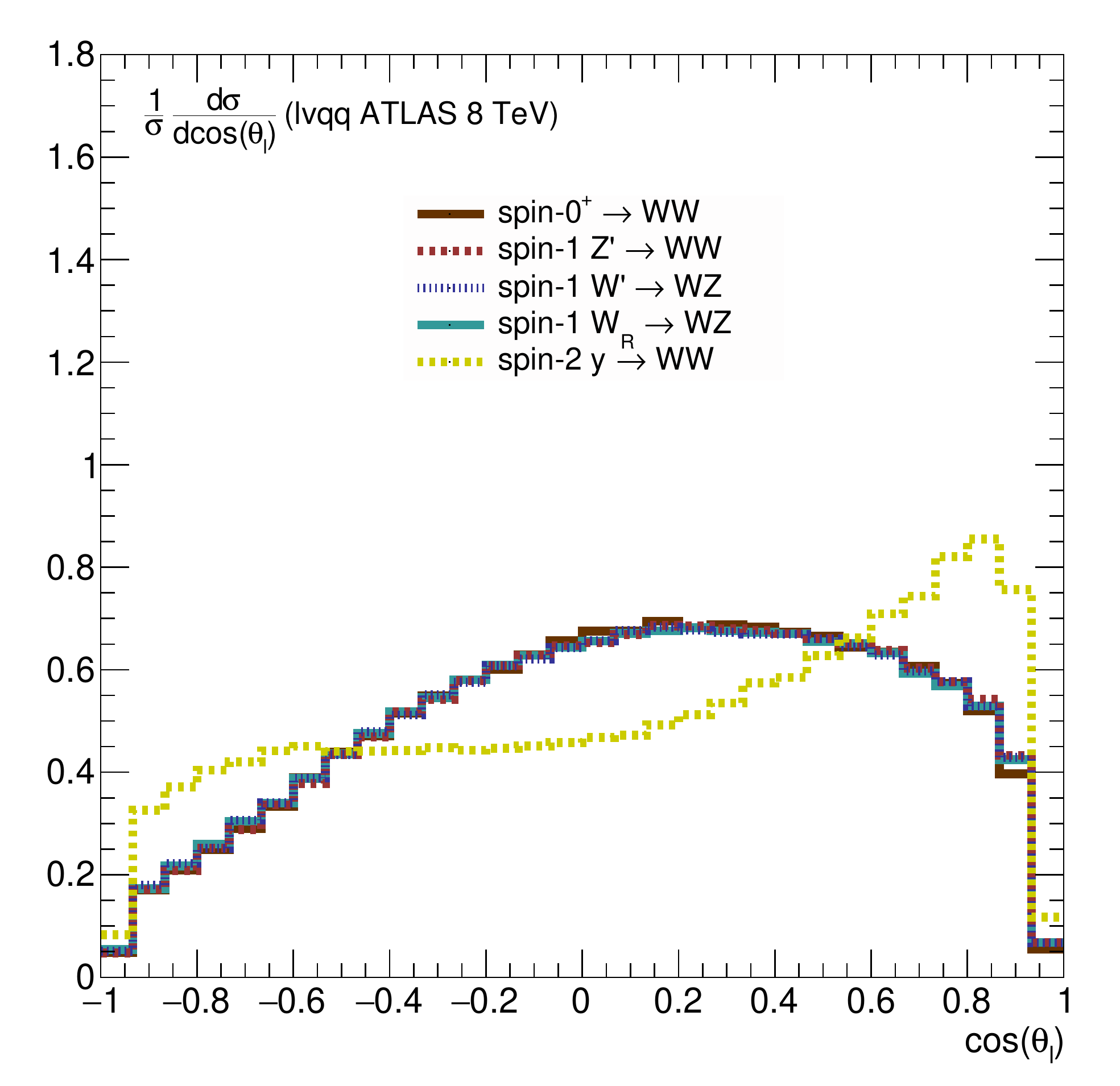} \hspace{-10px}
\includegraphics[scale=0.21]{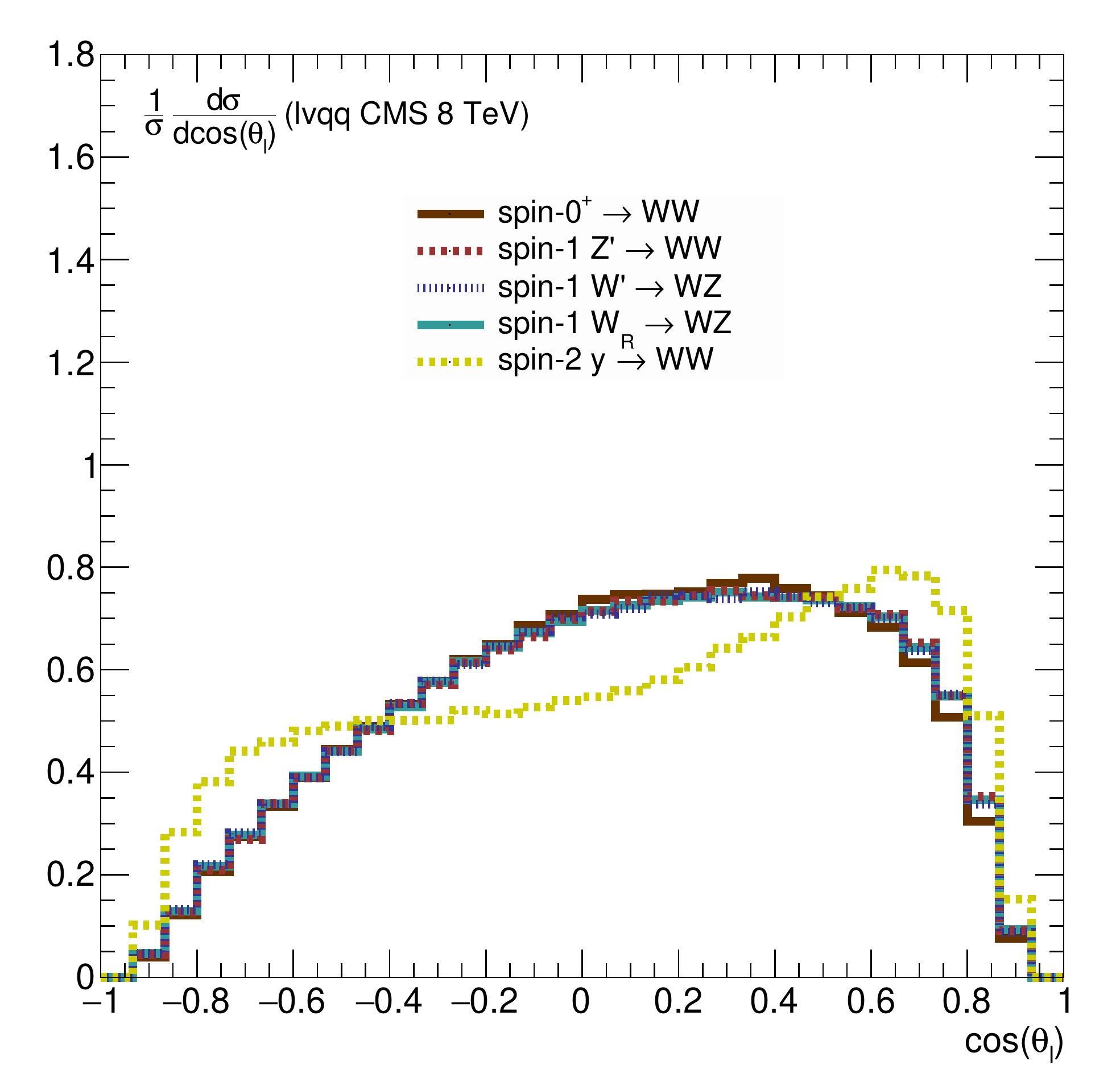} \hspace{-10px}
\includegraphics[scale=0.21]{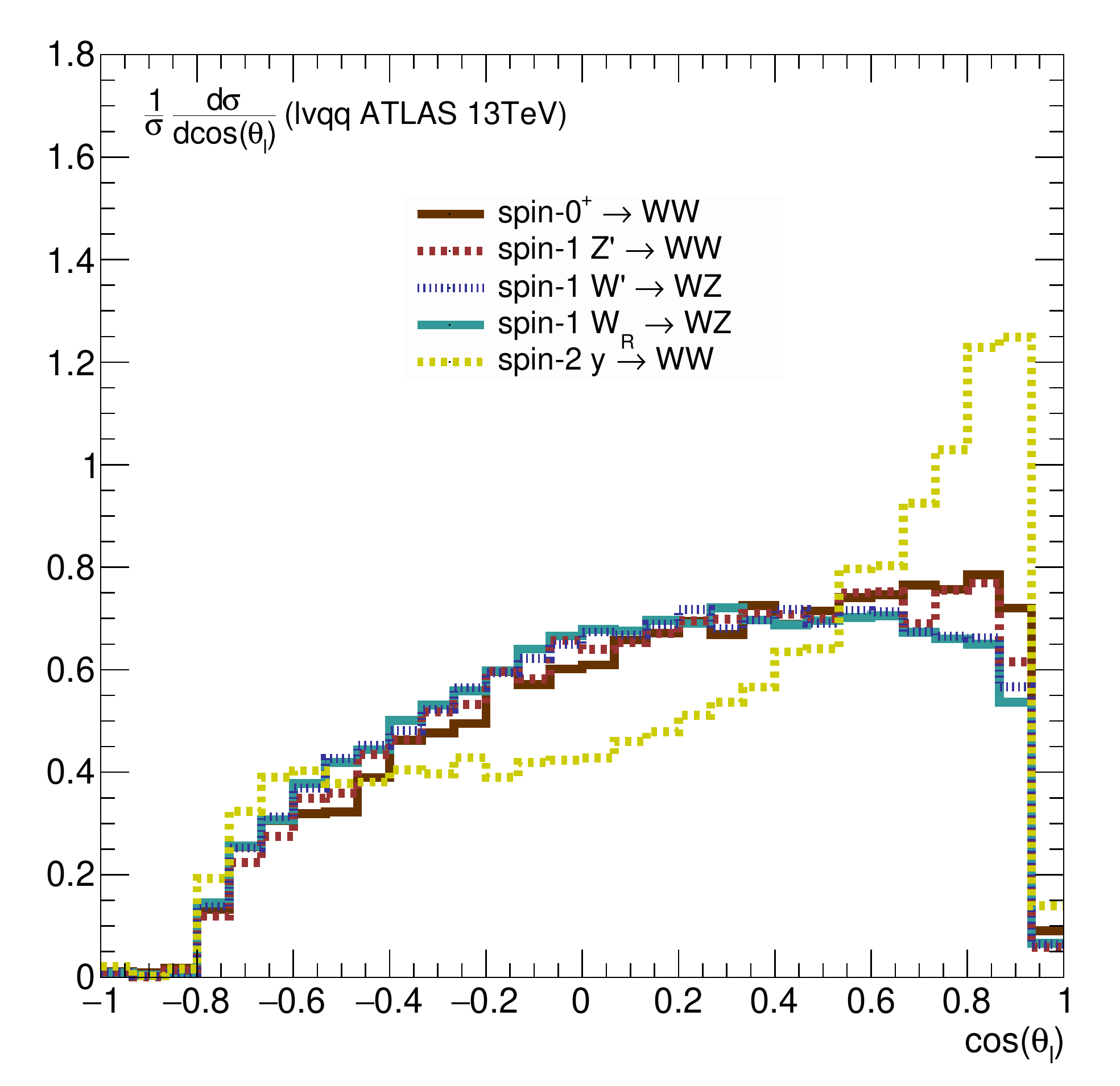} \hspace{-10px}
\includegraphics[scale=0.21]{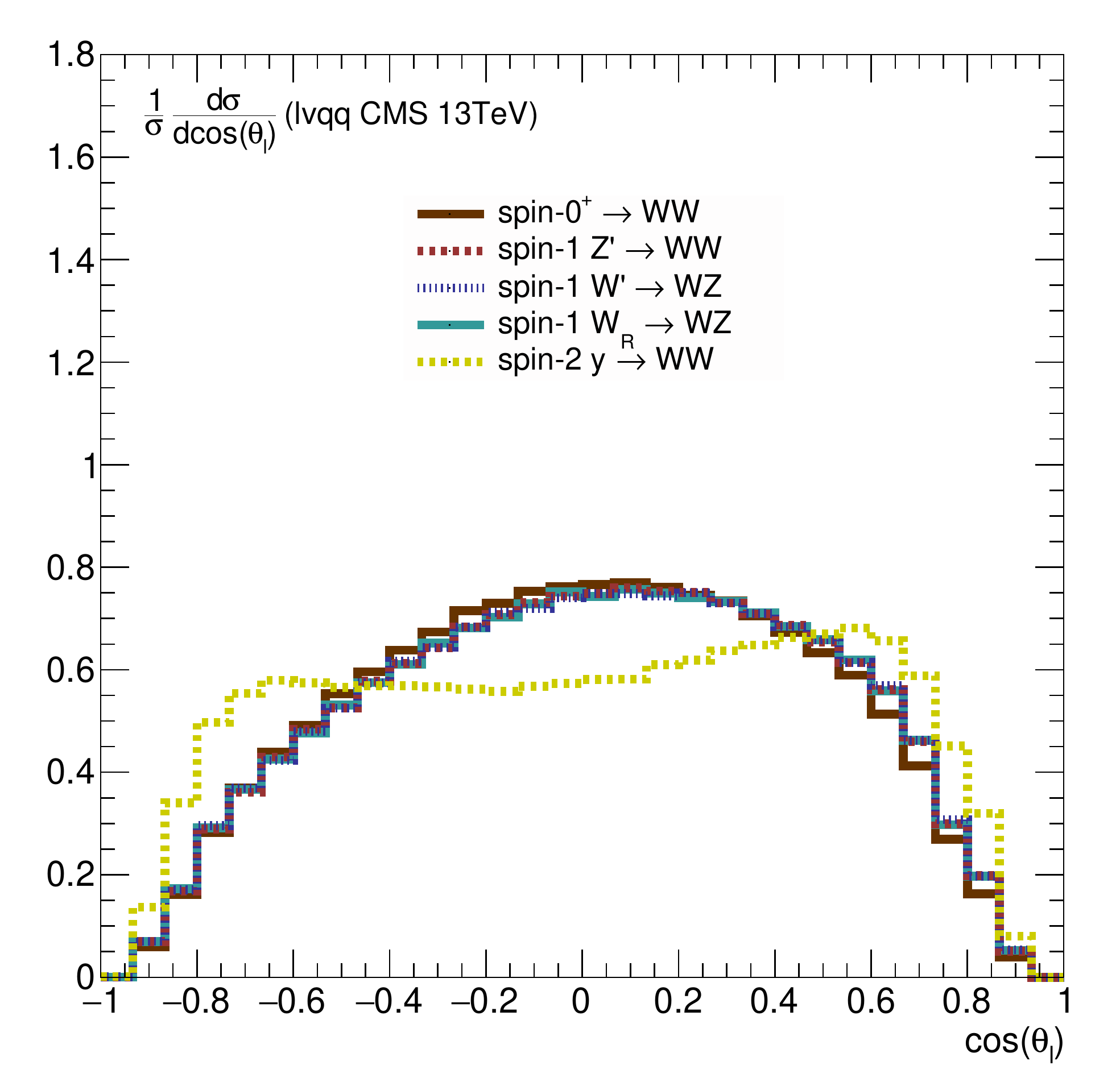}
\caption{Normalized differential distributions for $\cos \theta^*$
  (top row), $\cos \theta_q$ (middle row), and $\cos \theta_l$ (bottom
  row), for the semi-leptonic final state $\ell \nu qq$, after
  imposing the ATLAS 8~TeV (first column), CMS 8~TeV (second column),
  ATLAS 13~TeV (third column), and CMS 13~TeV (last column) analysis
  cuts.  We omit the $\Psi$ angle as it does not have any significant
  discrimination power.}
\label{fig:MCLepto_LVQQ}
\end{center}
\end{figure}

In Fig.~\ref{fig:MCLepto_LVQQ}, we show the normalized distributions
for $\cos \theta^*$, $\cos \theta_q$, and $\cos \theta_l$ for the
ATLAS and CMS 8~TeV and 13~TeV analyses.  We remark that $\Psi$ has no
discriminating power between the signal hypotheses, so we omit it from
the figure.  The $\cos \theta_q$ distributions are similar to those
from before, while the $\cos \theta_l$ shows a novel asymmetry.  

The asymmetries in the $\cos \theta_l$ distributions are the result of
contamination by leptonic $\tau$ decays.  In particular, the extra
neutrinos from the $\tau \to e \nu \nu$ and $\tau \to \mu \nu \nu$
decays skew the reconstruction of the leptonic decay of the $W^\pm$,
where the additional neutrinos result in a false reconstruction of the
rest frame of the $W^\pm$.  This incorrect rest frame preferentially
groups the charged lepton used for the $\cos (\theta_l)$ calculation
closer to the boost vector needed to move to the $W^\pm$ rest frame,
skewing the $\cos \theta_l$ distribution toward the $+1$ edge.

We also note, analogous to the $\ell \ell qq$ final state, the clear
cliffs in the $\cos \theta^*$ distribution evident in the ATLAS 13~TeV
analysis.  These cliffs again arise from the $p_T^{\ell \nu} > 0.4
m_{\ell \nu J}$ and $p_T^J > 0.4 m_{\ell \nu J}$ cuts, which
effectively enforce a $|\cos \theta^*| = 0.6$ maximum, as discussed
before.
\clearpage

\section{Projections for model discrimination from $4q$ final state}
\label{sec:projections}
We now quantify the discrimination power between the different spin
scenarios using the $CL_s$ method~\cite{Read:2002hq} to test one
signal against another in the $4q$ final state.  We define one signal
resonance plus dijet background as a signal hypothesis, whereas the
test hypothesis is a different spin resonance plus the same dijet
background.  We use the differential shapes $|\cos \theta^*|$, $|\cos
\theta_q|$, and $|\Psi|$ as individual discriminators as well as a
likelihood combination using all three observables.

We perform the pairwise signal hypothesis tests first using shape
information alone and second using both shape and rate information.
The normalized differential distributions serve as a first test for
signal comparisons, because, by construction, different models for a
newly discovered resonance will have the same fiducial signal cross
section in order to match the observed excess.  Hence, even if the
2~TeV excess seen by ATLAS with 8~TeV data is not confirmed by the
13~TeV dataset, our shape-only spin comparisons are indicative of the
expected performance of different observables at the initial discovery
stage.  On the other hand, if data from two different $\sqrt{s}$
working points is available, then the expected scaling from changes in
parton distribution functions (PDFs) on various signal rates would be
an additional handle to discriminate between models.

Since we adopt the ATLAS 2~TeV diboson excess as our case study, we
first normalize the respective differential shapes to this excess.  In
a 300~GeV window centered at $m_{JJ} = 2$~TeV, the ATLAS collaboration
observed an excess of 8 events over an expected background of 8.94
events~\cite{Aad:2015owa}, where we quote the inclusive diboson
tagging requirements.  We use this normalization factor, our simulated
signal efficiencies, and our simulated PDF rescaling factors to
determine the expected number of signal events for each of the other
experimental analyses.  In the shape only comparisons, the test
hypothesis is always normalized to the null hypothesis.  The
corresponding background expectations, again for inclusive diboson
selection cuts, are gleaned from each ATLAS and CMS analysis, albeit
with slightly shifted mass windows around the $X$ mass.\footnote{We
  use the following invariant mass bins: $[1850, 2150]$~GeV for ATLAS
  at 8~TeV, 
  $[1800, 2200]$~GeV 
  for ATLAS at 13~TeV, and $[1852.3, 2136.4]$~GeV for CMS at 13~TeV.}
Since the current ATLAS 13~TeV analysis does not show event counts for
an inclusive diboson selection, we estimate the inclusive background
expectation from their available data, which we detail in
Appendix~\ref{app:ATLAS13TeV_bkgd}.

\begin{figure}[tb]
\begin{center}
\includegraphics[scale=0.79]{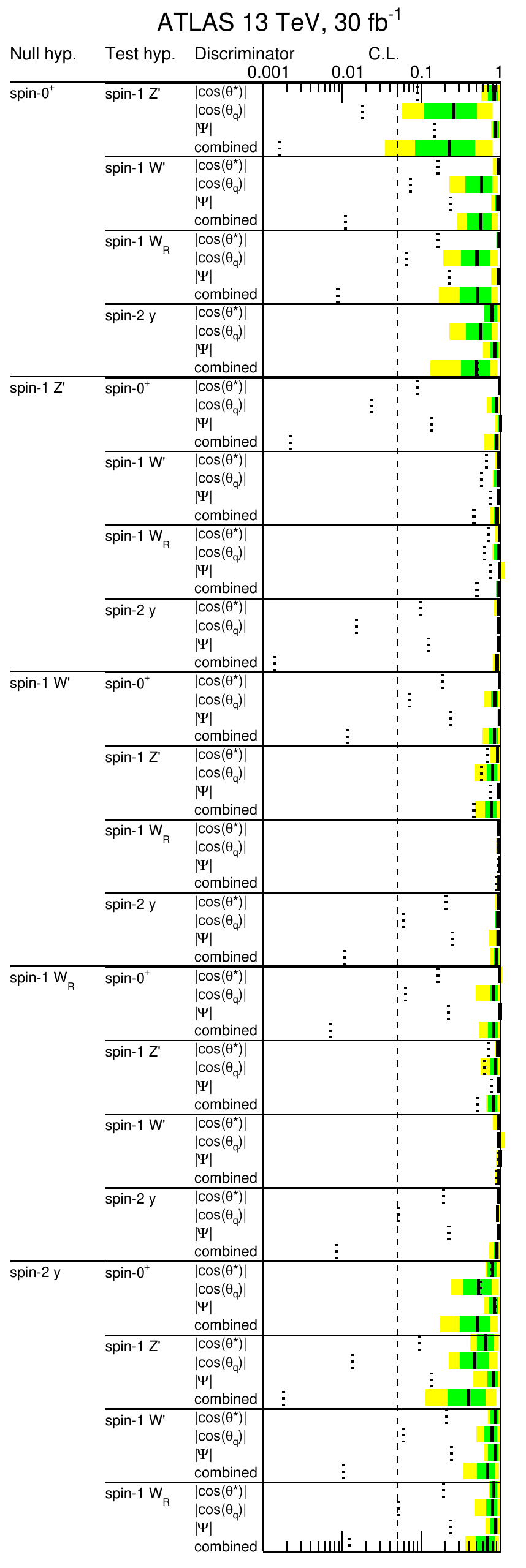}
\includegraphics[scale=0.79]{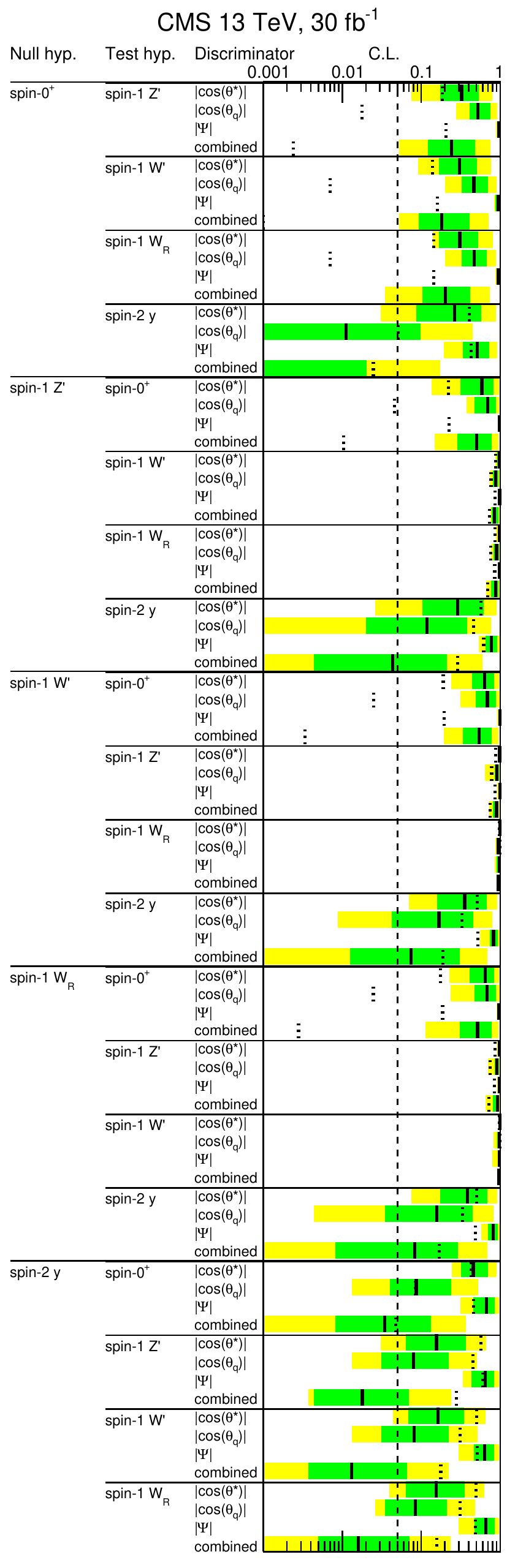}
\vspace{-0.2in}
\caption{Projected spin sensitivity for the 13~TeV ATLAS (left) and
  CMS (right) analyses with 30 fb$^{-1}$ integrated luminosity.  The
  long vertical dashed line indicates the 95\% exclusion C.L.  Within
  each row, the solid black line and the green and yellow shaded areas
  denote the central expected exclusion and the 68\% and 95\%
  likelihood expected exclusion intervals, using only shape
  information.  The dotted black line in each row shows the central
  expected exclusion limit including rate information, using the 2~TeV
  excess as the normalization of the respective signal hypotheses.}
\label{fig:Limits_13TeV_30fb}
\end{center}
\end{figure}

Not surprisingly, the current discrimination power between different
resonance spins is low given the small signal statistics of the 8~TeV
and 13~TeV analyses.  This situation is expected to dramatically
improve, however, with 30 fb$^{-1}$ luminosity of 13~TeV data.  In
Fig.~\ref{fig:Limits_13TeV_30fb}, we show the $CL_s$ values for a
given null hypothesis and various test hypotheses using the current
ATLAS 13~TeV and CMS 13~TeV analyses~\cite{ATLAS-CONF-2015-073,
  CMS-PAS-EXO-15-002} rescaled for 30 fb$^{-1}$ of luminosity.  We
assume a $25\%$ systematic uncertainty on the signal, and $30\%$ on
the dijet background.  In each row of each figure, the central
exclusion limit using only differential distributions is shown as a
solid black line, and the corresponding 68\% and 95\% expected
C.L.~exclusion limits are shown as the yellow and green bands.  The
dotted line in each row shows the shift in the central expected
exclusion limit if rate information is also added in the signal
hypothesis test.  These C.L.~results are not symmetric under
interchange of null hypothesis and test hypotheses, because in the
shapes-only $CL_s$ analysis, the test hypothesis is always scaled to
the null hypothesis, and thus the $S/B$ measure is not equal under the
interchange.  When rates are included, the Poisson errors are not
equal under interchange, and so the resulting C.L.~expectations are
again not equal.

We see that the most discrimination power comes between the spin-0 and
spin-1 cases vs.~the spin-2 case, which is expected from the clear
distinctions in angular correlations from
Fig.~\ref{fig:MCRecon_CosStar} for $\cos \theta^*$,
Fig.~\ref{fig:MCRecon_Cosq} for $\cos \theta_q$, as well as
Fig.~\ref{fig:MCRecon_Psi} for $\Psi$.  In particular, the $\cos
\theta_q$ observable provides significant discrimination, as the
spin-2 concavity in the reconstructed differential distribution is
opposite that of the spin-0 and spin-1 resonances.  We also remark
that the $\cos \theta_q$ observable has twice the statistical power of
the other $\cos \theta^*$ and $\Psi$ distributions because each event
gives two reconstructed vector boson candidates, and each vector boson
candidate contributes one entry to the $\cos \theta_q$ distribution.

We also see that CMS generally has stronger projected sensitivity than
ATLAS, which is a direct result of the different substructure analyses
employed by each experiment.  In particular, the ATLAS 13~TeV analysis
clusters large radius anti-$k_T$ jets with $R = 1.0$ and trims these
jets using a $k_T$ algorithm with $R = 0.2$ and hardness measure
$z_{\text{min}} = 0.05$.  We have seen from
Fig.~\ref{fig:Correlations} that the bulk of the quark pairs from $X
\to VV \to 4q$ decays lie within $\Delta R = 0.2$, which causes many
of the nominal subjets to be merged at the trimming stage.  
 
As a result, the efficiency for the ATLAS 13~TeV analysis to identify two
distinct subjets is significantly lower than the corresponding CMS
13~TeV analysis, causing the overall sensitivity to distinguishing
spin hypotheses to suffer.
\clearpage
The inclusion of rate information shows strong discrimination between
the spin-0 null hypothesis compared to the spin-1 hypotheses.  This
simply follows from the fact that our ad-hoc, gluon-fusion induced,
spin-0 diboson resonance enjoys a significant PDF rescaling factor
when going from 8~TeV to 13~TeV.  In contrast, the $qq'$-initiated
$Z'$, $W'$, and $W_R$ spin-1 signals are all largely indistinguishable
when only considering the $4q$ final state.  All of these spin-1
bosons couple to the SM electroweak bosons using the same tree-level
Lagrangian structure, which makes it very difficult to disentangle by
only considering the $4q$ excess.  The small sensitivity afforded by
shape and rate information in distinguishing a $Z'$ from a $W'$ or
$W_R$ explanation comes from the different PDF scaling from 8~TeV to
13~TeV between $qq'$ vs.~$q\bar{q}$ initial states.  We also note that
our signal and background events use inclusive $WW$, $WZ$, and $ZZ$
hadronic diboson tags, and thus additional sensitivity to $W'$ or
$W_R$ discrimination from a $Z'$ signal would come from separating
these diboson tagging categories.

In some cases, however, the inclusion of rate information decreases
the overall discrimination power between signal hypotheses.  This is
because the shapes-only test magnifies the importance of low event
count bins where the signal to background ratio is high, whereas the
shapes and rates test loses discrimination power by having an overall
lower significance for the given signals.  In particular, the linear
rescaling we use for matching the signal rates in the rates-only tests
overcomes the Poisson statistics governing the low-count bins that is
otherwise dominant in the rates and shapes test.

Overall, we see that the spin-2 signal hypothesis will be tested at
95\% C.L. using CMS 13~TeV cuts with 30 fb$^{-1}$ luminosity.  We also
project 95\% C.L. sensitivity between spin-0 and other spin scenarios
by combining rate information with the differential distributions.  If
a new diboson resonance appears, however, the shape information alone
from the current 13~TeV analyses would be insufficient to distinguish
spin-0 from spin-1 possibilities.

We conclude this section by discussing the possible improvements to
jet substructure analyses that could significantly help the prospects
of signal discrimination in a fully hadronic diboson final state.  We
have seen how the maximum $\Delta \eta_{JJ}$ cut introduces cliffs in
$\cos \theta^*$ that significantly cut away parts of phase space that
would tell a spin-1 signal from other possibilities.  Allowing a
looser $\Delta \eta_{JJ}$ cut, up to $\Delta \eta_{JJ} \leq 2.2$, for
example, would ensure that the extra sinuisoidal oscillation in the
spin-2 hypothesis would be more easily distinguished compared to the
spin-0 hypothesis and the dijet background, as seen in
Fig.~\ref{fig:MCRecon_CosStar}.  Although such a loose cut would lead
to an immense increase in multijet background, even intermediate
values of $\Delta \eta_{JJ} > 1.3$ would already aid discrimination
power between the different spin hypotheses.  We have also seen that
the minimum subjet $p_T$ balance requirement removes events above
$|\cos \theta_q| \approx 0.66$--$0.90$, depending on the
$y_{\text{min}}$ cut.  These events would have the best discrimination
power between spin-2 signals and other possibilities.

The most pernicious effect, however, comes from using a hard angular
scale, such as the $k_T$ reclustering with $R = 0.2$ inherent in the
trimming procedure used by ATLAS 13~TeV analysis.  This hard angular
scale not only causes distinct parton-level decays to merge into
single subjets, it also quashes the viability of a post-discovery
analysis that builds angular correlations from multiple subjets and
introduces significant sculpting effects in $\cos \theta^*$ and $\cos
\theta_q$ distributions.  For our 2~TeV case study, the efficiency to
find four distinct subjets would increase significantly if a smaller
reclustering radius of $R = 0.15$ were used, as seen in
Figure~\ref{fig:Correlations}, but the minimum radius for a given
resonance mass hypothesis with mass $m_X$ can be estimated from
$R_\text{min} \lesssim 2 m_{W/Z} / p_{T, X} \sim m_{W/Z} / m_{X}$.

A jet substructure method optimized for both signal discovery and
post-discovery signal discrimination would ameliorate these negative
effects.  The subjet $p_T$ balance requirement and alternate
reclustering methods that do not introduce a hard angular scale are
thus the most motivated details to modify for a spin-sensitive jet
substructure optimization.  We reserve a study to address these
questions for future work.

\section{Conclusion}
\label{sec:conclusion}

We have performed a comprehensive study of how angular correlations in
resonance decays to four quarks can be preserved, albeit distorted,
after effects from hadronization and showering, detector resolution,
jet clustering, and $W$ and $Z$ tagging via currently employed jet
substructure techniques.  We have connected the observed cliffs in
$\cos \theta^*$ to cuts on the maximum pseudorapidity difference
between the parent fat jets, the deficit of events around $\cos
\theta^*, \cos \theta_q \approx 0$ to the hard angular scale used in
the reclustering of subjets, and the removal of events above $\cos
\theta_q \approx 0.66$--$0.90$ to the subjet $p_T$ balance requirement
employed by the various analyses.  We have also emphasized the
importance of small angular scales for jet substructure reclustering,
having seen how large reclustering radii merge distinct decay products
of highly boosted vector parents and resulting sensitivity to spin
discrimination is greatly reduced.

We recognize that spin discrimination of a new resonance in diboson
decays is one facet of a possible post-discovery signal
characterization effort.  In particular, some of the degeneracies
among the various spin-1 signal hypotheses can only be distinguished
by observing semi-leptonic diboson decays as well as additional direct
decays to fermions.  The rates for the latter decays are model
dependent features of each given signal hypothesis.  In the special
case of the 2~TeV excess seen by ATLAS in 8~TeV data, additional
discrimination power between possible new physics resonances is
afforded by the simple fact that the LHC is now operating at 13~TeV.
The different production modes for spin-0, spin-1 neutral, spin-1
charged, and spin-2 resonances obviously scale differently going from
$\sqrt{s} = 8$ TeV to $\sqrt{s} = 13$ TeV, which establishes benchmark
expected significances for the different signals as a function of
luminosity. 

Our work, however, addresses the more general question about the
feasibility of using an analysis targetting a resonance in a fully
hadronic diboson decay for spin and parity discrimination.  It also
provides a method for distinguishing longitudinal versus transverse
polarizations of electroweak gauge bosons, which is an intrinsic
element of analyses aimed at probing unitarity of electroweak boson
scattering.  A future work will tackle the question of an optimized
jet substructure analysis that avoids introducing significant
distortions in angular observables and hence enhances the possible
spin sensitivity beyond the projections shown in
Fig.~\ref{fig:Limits_13TeV_30fb}.  We also plan to investigate angular
correlations in fully hadronic final states with intermediate new
physics resonances, as well as the viability of angular observables
using Higgs and top substructure methods.  Even without any
improvement, a spin-2 explanation for the 2~TeV excess will be tested
at the 95\% C.L. from other spin hypotheses with 30 fb$^{-1}$ of
13~TeV luminosity using only shape information, while spin-0
vs.~spin-1 discrimination would come from the combination of rate and
shape information.

\section*{Acknowledgments}
\label{sec:acknowledgments}

We would like to thank Michael Baker, Ian Lewis, Adam Martin, Jesse
Thaler, Andrea Thamm, Nhan Tran, and Yuhsin Tsai, for useful
discussions, and Riccardo Torre, Andrea Thamm for use of the Heavy
Vector Triplets FeynRules model and Bogdan Dobrescu and Patrick Fox
for use of the right-handed $W_R$ model. This research is supported by
the Cluster of Excellence Precision Physics, Fundamental Interactions
and Structure of Matter (PRISMA-EXC 1098).  The work of MB is moreover
supported by the German Research Foundation (DFG) in the framework of
the Research Unit New Physics at the Large Hadron Collider" (FOR
2239).

\begin{appendix}
\section{ATLAS 13~TeV background extraction, inclusive diboson selection}
\label{app:ATLAS13TeV_bkgd}
For our projections on spin sensitivity at 13~TeV LHC, we require the
background estimate for inclusive diboson selection cuts.  As the
current ATLAS 13~TeV analysis~\cite{ATLAS-CONF-2015-073} only provides
$WW$, $WZ$, and $ZZ$ event counts, which are not exclusive selection
bins because of overlapping $W$ and $Z$ mass windows, we extract the
inclusive number of events as follows.

For the mass range $1.0\text{ TeV} < m_{JJ} < 2.5$~TeV, the ATLAS
analysis specifies that 38 events lie in the overlap region and
contribute to all three channels.  We thus assign $p \approx
\sqrt{38/N}$ as a flat probability for an event with a $W$-tag to also
be a $Z$-tagged event and vice versa, where $N$ is the number of
events passing inclusive diboson tagging requirements.  We can write
$N = N_{W^0Z^0} + N_{W^0W^0} + N_{Z^0Z^0}$, where each category is
defined exclusively and without overlap.  Then,
\begin{align}
N &= N_{WZ}+N_{WW}+N_{ZZ}\notag\\
  &- N_{W^0Z^0}\cdot\left[\mathcal{P}(Z\text{ in overlap region})+
                         \mathcal{P}(W\text{ in overlap region})+
		         2 \mathcal{P}(W\text{ and }Z\text{ in overlap region})
		         \right]\notag\\
  &- N_{W^0W^0}\cdot\left[\mathcal{P}(\text{one }W\text{ in overlap region})+
                         2\mathcal{P}(\text{both }W\text{ in overlap region})
                         \right]\notag\\
  &- N_{Z^0Z^0}\cdot\left[\mathcal{P}(\text{one }Z\text{ in overlap region})+
                         2\mathcal{P}(\text{both }Z\text{ in overlap region})
                         \right] \ ,
\label{eq:N1}
\end{align}
where factors of 2 in Eq.~\ref{eq:N1} reflect the fact that this
particular event contributes to all three categories and therefore two
events need to be subtracted from the total sum.  From the ATLAS
analysis~\cite{ATLAS-CONF-2015-073}, we have
$N_{WZ} + N_{WW} + N_{ZZ} = 300$, thus
\begin{align}
  N = 300 &- N_{W^0Z^0} \left[ p(1-p)+p(1-p)+2p^2 \right] \notag\\
          &- N_{W^0W^0} \left[ 2p(1-p)+2p^2 \right] \notag\\
          &- N_{Z^0Z^0} \left[ 2p(1-p)+2p^2 \right] \notag\\
    = 300 &- 2Np \ .
\end{align}
Using $p \approx \sqrt{38 / N}$, and solving for $N$, we obtain $N
\approx 149$, and thus 75 events fall into two diboson categories and
38 events are triply counted, which is very similar to the breakdown
of double and triple counted events in the ATLAS 8~TeV
analysis~\cite{Aad:2015owa}.  We use this fraction to estimate the
expected number of background events passing the inclusive diboson
tagging requirements.
\end{appendix}

\bibliographystyle{apsrev4-1}
\bibliography{referencelist}

\end{document}